\documentclass[a4paper,11pt]{article}
\pdfoutput=1 
\usepackage{jheppub} 
\usepackage[T1]{fontenc} 
\usepackage[numbers,sort&compress]{natbib}
\usepackage{graphicx}  
\usepackage{float}
\usepackage{dcolumn}   
\usepackage{bm}        
\usepackage{amssymb}   
\usepackage{slashed}   
\usepackage{amsmath}   
\usepackage{verbatim}  
\usepackage{hyperref}
\usepackage{multicol}
\usepackage{multirow}
\usepackage{booktabs}
\usepackage{subfigure}
\usepackage{subcaption}
\usepackage{color} 
\usepackage{float}

\definecolor{Blue}{rgb}{0, 0.1, 0.5}

\newcounter{GYC}

\newcounter{RMQ}

\graphicspath{{Figures/}}
\begin{document}
\title{Search for Neutral Triple Gauge Couplings with $ZZ$ Production at Future Electron Positron Colliders}
\author[a,b]{Yu-Chen Guo}
\author[a]{Chun-Jing Pan}
\author[c,d]{Man-Qi Ruan}
\author[a,b,1]{Ji-Chong Yang,\note{Corresponding author.}}


\affiliation[a]{Department of Physics, Liaoning Normal University, No. 850 Huanghe Road, Dalian 116029, China}
\affiliation[b]{Center for Theoretical and Experimental High Energy Physics, Liaoning Normal University, No. 850 Huanghe Road, Dalian 116029, China}
\affiliation[c]{
Institute of High Energy Physics, Chinese Academy of Sciences, 19B Yuquan Road, Shijingshan District, Beijing 100049, China}
\affiliation[d]{University of Chinese Academy of Sciences, 19A Yuquan Road, Shijingshan District, Beijing 100049, China}

\emailAdd{ycguo@lnnu.edu.cn}
\emailAdd{panchunjing2022@163.com}
\emailAdd{ruanmq@ihep.ac.cn}
\emailAdd{yangjichong@lnnu.edu.cn}

\abstract{
This study investigates Neutral Triple Gauge Couplings (nTGCs) through $ZZ$ production at future electron-positron colliders.
The impact of beam polarization on cross section is analyzed.
We compare the signals and backgrounds for five different $ZZ$ decay channels and present our event selection strategies for future $e^+e^-$ colliders.
The expected coefficient constraints for each decay channels are provided, and final expected constraints are derived by combining results from the different decay patterns.
Our analysis indicates that future electron-positron colliders will have significantly enhanced detection capabilities for nTGCs compared to the current LHC experiments, with expected improvements in constraints by one to two orders of magnitude.
}
\maketitle
\flushbottom

\section{Introduction}
\label{sec:intro}
In the absence of clear signals of new physics (NP) beyond the Standard Model (SM) from the Large Hadron Collider (LHC) experiments so far, the indirect detection of signs of new physics through precise measurements of SM processes has gained more attention.
When the mass of a new particle is significantly larger than the electroweak symmetry breaking scale, the Standard Model Effective Field Theory (SMEFT) provides a systematic parameterization of NP effects~\cite{weinberg,SMEFTReview1,SMEFTReview2,SMEFTReview3,SMEFTReview4}, with the advantage of not relying on any particular model.
The Lagrangian of SMEFT consists of operators of dimension-4, representing the SM itself, as well as higher-dimensional operators constructed from SM fields.
The Lagrangian of SMEFT then takes the form
\[\mathcal{L}_{\rm SMEFT}=\mathcal{L}_{\rm{SM}}+\sum^\infty_{d=5}\sum _j\frac{C^{(d)}_{j}}{\Lambda^{d-4}}\mathcal{O}^{(d)}_j,
\label{eq.1.1}
\]
where $\mathcal{O}^{(d)}$ is an operator of dimension $d$ invariant under the $SU(3)_c\times SU(2)_L\times U(1)_Y$ gauge group of the SM, $\Lambda$ represents the mass scale of new particles, with the power of $\Lambda$ in each term determined by dimensional analysis, and $C^{(d)}_{j}$ is the Wilson coefficient of the operator.
For any new physics model that satisfies the above assumptions, integrating out the new fields will produce a Lagrangian with this structure.

Depending on the gauge symmetry, triple gauge couplings (TGC) and quartic gauge couplings (QGC) exist in the SM.
Measurements of these couplings not only test the nature of spontaneous electroweak symmetry breaking (EWSB) but also provide a unique window for searching for new physics.
Anomalous gauge couplings beyond the SM do not appear in the dimension-4 terms of the SM Lagrangian, and first arise in the dimension-6 terms of SMEFT.
Neutral triple gauge coupling (nTGC)~\cite{ntgc1,ntgcfcchh,ntgc2,ntgc3} and the gluonic quadratic gauge coupling (gQGC)~\cite{gQGC2018} all arise first in gauge-invariant dimension-8 operators.
Any sign of their detection would provide direct prima facie evidence for the existence of new physics.
Additionally, it has been shown that dimension-8 operators play a crucial role from the convex geometry perspective within the SMEFT space \cite{positivity1}.
Many researchers have emphasised the importance of dimension-8 operators \cite{d81,looportree,ssww,aqgcold,aqgcnew,positivity2,positivity3}.

In general, the contribution of dimension-6 operators to new physics is larger than that of dimension-8 or higher operators.
Consequently, there are fewer experimental probes and constraint studies on dimension-8 SMEFT interactions.
However, in some cases, dimension-8 operators can make leading contributions to NP processes.
Examples include dimension-8 operators contributing to light-by-light scattering \cite{Ellis:2017edi}, vector boson scattering (VBS)~\cite{Guo:2019agy,Guo:2020lim,Jiang:2021ytz,Yang:2021pcf,Yang:2021ukg,Yang:2022fhw,Ari:2021rmx,Yang:2023gos,Zhang:2023khv}, di-boson productions~\cite{Ellis:2018cos,Ellis:2021dfa,Ellis:2019zex,Ellis:2020ljj,Ellis:2022zdw,Ellis:2023ucy,Liu:2024tcz,Fu:2021mub,Yang:2021kyy}, and tri-boson productions~\cite{Yang:2020rjt,Dong:2023nir,Zhang:2023yfg,Zhang:2023ykh}.
Many experimental measurements have been made at the LHC~\cite{sswwexp1,sswwexp2,zaexp1,zaexp2,zaexp3,waexp1,zzexp1,zzexp2,wzexp1,wzexp2,wwexp1,wwexp2,wvzvexp,waexp2,zzexp3}.

At dimension-8, there are four CP-conserving operators contributing to nTGCs, they are~\cite{ntgc1,ntgc2}
\begin{equation}
\begin{split}
\mathcal{L}_{\rm nTGC}=&\frac{{\rm sign}(c_{\tilde{B}W})}{\Lambda_{\tilde{B}W}^4}\mathcal{O}_{\tilde{B}W}+\frac{{\rm sign}(c_{B\tilde{W}})}{\Lambda_{B\tilde{W}}^4}\mathcal{O}_{B\tilde{W}} \\
&+\frac{{\rm sign}(c_{W\tilde{W}})}{\Lambda_{\tilde{W}W}^4}\mathcal{O}_{\tilde{W}W}+\frac{{\rm sign}(c_{B\tilde{B}})}{\Lambda_{\tilde{B}B}^4}\mathcal{O}_{\tilde{B}B},\\
\end{split}
\label{eq.1.2}
\end{equation}
with
\begin{equation}
\begin{split}
&\mathcal{O}_{\tilde{B}W}=i H^{+}\tilde{B}_{\mu \nu} W^{\mu \rho} \left\{D_{\rho },D^{\nu }\right\}H+h.c.,\;\;\\
&\mathcal{O}_{B\tilde{W}}=i H^{+} B_{\mu \nu} \tilde{W}^{\mu \rho} \left\{D_{\rho },D^{\nu }\right\}H+h.c.,\\
&\mathcal{O}_{\tilde{W}W}=i H^{+}\tilde{W}_{\mu \nu} W^{\mu \rho} \left\{D_{\rho },D^{\nu }\right\}H+h.c.,\;\;\\
&\mathcal{O}_{\tilde{B}B}=i H^{+}\tilde{B}_{\mu \nu} B^{\mu \rho} \left\{D_{\rho },D^{\nu }\right\}H+h.c.,\\
\end{split}
\label{eq.1.3}
\end{equation}
where $\tilde{B}_{\mu \nu}\equiv \epsilon_{\mu\nu\alpha\beta} B^{\alpha\beta}$, $\tilde{W}_{\mu \nu}\equiv \epsilon_{\mu\nu\alpha\beta} W^{\alpha\beta}$ and $W_{\mu\nu}\equiv W_{\mu\nu}^a \sigma^{a}/2$ where $\sigma^a$ are Pauli matrices.
For simplicity, we define $f_X\equiv {\rm sign}(c_X)/\Lambda ^4_X$.

On the theoretical side, anomalous tensorial structures can be generated within a renormalizable theory at higher loop orders. For example, these structures can arise through a fermion triangle diagram in the Standard Model (SM). Loop-generated anomalous couplings have been studied in the Minimal Supersymmetric Standard Model (MSSM)~\cite{Gounaris:2000tb,Choudhury:2000bw,Yue:2021snv} and the Little Higgs model~\cite{Dutta:2009nf}.
A non-commutative extension of the SM (NCSM)~\cite{Deshpande:2001mu,Deshpande:2011uk} can also provide an anomalous coupling structure in the neutral sector, with the possibility of a trilinear $\gamma\gamma\gamma$ coupling~\cite{Deshpande:2001mu}.
Additionally, the effective couplings $ZZZ$ and $ZZ\gamma$ can be observed as deviations from the SM prediction in the measured $Z$-pair cross section, which have been proposed in the context of two-Higgs-doublet models~\cite{2HDM,Yang:2015hsa} and in low scale gravity theories~\cite{Agashe:1999qp}.

Future $e^+e^-$ colliders have great physics potential and provide one of the rare direct windows to NP at dimension-8.
Previous studies have thoroughly studied probing nTGCs by measuring $Z\gamma$ production at future $e^+e^-$ colliders \cite{Ananthanarayan:2011fr,Ananthanarayan:2014sea,Ellis:2019zex,Ellis:2020ljj,Fu:2021mub,Yang:2021kyy,Spor:2022zob,Liu:2024tcz} and $pp$ colliders \cite{SENOL2018365,Ellis:2022zdw,Ellis:2023ucy}.
In this work, we study how nTGCs can be probed by measuring the $e^+e^-\to ZZ$ process with decays at different versions of circular and linear $e^+e^-$ colliders, including the CEPC~\cite{CEPC}, ILC~\cite{ILC}, and CLIC~\cite{CLIC}.

The plan of this paper is as follows.
In Section \ref{sec2}, we discuss the beam polarization effects on cross section of $ZZ$ production at $e^+e^-$ colliders.
In Section \ref{sec3}, we compare the signals and backgrounds of five $ZZ$ decay channels and present our event selection strategy for $e^+e^-$ colliders with different energies and luminosities.
The expected constraints on the coefficients with combined results are also presented in Section \ref{sec3}.
Finally, we summarize our conclusions in Section \ref{sec4}.
We compare the results obtained with experimental measurements of nTGCs from the LHC~\cite{ATLAS:2018nci}. Depending on the capabilities of different future lepton colliders to detect nTGCs, their sensitivity could exceed that of the current LHC.
The $e^+e^-\to ZZ$ process will thus provide a unique and promising channel for the detection of new physics at future lepton colliders.

\section{\texorpdfstring{$ZZ$}{ZZ} production at \texorpdfstring{$e^+e^-$}{ee} colliders}
\label{sec2}

At future electron-positron colliders, the di-boson pair production can be an interesting topic and ideal motivation.
In this Section, we first discuss the cross section of the $e^+e^- \to ZZ$ process induced by nTGCs and the contribution of the Standard Model to this process.
We then analyze the influence of initial electron polarization on the cross section.

\begin{figure}[htbp]
\centering
\includegraphics[height=0.19\hsize]{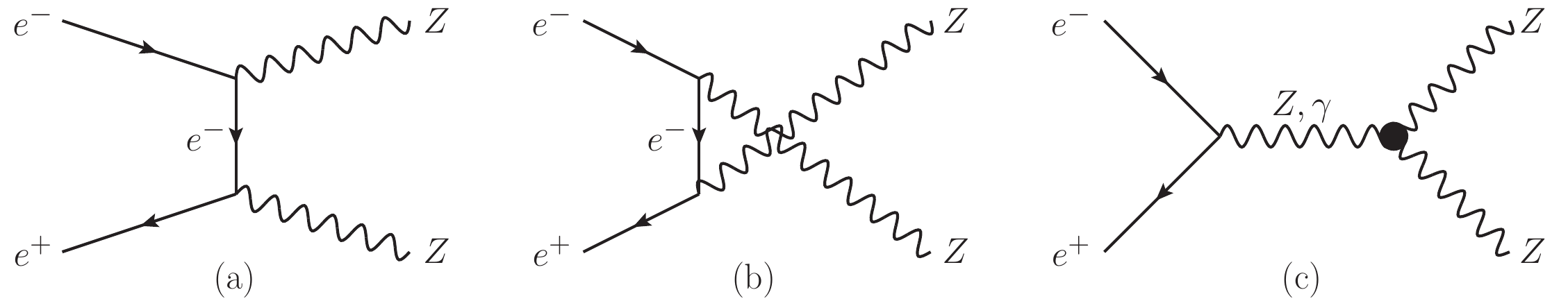}
\caption{Feynman diagrams of the \(e^+e^-\to ZZ\) production.
Diagrams (a) and (b) are the SM contributions. Diagram (c) corresponds to contributions from the nTGCs.}
\label{fig:2-1}
\end{figure}

In the SM, the process $e^+e^- \to ZZ$  proceeds via the $t$- and $u$-channel exchange diagrams at tree level.
The presence of nTGCs would lead to a sizable enhancement of $ZZ$ final states via $s$-channel \(e^+e^-\) scattering.
The Feynman diagrams are shown in Fig.~\ref{fig:2-1}.

The process of $e^+e^- \to ZZ$ can be affected by dimension-8 operators through the $ZZZ$ and $ZZ\gamma$ couplings.
In the case of $ZZV$ coupling with the $Z$ boson and $\gamma$ on-shell, there is only one independent operator because the $\mathcal{O}_{B\tilde{W}}$ operator is equivalent to the $\mathcal{O}_{\tilde{B}W}$ operator, and the $\mathcal{O}_{\tilde{W}W}$ and $\mathcal{O}_{\tilde{B}B}$ operators do not contribute~\cite{ntgc2}.
We focus on the $\mathcal{O}_{\tilde{B}W}$ operator for this nTGC study.
At the leading order of $s$, the analytical cross sections are given by

\begin{equation}
\begin{split}
&\sigma_{\rm{nTGC}}(ZZ)=\frac{f_{\tilde{B}W}^2c_W^2 M_Z^2 s_W^2 \left(s-4 M_Z^2\right)^{\frac{5}{2}}}{24 \pi \sqrt{s}} ,\\
&\sigma_{\rm{int}}(ZZ)=\frac{f_{\tilde{B}W}e^2 M_Z^2 \left(4 s_W^2-1\right) \sqrt{s-4 M_Z^2}}
{32 \pi  c_W s^\frac{3}{2} s_W \sqrt{s \left(s-4 M_Z^2\right)}}\\
&\quad \quad \times\left[\left(2 M_Z^2+s\right) \sqrt{s \left(s-4 M_Z^2\right)}-4 M_Z^2 \left(s-M_Z^2\right)  \tanh ^{-1}\left(\frac{\sqrt{s \left(s-4 M_Z^2\right)}}{s-2 M_Z^2}\right)\right],\\
&\sigma_{\rm{SM}}(ZZ)=\frac{e^4 \sqrt{s-4 M_Z^2}\left(32 c_W^8-96 c_W^6+120 c_W^4-72 c_W^2+17\right)}{128 \pi  c_W^4 s^2 {s^4_W} \left(2 M_Z^2-s\right) \sqrt{s-4 M_Z^2}} \\
&\quad \quad\times \left[\sqrt{s} \left(s-2 M_Z^2\right) \sqrt{s-4 M_Z^2}-\left(4 M_Z^4+s^2\right) \tanh ^{-1}\left(\frac{\sqrt{s} \sqrt{s-4 M_Z^2}}{s-2 M_Z^2}\right)\right].
\end{split}
\label{eq.2.1}
\end{equation}

The anomalous $ZZZ$ and $ZZ\gamma$ couplings are allowed by electromagnetic gauge invariance and Lorentz invariance for on-shell $Z$ and $\gamma$ bosons.
The $ZZ\gamma$ coupling is absent at tree level and is highly suppressed when allowed by internal particle loops in the SM, thus forbidding the $s$-channel production of $ZZ$. Therefore, any deviation from the tree-level SM predictions will indicate the presence of beyond SM (BSM) physics. Although the interference of the dimension-8 operator with the SM contributes to the amplitude at $\mathcal{O}(1/\Lambda^4)$, the quadratic contribution of the dimension-6 operators does not need to be considered for the process of $e^+e^- \to ZZ$, as no dimension-6 operator contributes to the nTGCs.

The scattering process at the positron collider can be enhanced or reduced by the initial beam polarization. The hard bremsstrahlung cross section for polarized $e^+e^- \to ZZ$ scattering can be obtained as,
\begin{equation}
\begin{split}
&\sigma_{\rm{pol}}^{\rm{SM}}(ZZ)=-\left(16 s_W^8 (P_{e^+}+1) (P_{e^-}-1)+\left(1-2 c_W^2\right)^4 (P_{e^+}-1) (P_{e^-}+1)\right)\\
&\quad\quad\times\frac{e^4 \left(\sqrt{s} \left(s-2 M_Z^2\right) \sqrt{s-4 M_Z^2} - \left(4 M_Z^4+s^2\right) \tanh ^{-1}\left(\frac{\sqrt{s} \sqrt{s-4 M_Z^2}}{s- 2 M_Z^2}\right)\right)}{128 \pi  c_W^4 s^2 s_W^4 \left(2 M_Z^2-s\right)},\\
&\sigma_{\rm{pol}}^{\rm{int}}(ZZ)=\left(4 \left(c_W^2-1\right) (P_{e^+}+1) (P_{e^-}-1) s_W^2+\left(1-2 c_W^2\right)^2 (P_{e^+}-1) (P_{e^-}+1)\right)\\
&\quad\quad\times\frac{f_{\tilde{B}W} e^2 M_Z^2 \left(\sqrt{s} \left(2 M_Z^2+s\right) \sqrt{s-4 M_Z^2}-4 M_Z^2 \left(s-M_Z^2\right) \tanh ^{-1}\left(\frac{\sqrt{s} \sqrt{s-4 M_Z^2}}{s-2 M_Z^2}\right)\right)}{32 \pi c_W s^2 s_W },\\
&\sigma_{\rm{pol}}^{\rm{nTGC}}(ZZ)=-\frac{f_{\tilde{B}W}^2 c_W^2 M_Z^2 s_W^2 \left(s-4 M_Z^2\right)^\frac{5}{2} (P_{e^-}P_{e^+}-1)}{24 \pi \sqrt{s}}.\\
\end{split}
\label{eq.2.3}
\end{equation}
where $P_{e^-}$ and $P_{e^+}$ denote the polarization parameter of electron and positron respectively, ranging from $-1$ to $+1$, with $+1$ ($-1$) for the left (right) handed helicity.

\begin{table}[H]
\centering	
\caption{
Comparison of the hard bremsstrahlung cross section (fb) of the SM and nTGCs contributions for 100\% polarized $e^- e^+\to Z Z$ scattering at the characteristic c.m. energies of CEPC, ILC, and CLIC.
\label{polcs}}
\begin{tabular}{lcccccc}
	\hline
	\hline
	$P_{e^-},P_{e^+}$  &	  & 0, 0   &  $-1$, $-1$  & $-1$, 1  & 1, $-1$ & 1, 1\\
	\hline
	$\sqrt{s}=250$ GeV & nTGCs & $4.50\times 10^{-3}$ & 0 & $8.99\times10^{-3}$ & $8.99\times10^{-3}$ & 0\\
 	                   &  SM  & 1050 & 0 & 2637 & 1564 & 0\\
 	\hline
        $\sqrt{s}=360$ GeV & nTGCs & 0.06 & 0 & 0.12 & 0.12 & 0\\
 	                   &  SM  & 627 & 0 & 1574 & 933 & 0\\
        \hline
	$\sqrt{s}=500$ GeV & nTGCs & $0.336$ & 0 & $0.673$ & $0.673$ & 0\\
 	                   &  SM  & 397 & 0 & 996 & 591 & 0\\
	\hline
	$\sqrt{s}=1$ TeV & nTGCs  & $7.07$ & 0 & $14.1$ & $14.1$ & 0\\
 	                    &  SM   & 145 & 0 & 363 & 215 & 0\\
	\hline
    $\sqrt{s}=3$ TeV & nTGCs  & 617 & 0 & 1234 & 1234 & 0\\
 	                    &  SM   & $24.9$ & 0 & $62.5$ & $37.1$ & 0\\
	\hline
\end{tabular}
\end{table}

Numerical results are evaluated for two energy points and the following helicity fraction of the electron ($P_{e^-}$) and positron ($P_{e^+}$)
beam polarization:
\begin{eqnarray}
(P_{e^-},P_{e^+})=
(0,0),(-1,-1),(-1,+1),(+1,-1),(+1,+1).
\label{SetPolarization}
\end{eqnarray}
It is convenient to make a comparison of the cross sections induced by the SM couplings and nTGCs with different values of polarization in Table \ref{polcs}.
The results are calculated with $f_{\tilde{B}W} = 1 \ (\rm{TeV}^{-4})$ as the characteristic parameter.
The $e^+e^-$ colliders will be constructed in several centre-of-mass (c.m.) energy stages, 250 GeV, 360 GeV, 500 GeV, 1 TeV and 3 TeV for two stages of CEPC and ILC, as well as CLIC, respectively.
For each energy, the cross section of contributions of nTGCs $\sigma_{--,++}$ is 0, and $\sigma_{+-}$ or $\sigma_{-+}$ is all about 2 times larger than the unpolarized cross section. For the SM contributions, $\sigma_{+-}$ is about 1.5 times larger than the unpolarized cross section.

\begin{figure}[H]
\centering
\subfigure[250 GeV]{\includegraphics[height=0.35\hsize]{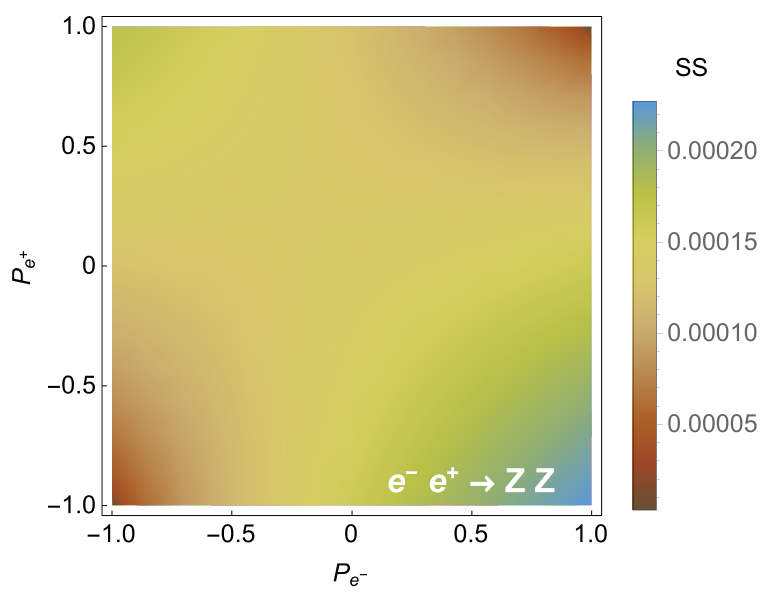}}
\subfigure[360 GeV]{\includegraphics[height=0.35\hsize]{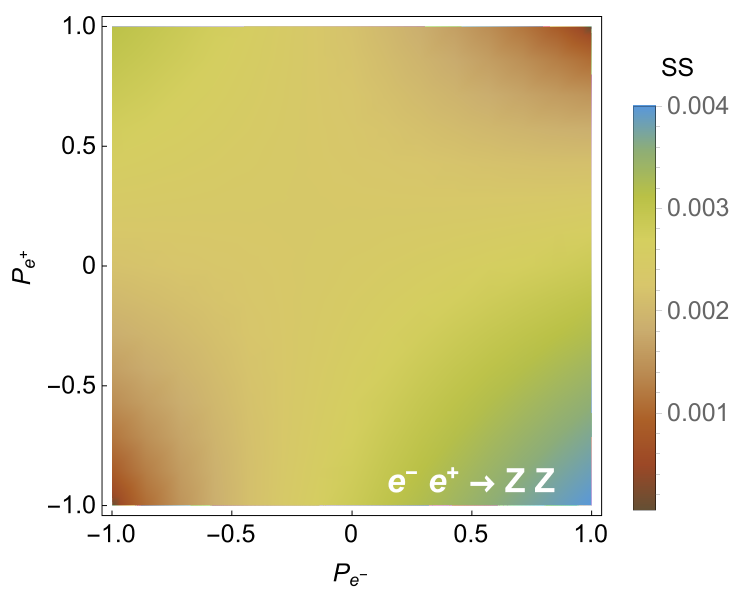}}
\subfigure[500 GeV]{\includegraphics[height=0.35\hsize]{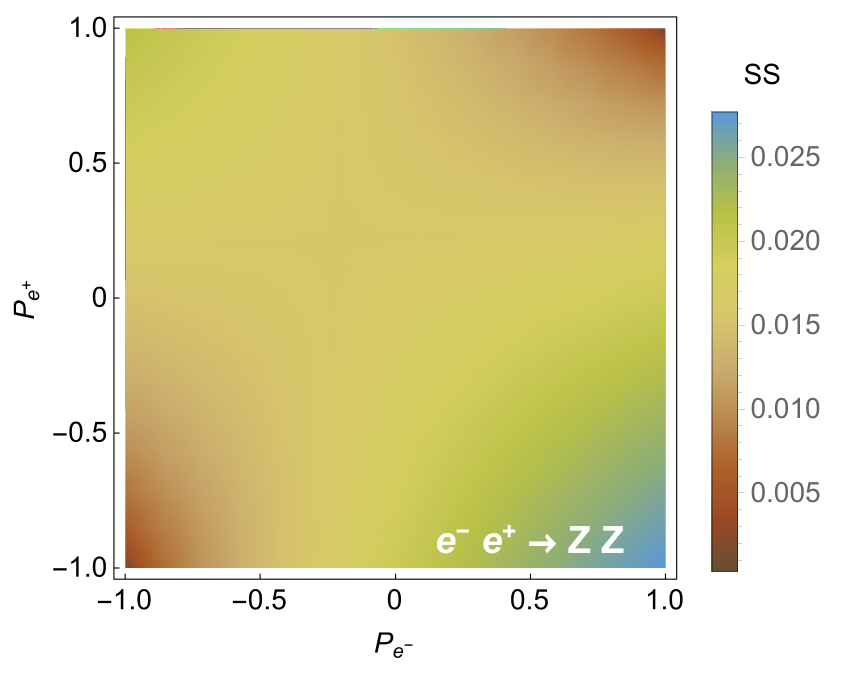}}
\subfigure[1 TeV]{\includegraphics[height=0.35\hsize]{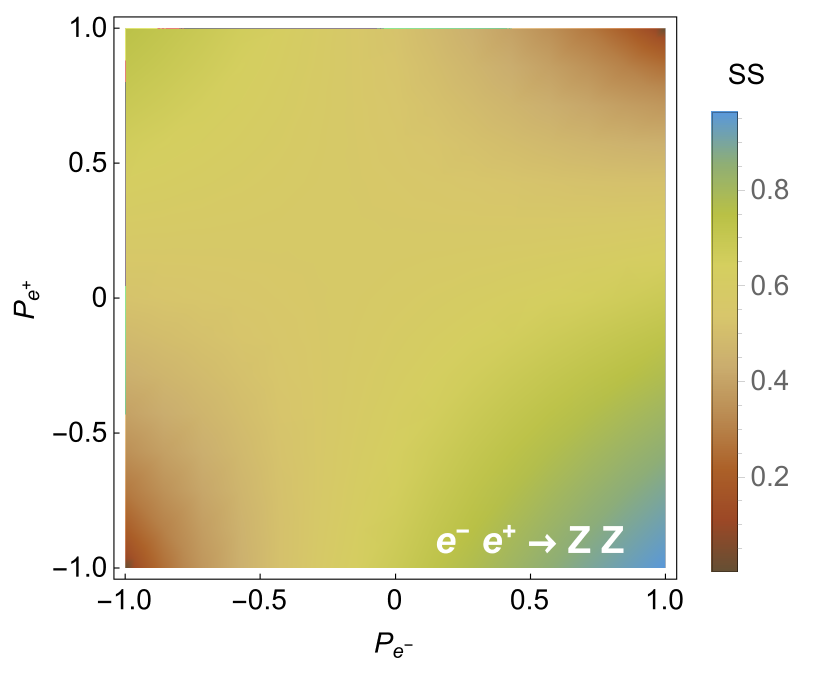}}
\subfigure[3 TeV]{\includegraphics[height=0.35\hsize]{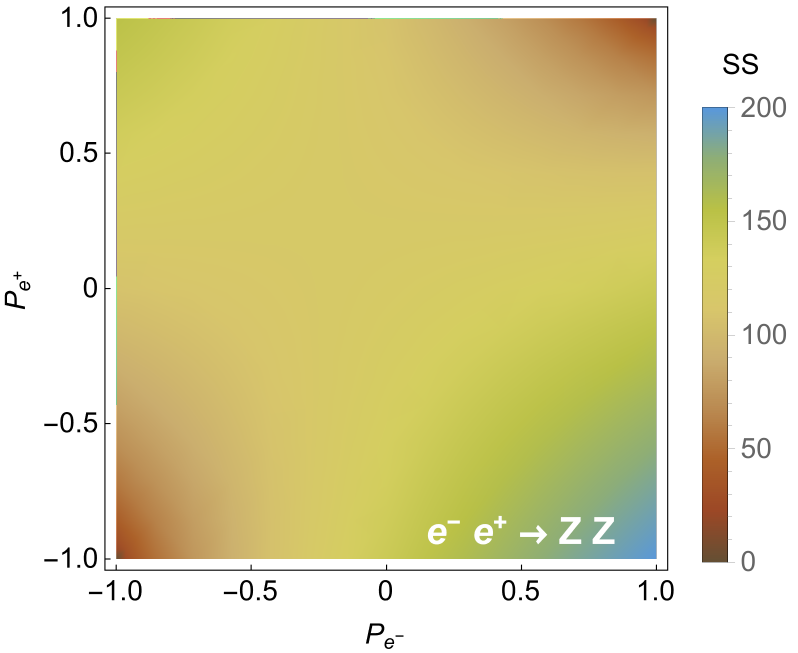}}
\caption{The statistical significance with $f_{\tilde{B}W} = 1 \ (\rm{TeV}^{-4})$ and different beam polarization parameter for $\sqrt{s}=$ 250 GeV, 360 GeV, 500 GeV, 1 TeV and 3 TeV, respectively.}
\label{fig:2-2}
\end{figure}

Figure \ref{fig:2-2} shows the effects of initial beam polarization on cross sections at the various proposed \(e^+e^-\) colliders, based on a simple signal significance analysis with
\begin{eqnarray}
S_{stat}(ZZ)=\frac{\sigma^{\rm nTGC}_{\rm pol}+\sigma^{\rm int}_{\rm pol}}{\sqrt{\sigma^{\rm nTGC}_{\rm pol}+\sigma^{\rm int}_{\rm pol}+\sigma^{\rm SM}_{\rm pol}}}.
\end{eqnarray}
It shows that beam polarization can improve significance by a factor of $1.4\sim2$ compared to the unpolarized process at the electron-positron colliders with a range of c.m. energy from 250 GeV to 3 TeV.
The results have demonstrated the significance of polarization effects.

\section{Probing nTGCs via \texorpdfstring{$ZZ$}{ZZ} production at the future \texorpdfstring{$e^+e^-$}{e+e-} colliders}
\label{sec3}
$ZZ$ production processes can be further divided into five decay modes in the final states. In this analysis we study the following subprocesses:
\begin{equation}
e^+e^-\to Z_1(\ell^+\ell^-)Z_2(\ell^+\ell^-),
\end{equation}
\begin{equation}
e^+e^-\to Z_1(jj)Z_2(jj),
\end{equation}
\begin{equation}
e^+e^-\to Z_1(\ell^+\ell^-)Z_2(jj),
\end{equation}
\begin{equation}
e^+e^-\to Z_1(\ell^+\ell^-)Z_2(\nu\bar{\nu}),
\end{equation}
\begin{equation}
e^+e^-\to Z_1(jj)Z_2(\nu\bar{\nu}),
\end{equation}
where $\ell$ is either $e$ or $\mu$. All processes are generated using \verb"MadGraph5_aMC@NLO" toolkit~\cite{madgraph,feynrules} with the LHCO output format; hadronization is performed by \verb"PYTHIA8" \cite{pythia}.
A fast detector simulation is applied using \verb"Delphes" \cite{delphes} with the CEPC, ILC and two stages of CLIC collider cards. Finally, the events are reconstructed with the \verb"MLAnalysis"~\cite{Guo:2023nfu} package.
All events simulated at future $e^+e^-$ colliders will use an electron positron beam longitudinal polarisation of (+0.8, -0.8).
Table~\ref{tab:luminosities} shows the integrated luminosities $\mathcal{L}$ for the CEPC, ILC and CLIC at different stages of  $\sqrt{s}$.

\begin{table}
\centering
\caption{Summary of center-of-mass energies ($\sqrt{s}$) and corresponding integrated luminosities ($\mathcal{L}$) for different colliders.}
\begin{tabular}{|c|c|c|}
\hline
Collider & Center-of-Mass Energy ($\sqrt{s}$) & Integrated Luminosity ($\mathcal{L}$) \\ \hline
CEPC             & 250 GeV                                  & 5 $\rm{ab^{-1}}$                                 \\ \hline
CEPC             & 360 GeV                                  & 1 $\rm{ab^{-1}}$                                 \\ \hline
ILC              & 500 GeV                                  & 500 $\rm{fb^{-1}}$                               \\ \hline
ILC             & 1 TeV                                    & 1 $\rm{ab^{-1}}$                                 \\ \hline
CLIC             & 3 TeV                                    & 5 $\rm{ab^{-1}}$                                 \\ \hline
\end{tabular}
\label{tab:luminosities}
\end{table}

\subsection{Selection of \texorpdfstring{$e^+e^-\to 2\ell2\ell'$}{4l}}
\label{sec3.1}

\begin{figure}[H]
\centering
\includegraphics[height=0.19\hsize]{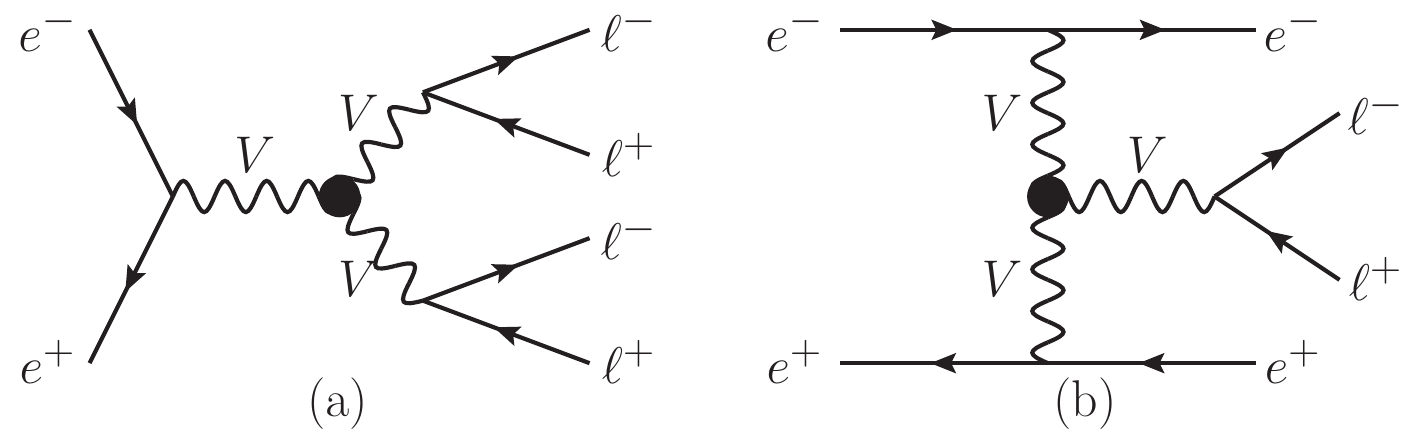}
\caption{The Feynman diagrams of the nTGC contributions for the process $e^+e^-\to 2\ell 2\ell'$, where $VVV$ stands for $ZZZ$, $ZZ\gamma$ or $Z\gamma\gamma$ couplings.}
\label{fig:4l-sig}
\end{figure}

\begin{figure}[H]
\centering
\includegraphics[width=0.6\hsize]{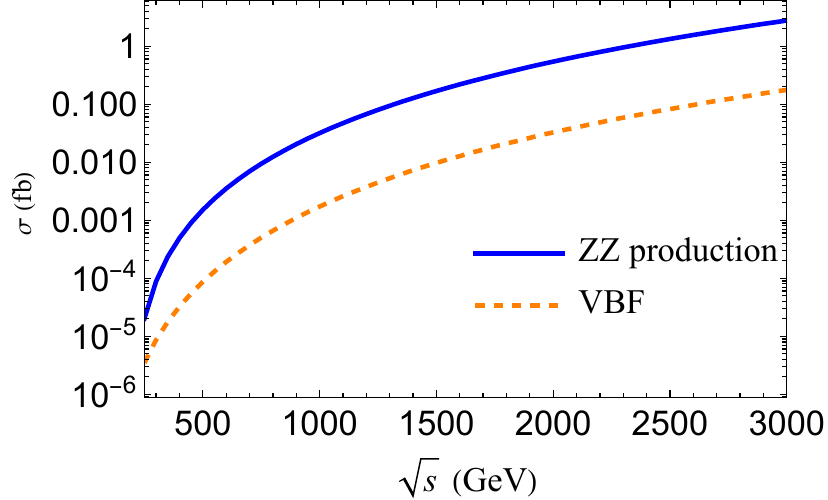}
\caption{A comparison of $\sigma_{\rm diboson}$ and $\sigma_{\rm VBF}$ for $e^+e^-\to 2\ell2\ell'$ as a function of $\sqrt{s}$ with $f_{\tilde{B}W} = 1 \ (\rm{TeV}^{-4})$.}
\label{fig:s-channel&VBF}
\end{figure}

\begin{figure}[H]
\centering
\includegraphics[height=0.4\hsize]{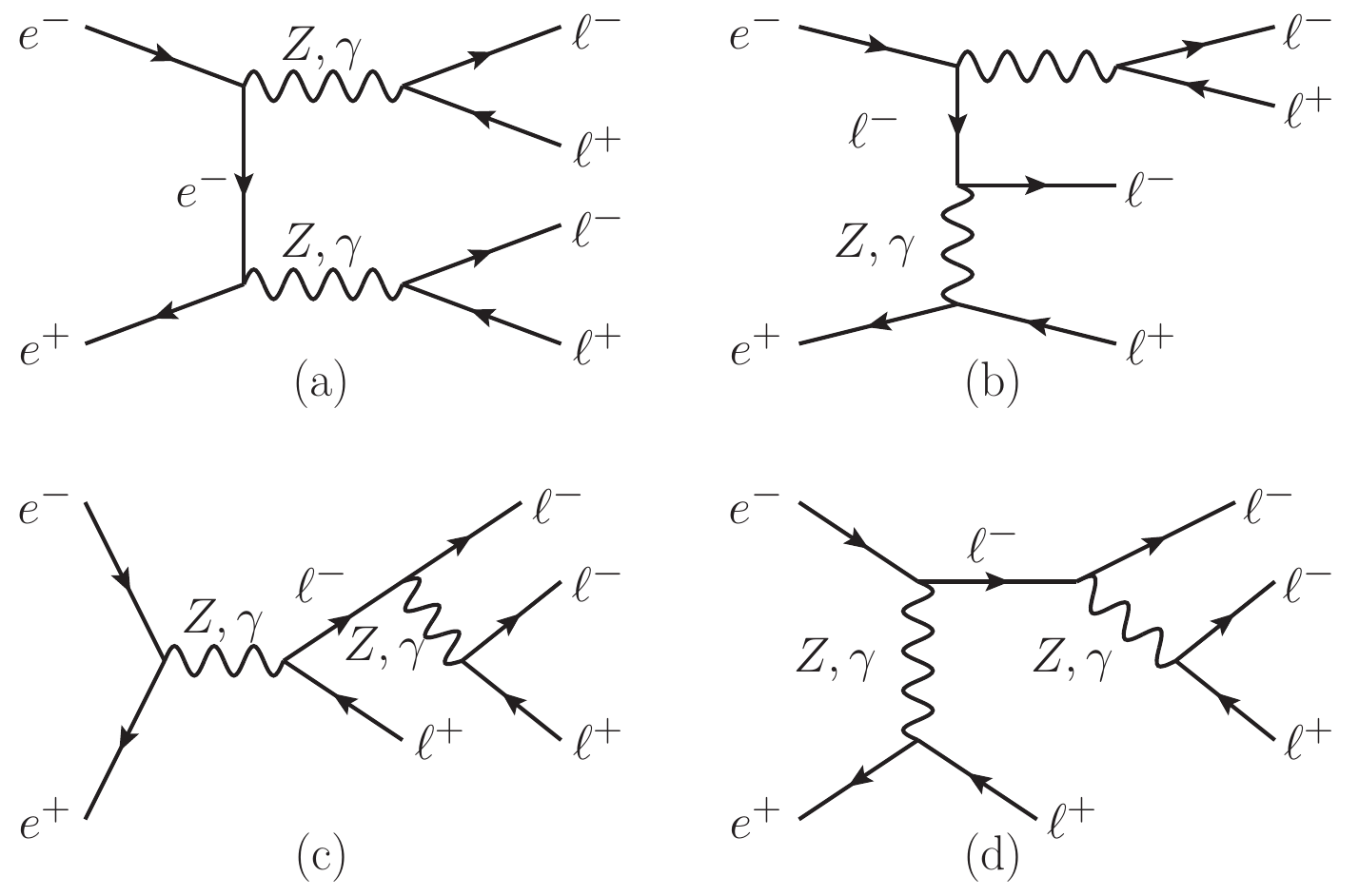}
\caption{Typical Feynman diagrams of the SM backgrounds for the process $e^+e^-\to 2\ell 2\ell'$.}
\label{fig:4l-bg}
\end{figure}

$Z$-pairs decaying to final states with four charged leptons ($\ell=e, \mu$) produce low multiplicity events with a clear topological signature that is exploited to maximize the selection efficiency.
In addition to $ZZ$ production, nTGCs can contribute to the $e^+e^-\to 2\ell 2\ell'$ process via di-boson production ($\gamma\gamma, Z\gamma$) and VBS processes, as shown in Fig.~\ref{fig:4l-sig}.
Therefore, we need to compare the contribution of nTGC in the annihilation process and the VBS process.
From Fig.~\ref{fig:s-channel&VBF}, di-boson production dominates the signal above 10 times over the vector boson fusion (VBF) contribution for $e^+e^-\to 2\ell2\ell'$.
The annihilation process is more sensitive to the dimension-8 operators, so in signal and background analysis, the kinematic distributions of the nTGC signal will mainly refer to the characteristics of the annihilation process.

The corresponding Feynman diagrams of the SM backgrounds are shown in Fig.~\ref{fig:4l-bg}.
The major contributions to the background are due to $Z$ production in association with two isolated leptons, as shown in Fig.~\ref{fig:4l-bg}(b, c, d). The four-lepton final state from the di-boson production in the SM comes from the $t$-channel diagrams as Fig.~\ref{fig:4l-bg}(a).

\begin{table}
\caption{\label{Tab:region}The ranges of coefficients $f_{\tilde{B}W} ({\rm TeV}^{-4}$) used for each energy point in the MC simulation.}
    \centering
    \begin{tabular}{lllll}
    \hline
    $\sqrt{s}$=250 GeV&$\sqrt{s}$=360 GeV&$\sqrt{s}$=500 GeV&$\sqrt{s}$=1000 GeV&$\sqrt{s}$=3000 GeV\\ \hline
    [$-100,100$]&[$-100,100$]&[$-10,10$]&[$-2.0,2.0$]&[$-0.15,0.15$]\\ \hline
    \end{tabular}
\end{table}

Variables used in this analysis include the transverse momentum $p_T=\sqrt{p^2_x+p^2_y}$ and the pseudorapidity defined in terms of the polar angle $\theta$ as $\eta = -\ln{[\tan{(\theta/2)}]}$. The basic cuts are set as same as the default settings of \verb"MadGraph5_aMC@NLO", except for $\Delta R_{\ell\ell}$ which is defined as $\sqrt{\Delta \eta_{\ell\ell} ^2+\Delta \phi_{\ell\ell}^2}$ where $\Delta \eta_{\ell\ell}$ and $\Delta \phi_{\ell\ell}$ are the differences between pseudo-rapidities and azimuth angles of the leptons, respectively.
In the basic cuts we use $\Delta R_{\ell\ell}>0.2$~\cite{ATLAS:2017vqm} as configurations with collinear leptons are excluded.
In the MC simulation, we use the coefficients in the ranges listed in Table~\ref{Tab:region}.

To study the kinematic features of the signal and background, We adopt the conventional approach of considering one operator at a time, as is widely used in analyses within the SMEFT framework. The largest coefficients in Table~\ref{Tab:region} are adopted. We require at least four leptons in the $4\ell$ signal mode. All the numerical results are presented after lepton number cut $N_\ell \geq 4$ (denoted as $N_\ell$ cut).

\begin{figure}[htbp]
\centering
\includegraphics[width=0.45\hsize]{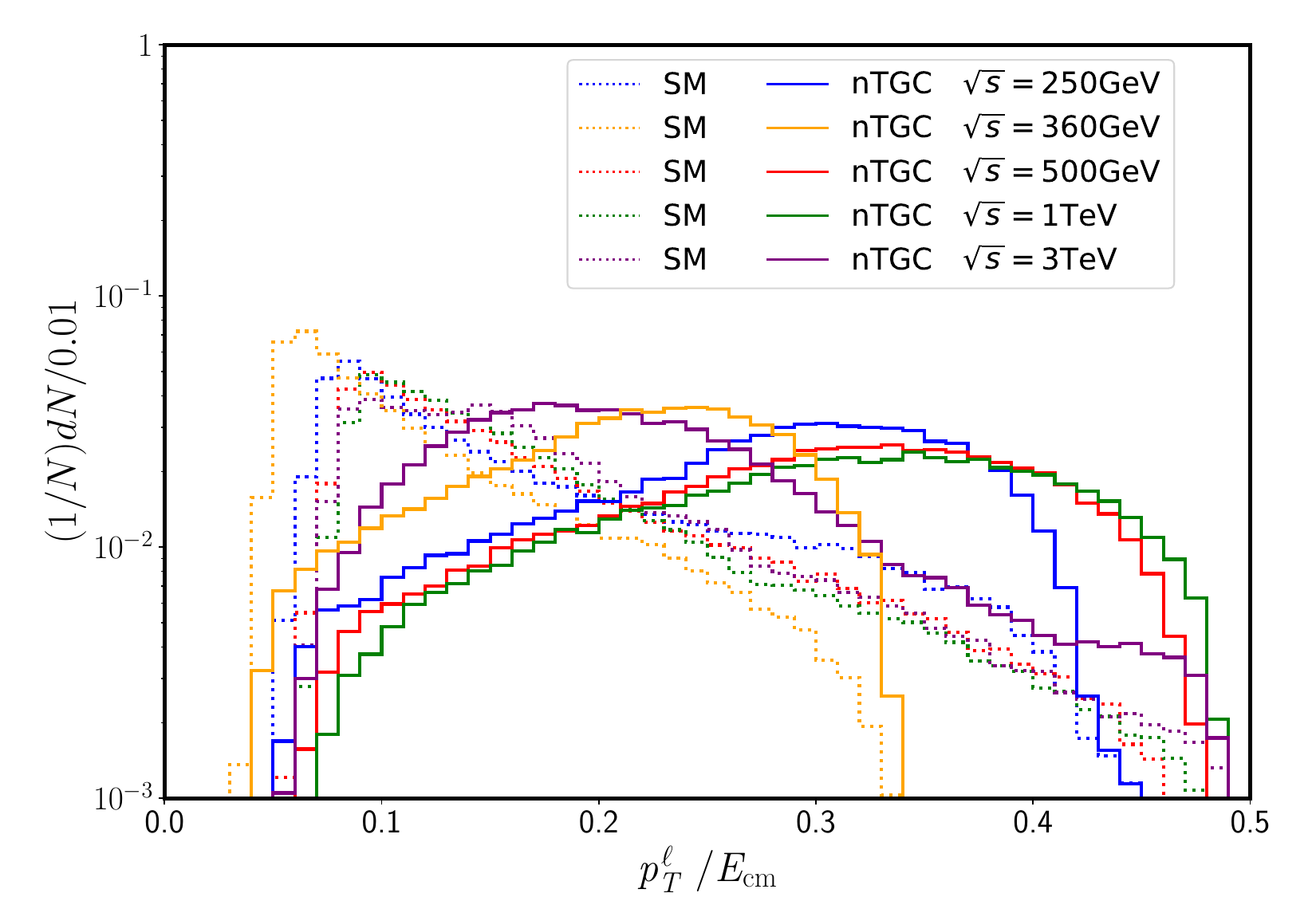}
\includegraphics[width=0.45\hsize]{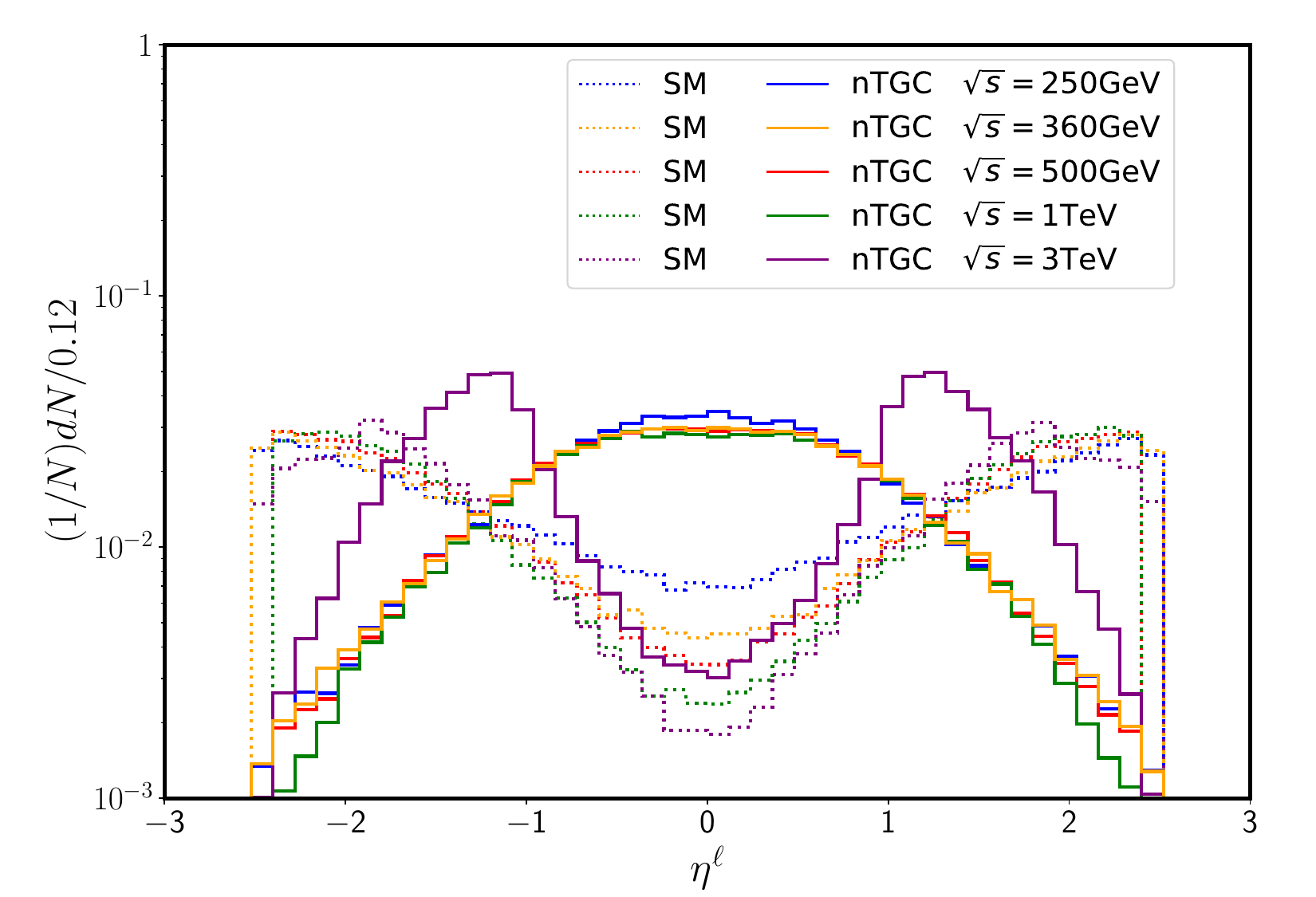}
\includegraphics[width=0.45\hsize]{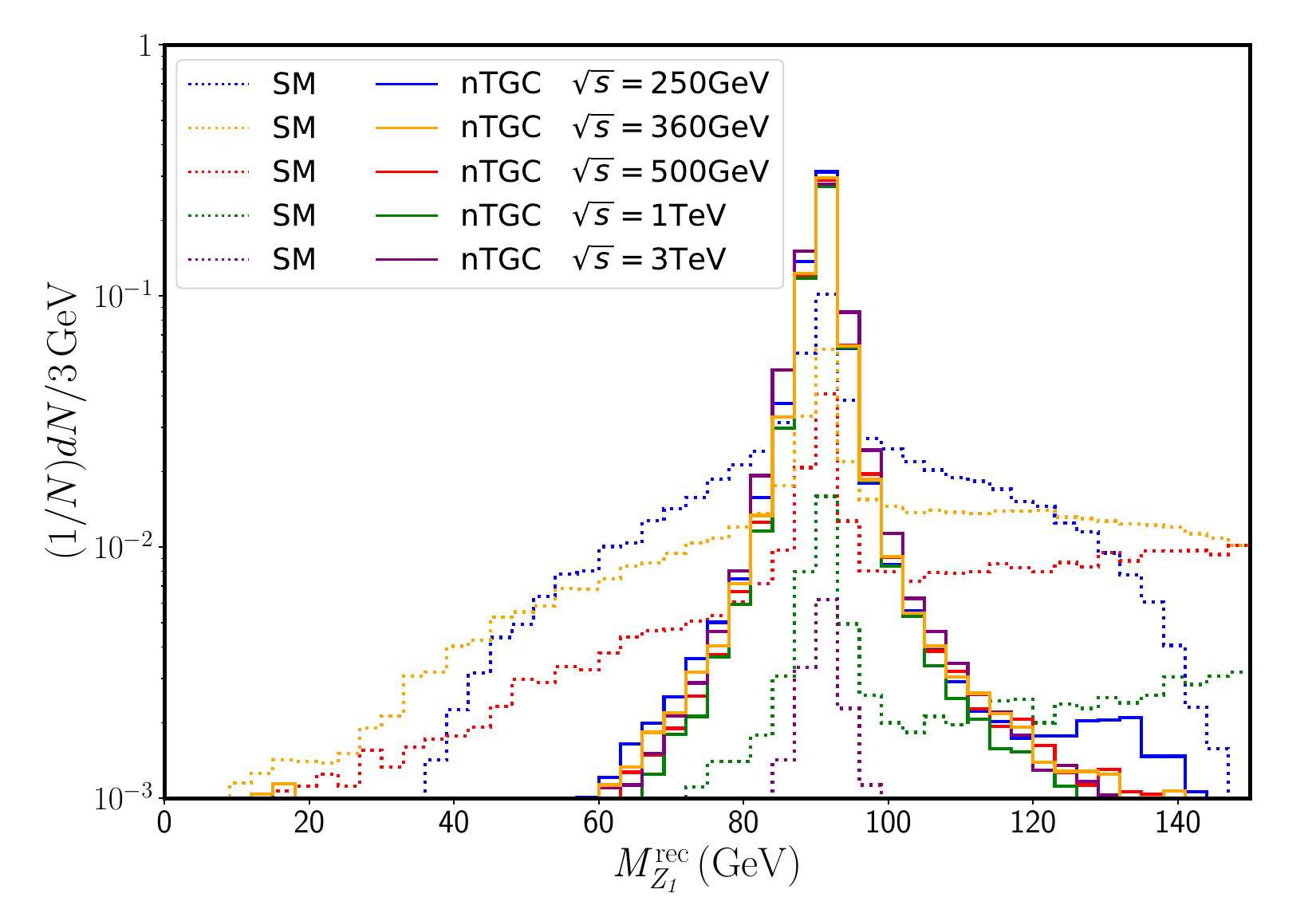}
\includegraphics[width=0.45\hsize]{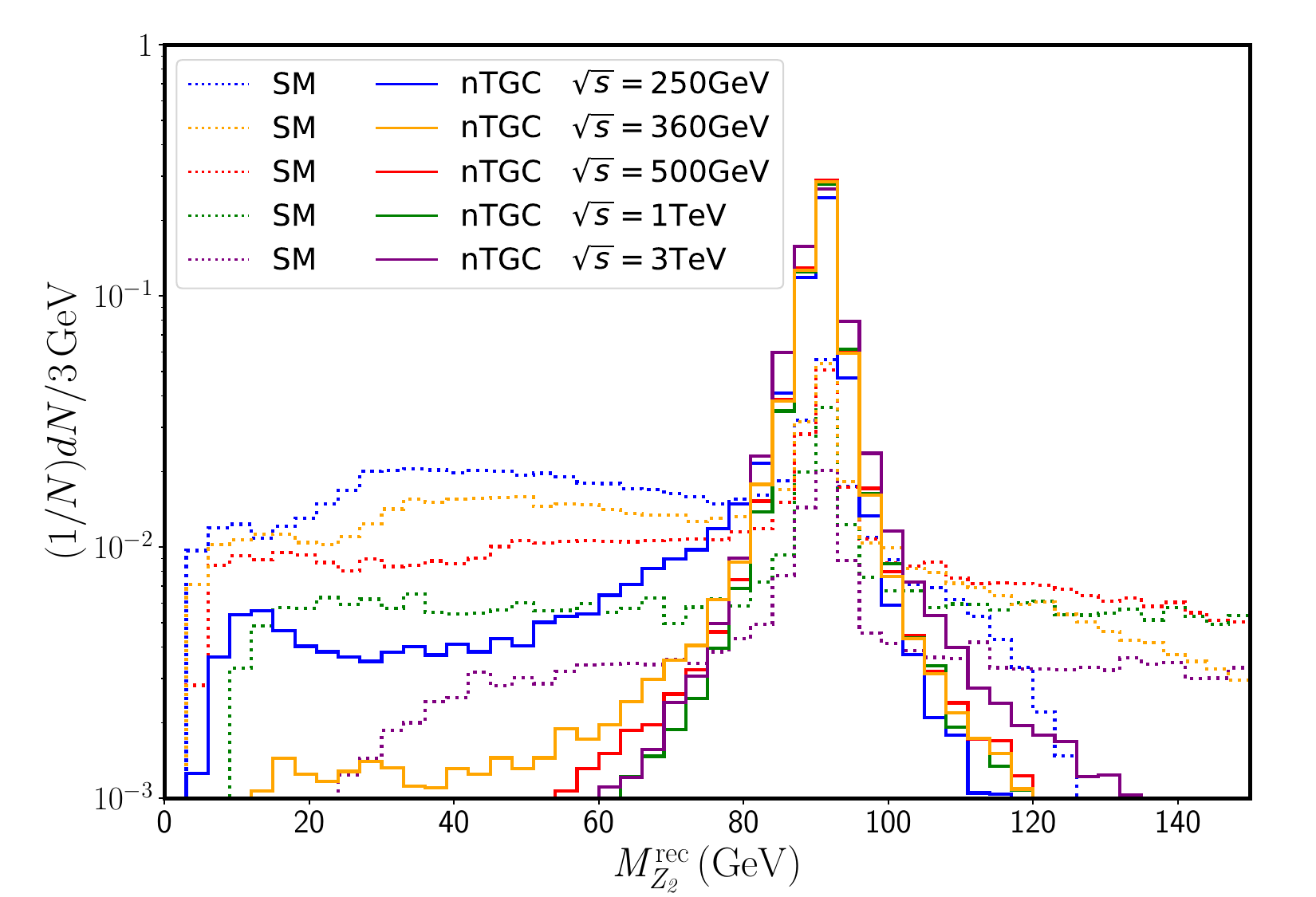}
\includegraphics[width=0.45\hsize]{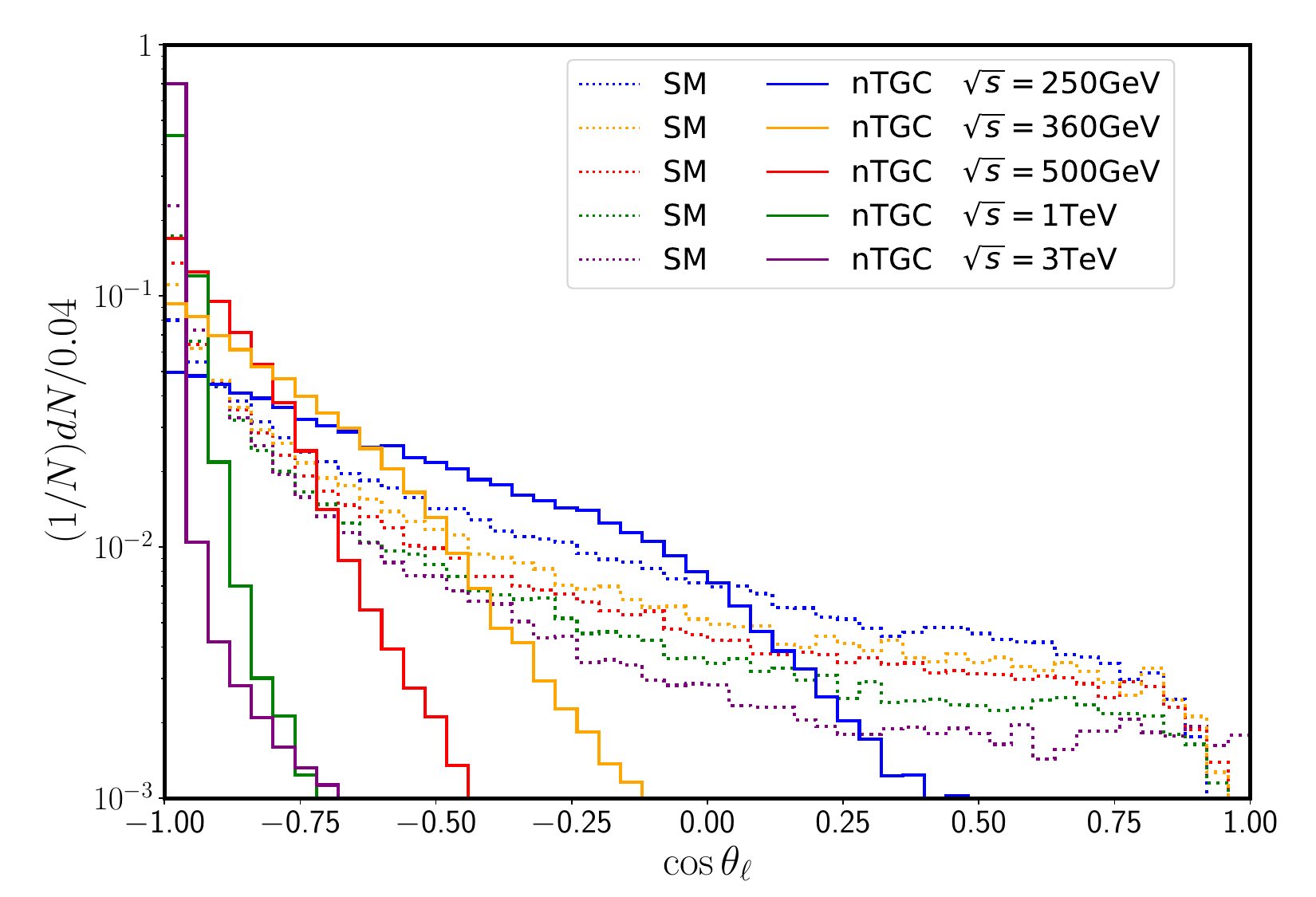}
\caption{The normalized distributions of $p_T^{\ell}/E_{\rm{cm}}$, $\eta^{\ell}$, $M_{Z_1}^{\rm{rec}}$, $M_{Z_2}^{\rm{rec}}$ and  $\cos{\theta_\ell}$ for $e^+e^-\to 2\ell 2\ell'$.}
\label{fig:distribution-4l}
\end{figure}

For the SM diagram in Fig.~\ref{fig:4l-bg}(a), we expect that the distribution of SM background events is dominantly around small $t$ and thus the transverse momentum of one of the vectors is small.
Denoting $p_T^\ell$ as the transverse momentum of the hardest lepton and $E_{\rm{cm}}$ as the c.m. energy, the normalized distributions of $p_T^{\ell}/E_{\rm{cm}}$ for the signal of $\mathcal{O}_{\tilde{B}W}$ and the background are displayed in Fig.~\ref{fig:distribution-4l}. The leptons from the SM background are softer than those from the nTGC signal.
In addition, we also checked the $\eta^\ell$ distribution of the final state leptons.
At low energies, the scattering angle of low-energy particles is smaller, and the direction of motion of the final-state leptons is more forward.
At high energies, the scattering of high-energy particles will cause the final-state leptons to be more widely distributed in different directions.
The distribution of $\eta^\ell$ of the signal will approach the SM background with the increase of the collision energy. Therefore, we did not use $\eta^\ell$ as the cut in the signal background analysis for CLIC.
The angle between the same sign leptons $\cos{\theta_\ell}$ is used at high energy.
At high energies, the direction of the lepton is approximately consistent with that of the hightly boosted $Z$, so the angle between leptons of the same sign is approximately the angle between two parent particles.
The dominate contribution of nTGCs is through $s$-channel of the di-boson diagrams.
Therefore, the angular distributions of di-boson production induced by nTGCs and the SM should be different.

\begin{table}
    \centering
    \caption{The event selection strategy and cross sections (fb) after cuts for each energy point. The results of nTGCs are obtained using the upper bounds of the coefficients in Table~\ref{Tab:region}.}
    \label{tab:cut-4l}
    \footnotesize
    \scalebox{0.95}{
    \begin{tabular}{lllllllllll}
    \hline
    \multirow{2}{*}{~}&\multicolumn{2}{c}{$\sqrt{s}$=250 GeV}&\multicolumn{2}{c}{$\sqrt{s}$=360 GeV}&\multicolumn{2}{c}{$\sqrt{s}$=500 GeV}&\multicolumn{2}{c}{$\sqrt{s}$=1 TeV}&\multicolumn{2}{c}{$\sqrt{s}$=3 TeV}\\
    \cmidrule(r){2-3}  \cmidrule(r){4-5} \cmidrule(r){6-7} \cmidrule(r){8-9} \cmidrule(r){10-11}
    \noalign{\smallskip}
    \multirow{2}{*}{~}&SM&NP&SM&NP&SM&NP&SM&NP&SM&NP\\ \hline
    Basic Cuts &22.83&0.254&16.50&3.015&11.64&0.161&5.301&0.109&1.336&0.022\\
    $p_T^{\ell}/{E_{\rm{cm}}}>0.15$&11.01&0.228&7.387&2.719&~&~&~&~&~&~\\
    $p_T^{\ell}/{E_{\rm{cm}}}>0.2$&~&~&~&~&3.584&0.133&1.555&0.092&~&~\\

    $\left|{\eta^\ell}\right|<1.5$&9.244&0.225&5.712&2.652&3.519&0.133&1.502&0.092&~&~\\

    $\cos{\theta_\ell}<0.4$&8.882&0.225&~&~&~&~&~&~\\
    $\cos{\theta_\ell}<-0.2$&~&~&4.909&2.620&~&~&~&~\\
    $\cos{\theta_\ell}<-0.6$&~&~&~&~&2.534&0.130&1.102&0.092&1.060&0.022\\
    $|M_{Z_1,Z_2}^{\rm{rec}}-M_Z|<10$ GeV&2.460&0.157&0.976&2.074&0.361&0.104&0.079&0.073&0.014&0.016\\
    Efficiency $\epsilon$ &6.8\%&40\%&3.5\%&43\%&1.8\%&40\%&0.8\%&40\%&0.6\%&53\%\\
    \hline
    \end{tabular}}
\end{table}

Since the main contribution to the signal comes from the di-boson production, the SM background from mono $Z$ or $\gamma$ production can be suppressed by requiring the event to contain two reconstructed $Z$ bosons.
$M_{Z_1}^{\rm{rec}}$ and $M_{Z_2}^{\rm{rec}}$ represent individually the reconstructed invariant mass of the first pair and the second pair of leptons.
The distributions of $M_{Z_1}^{\rm{rec}}$ and $M_{Z_2}^{\rm{rec}}$ in Fig.~\ref{fig:distribution-4l} exhibit peaks around the mass of the $Z$ boson for the signal.
Most background events reconstruct $M_{Z_1}^{\rm{rec}}$, but do not work well for reconstructing $M_{Z_2}^{\rm{rec}}$.
The joint requirement of the invariant mass of the reconstructed $Z_{1,2}$ is required to satisfy $80 < M_{\ell\ell} < 100$ GeV, which can effectively suppress the background events from Fig.~\ref{fig:4l-bg}(b,c,d). However, other kinematic observable measurements are needed to further optimize the event screening strategy to eliminate the background from Fig.~\ref{fig:4l-bg}(a).
Table~\ref{tab:cut-4l} shows the cross sections with event selection strategy and cut efficiencies obtained in all energy cases.

\begin{figure}[H]
\centering
\subfigure[250 GeV]{\includegraphics[width=0.45\hsize]{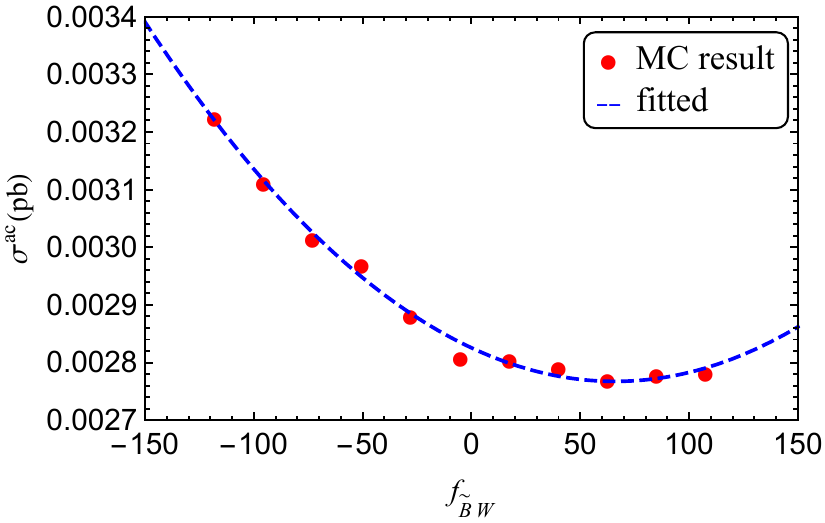}}
\subfigure[360 GeV]{\includegraphics[width=0.45\hsize]{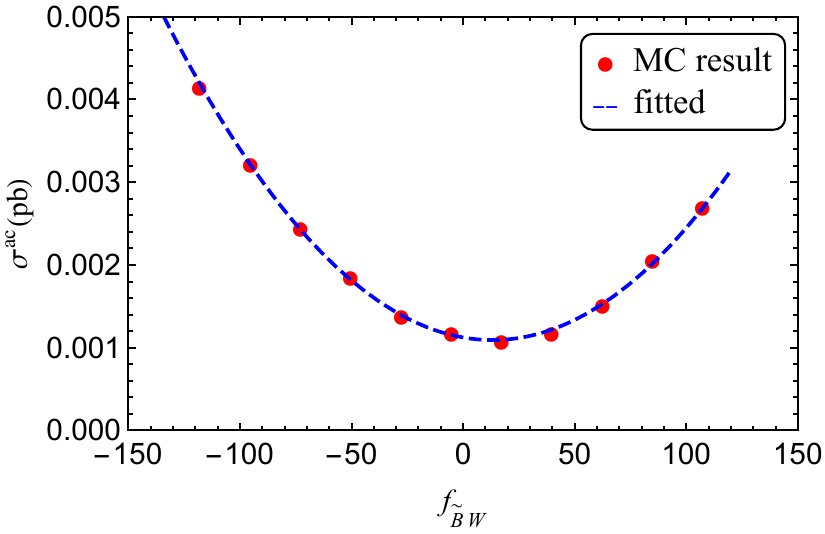}}
\subfigure[500 GeV]{\includegraphics[width=0.45\hsize]{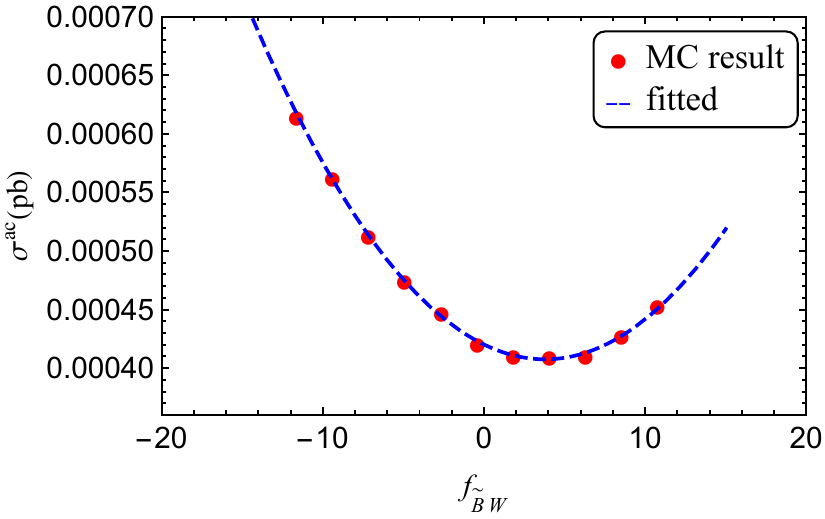}}
\subfigure[1 TeV]{\includegraphics[width=0.45\hsize]{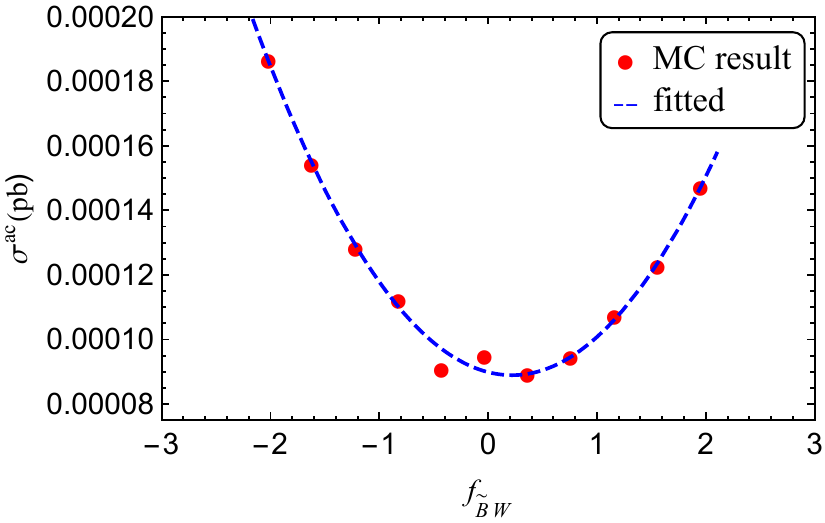}}
\subfigure[3 TeV]{\includegraphics[width=0.45\hsize]{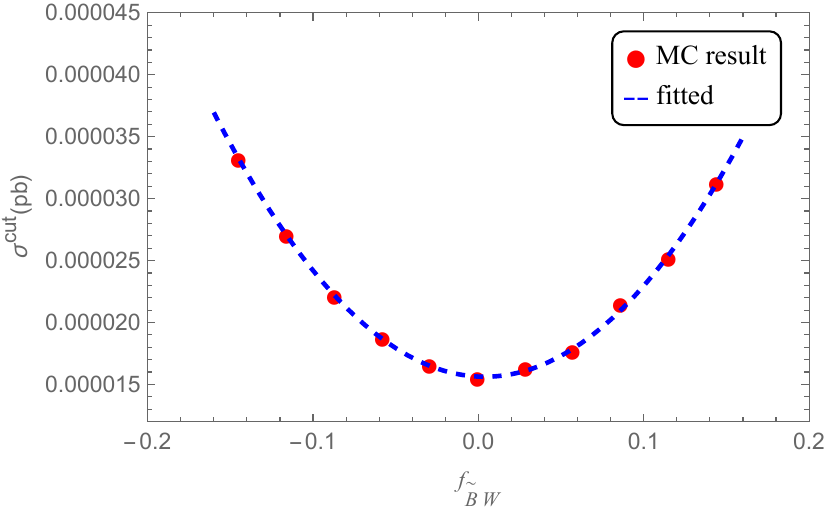}}
\caption{ Cross sections as function of $f_{\tilde{B}W}$~($\rm TeV^{-4}$) for $e^+e^-\to 2\ell2\ell'$.}
\label{fig:4l-csfit}
\end{figure}

The total cross sections after event selection strategy (denoted by $\sigma_{\rm nTGC}^{\rm ac}$) are obtained by scanning some values of coefficient with $\mathcal{O}_{\tilde{B}W}$, and shown in Fig.~\ref{fig:4l-csfit}.
Taking into account the impact of interference terms, the cross sections $\sigma_{\rm nTGC}^{\rm ac}$ can be fitted as bilinear functions of $f_{\tilde{B}W}$ ($\rm{TeV^{-4}}$), as
\begin{equation}
\begin{split}
&\sigma_{\rm nTGC}^{\rm ac} =\epsilon_{\rm SM}\sigma_{\rm SM}+ f_{\tilde{B}W}^2\sigma^{\rm ac}_{\rm NP}+ f_{\tilde{B}W} \hat{\sigma }_{\rm int} \;,
\end{split}
\label{eq.3.6}
\end{equation}
where $\epsilon_{\rm SM}$ and $\epsilon _{\rm NP}$ are the cut efficiencies of SM and nTGC events, respectively, which are listed in Table~\ref{tab:cut-4l}. $\sigma^{\rm ac}_{\rm NP}$ is NP contribution after cuts to be fitted and $\hat{\sigma }_{\rm int}$ is the interference parameter to be fitted. The numerical results of efficiencies are given in Table~\ref{tab:cut-4l}.
The fitted results of $\hat{\sigma }_{\rm int}$ can then be obtained and are given in Table~\ref{tab.4l-fit}.

\begin{table}[H]
\begin{center}
\caption{\label{tab.4l-fit}The fitted values of $\sigma^{\rm ac}_{\rm NP}$~($\rm fb\times TeV^8$) and $\hat{\sigma} _{\rm int}$~($\rm fb\times TeV^4$) for each energy point, where $\sigma^{\rm ac}_{\rm NP}$ and $\hat{\sigma} _{\rm int}$ is defined in Eq.~(\ref{eq.3.6}).}
\begin{tabular}{c|c|c|c|c|c}
\hline
$\sqrt{s}$  & 250 GeV & 360 GeV & 500 GeV & 1 TeV & 3 TeV \\
\hline
$\hat{\sigma} _{\rm int}$ & $-0.0018$ & $-0.0048$ & $-0.0067$  & $-0.0085$  & $-0.0062$ \\
\hline
$\sigma^{\rm ac}_{\rm NP}$ & $0.00001$ & $0.0002$ & $0.0008$  & $0.0194$  & $0.7942$ \\
\hline
\end{tabular}
\end{center}
\end{table}

The sensitivity of the futrue $e^-e^+$ colliders to the nTGC operators and the expected constraints on the coefficients can be obtained with respect to the significance defined as~\cite{Cowan:2010js,pdg}
\begin{eqnarray}
\label{eq:ss}
\mathcal{S}_{stat}=\sqrt{2 \left[(N_{\rm bg}+N_{s}) \ln (1+N_{s}/N_{\rm bg})-N_{s}\right]}\;,
\end{eqnarray}
where $N_s=N_{\rm nTGC}-N_{\rm SM}$ and $N_{\rm bg}=N_{\rm SM}$.
The likelihood function for $N_s$ is defined as
\begin{eqnarray}
L(N_s)=\frac{(N_{\rm bg}+N_{s})^n}{n!}e^{-(N_{\rm bg}+N_{s})}\;,
\end{eqnarray}
and one can find the maximum by setting $\partial {\rm ln}L/\partial N_s=0$. The approximate significance is then obtained by taking the square root of the likelihood ratio statistic $-2{\rm ln}(L(0)/L(N_s))$ after letting $n=N_{\rm bg}+N_{s}$~\cite{Cowan:2010js}.
The number of events $N_{\rm nTGC}$ is obtained by setting certain high-dimensional coefficient $f_{\tilde{B}W}$ in SMEFT.
The number of background events $N_{\rm SM}$ denotes the SM prediction with the high-dimensional coefficient vanishing.

As a result, we can obtain the projected sensitivity on $f_{\tilde{B}W}$ by taking $2\sigma$, $3\sigma$ or $5\sigma$ significance.
The results are shown in Table~\ref{tab.constraint-4l}.

\begin{table*}[hp]
    \centering
    \caption{The expected constraints on $f_{\tilde{B}W}$ ($\rm{TeV^{-4}}$) for $e^+e^-\to 2\ell2\ell'$ at each energy point of CEPC, ILC and CLIC with corresponding design luminosities.}
    \label{tab.constraint-4l}
    \begin{tabular}{llllll}
    \hline
    \multirow{2}{*}{$S_{stat}$}&$\sqrt{s}$ (GeV)&\\
    \cline{2-6} ~&250&360&500&1000&3000\\ \hline
     2&[$-23.1,155.0$]&[$-10.4,36.9$]&[$-5.4,13.0$]&[$-0.84,1.28$]&[$-0.088,0.096$]\\
     3&[$-32.8,164.6$]&[$-14.3,40.8$]&[$-7.3,14.9$]&[$-1.11,1.55$]&[$-0.114,0.121$]\\
     5&[$-49.9,181.8$]&[$-21.1,47.5$]&[$-10.7,18.2$]&[$-1.57,2.01$]&[$-0.160,0.169$]\\ \hline
    \end{tabular}
\end{table*}

\subsection{Selection of \texorpdfstring{$e^+e^-\to4j$}{4j}}
\label{sec3.2}
\begin{figure}[H]
\centering
\includegraphics[height=0.2\hsize]{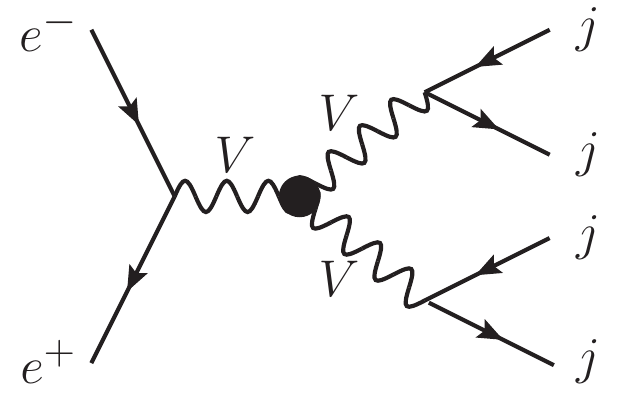}
\caption{The signal Feynman diagrams for the process $e^+e^-\to 4 j $, where $VVV$ stands for $ZZZ$, $ZZ\gamma$, $Z\gamma\gamma$, $ZWW$ or $\gamma WW$ couplings.}
\label{fig:4j-signal}
\end{figure}

\begin{figure}[H]
\centering
\includegraphics[height=0.45\hsize]{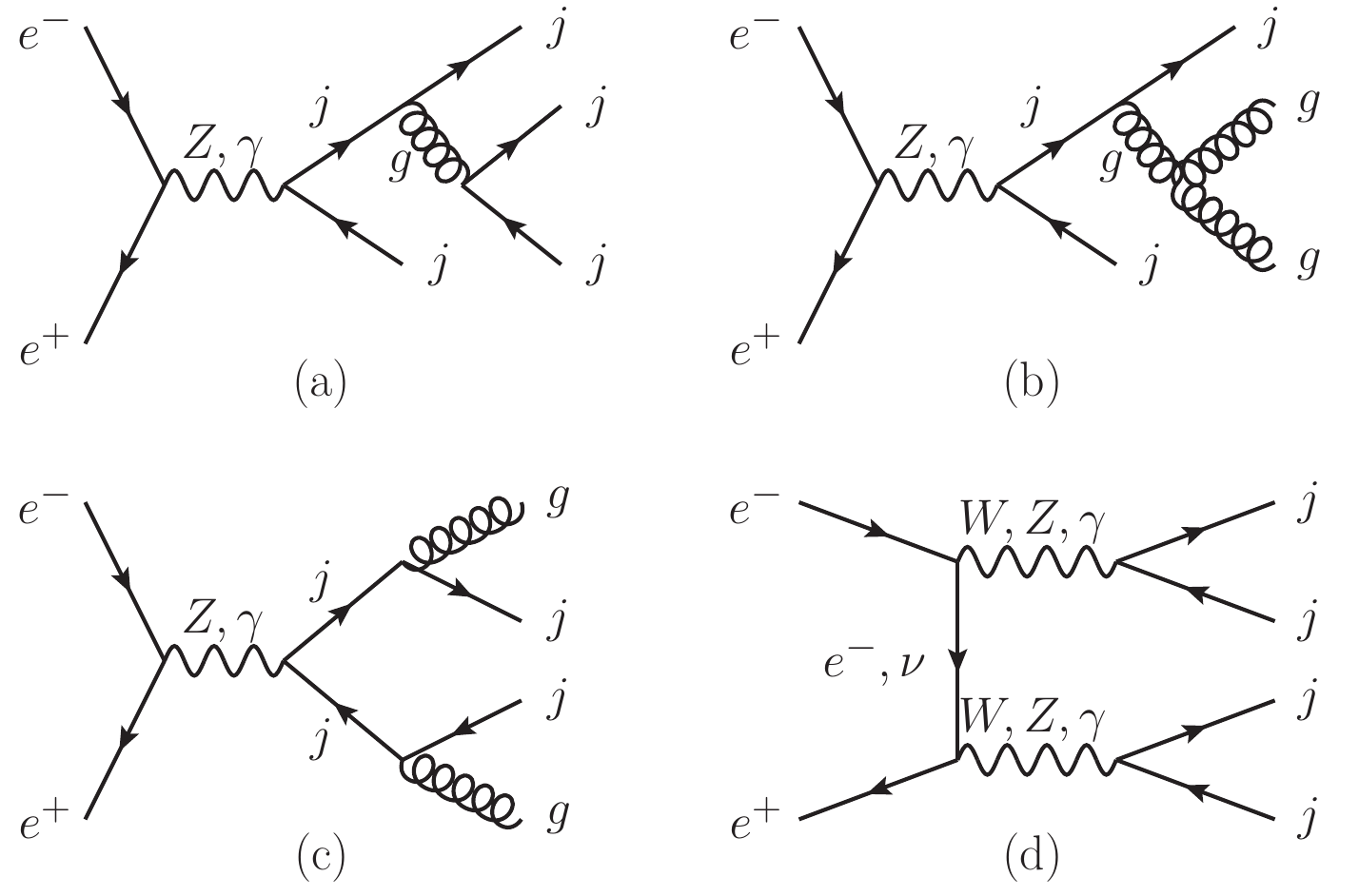}
\caption{The background Feynman diagrams for the process $e^+e^-\to 4 j$.}
\label{fig:4j-bg}
\end{figure}

The hadronic selection of $ZZ$ production is very important, which represents 49\% of the $ZZ$ decay topologies.
The signal is characterized by four or more jets in the final state, and the contribution to the signal is not only the pair production of $Z$ boson, but also the $W$ boson.
$e^+e^-\to 4 j$ channel has only one topology induced by 5 electroweak vertices $ZZZ$, $Z\gamma \gamma$, $ZZ \gamma$, $WWZ$ and $WW\gamma$.
At tree level, the Feynman diagram is shown in Fig.~\ref{fig:4j-signal}.
The background include reducible contributions from QCD processes and di-boson productions.
Fig.~\ref{fig:4j-bg} shows the typical Feynman diagrams.

\begin{figure}[H]
\centering
\includegraphics[width=0.45\hsize]{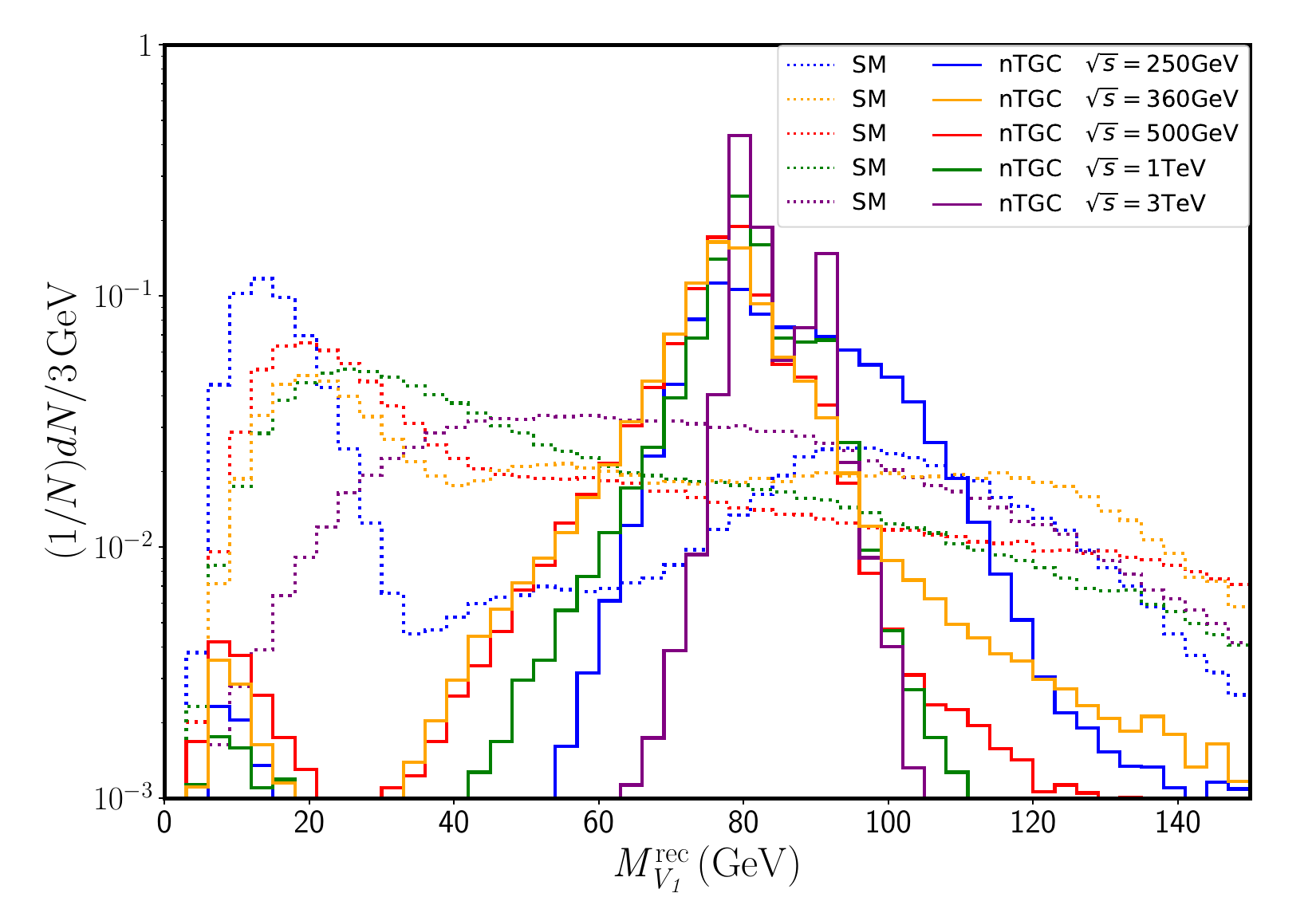}
\includegraphics[width=0.45\hsize]{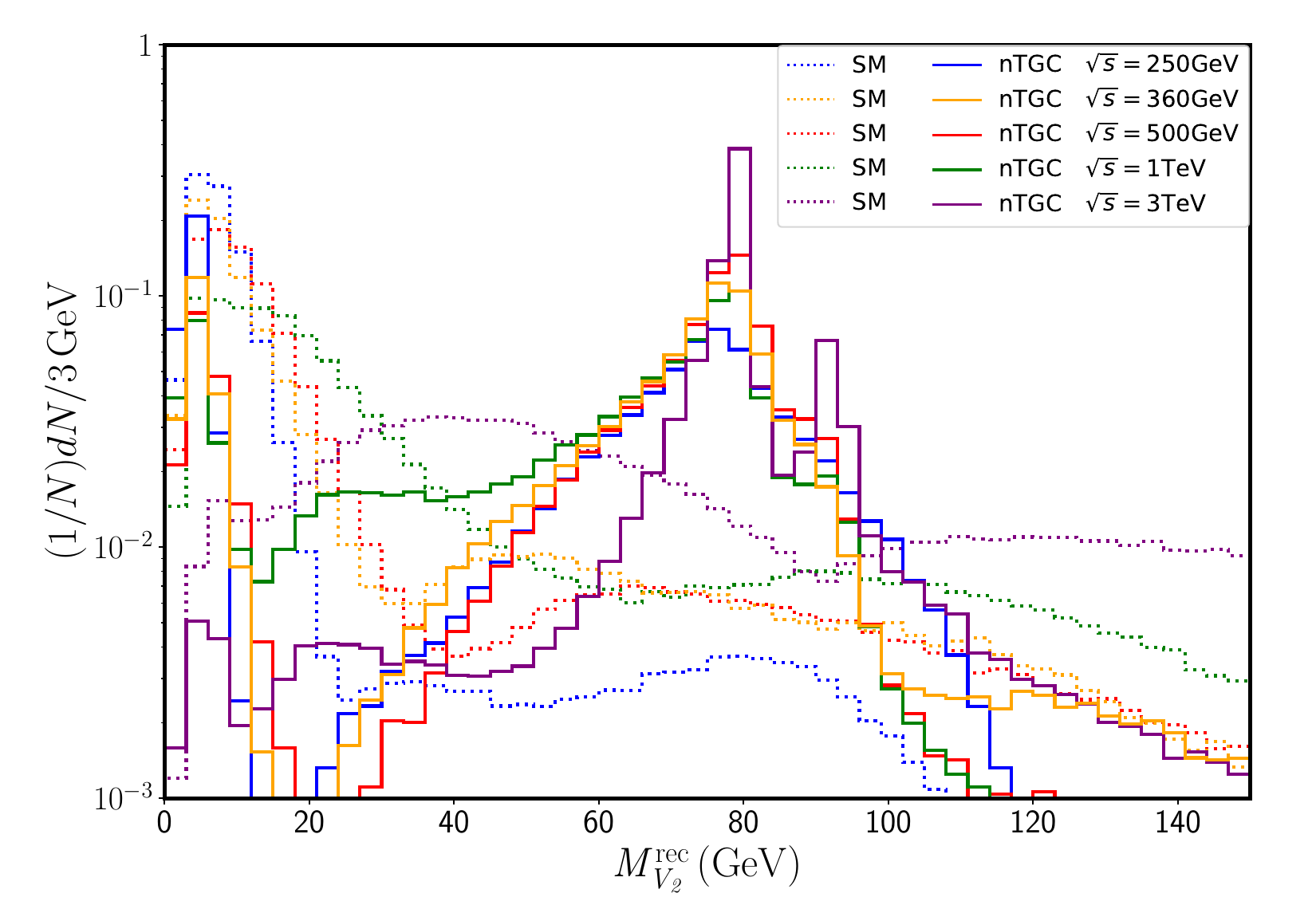}
\caption{The normalized distributions of $M_{V_{1}}^{\rm{rec}}$, $M_{V_{2}}^{\rm{rec}}$ for $e^+e^-\to 4 j$.}
\label{fig:4j-Mjj}
\end{figure}

The event selection strategy is optimized separately for each energy point of furture $e^+e^-$ colliders.
The basic cuts are the requirements of the isolated jet.
A four-jet event sample is formed by requiring the jet resolution parameter are transverse momentum $p_T^j > 20$ GeV and pseudo-rapidity $\eta_j < 5.0$. Consider collinear jets from the same pairing at high energy scale,  signal and background events required the number of jets of $N_j \geq 2$.
Furthermore, the signal events are required to have at least two $V \to jj$ candidate ($V = W^\pm, Z$), denoted by $V_1$ and $V_2$.
The invariant masses of the two sets of di-jets, $M_{V_1}$ and $M_{V_2}$, are reconstructed by the circular pairing, and then evaluating the mass difference $\Delta_{M}=|M_{jj}-M_V|$ between each pairing and the vector boson.
$\Delta_{M}$ will be calculated twice depending on whether $M_V$ takes $W$ mass or $Z$ mass,
The pairing scheme with the smallest sum of $\Delta_{M}$ for $V_1$ and $V_2$ is selected, and the mass window of $M_{V_1}$ and $M_{V_2}$ is used as the improved cuts to discriminate the QCD background events against the signal events of hadronic vector boson-pair events.
The invariant mass distributions of the signal exhibit distinct peaks within the mass range of the $W$ and $Z$ bosons, as shown in figures \ref{fig:4j-Mjj}. The distributions of $M_{V_1}$ and $M_{V_2}$ display small peaks around 15 GeV and 10 GeV, respectively, which could be attributed to $Z\gamma^*$ and $\gamma^*\gamma^*$ production.
Considering that $Z\gamma$ production constitutes a minor portion of the signal and that background events dominate the low-energy region for $M_{V_1}$ and $M_{V_2}$, we require that both $M_{V_1}$and $M_{V_2}$ of the signal event satisfy the invariant mass range of the massive vector boson.
We apply a mass window cut of $60 < M_{V_1,V_2} < 110$ GeV to ensure that the selected events consist of $4$ jets from $W$ and $Z$ boson pair production induced by nTGC. The invariant mass selection aims for an inclusive selection of fully  hadronic vector boson-pair decays.

\begin{figure}[htbp]
\centering
\subfigure[250 GeV]{\includegraphics[width=0.45\hsize]{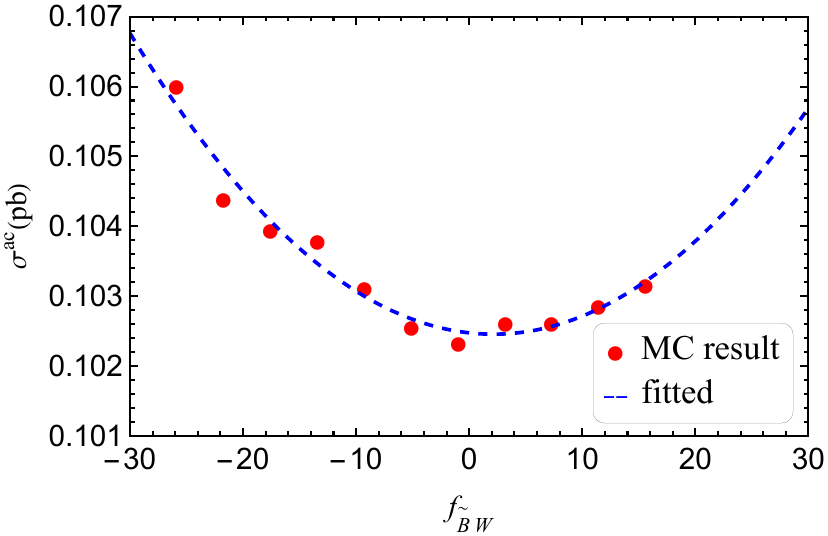}}
\subfigure[360 GeV]{\includegraphics[width=0.45\hsize]{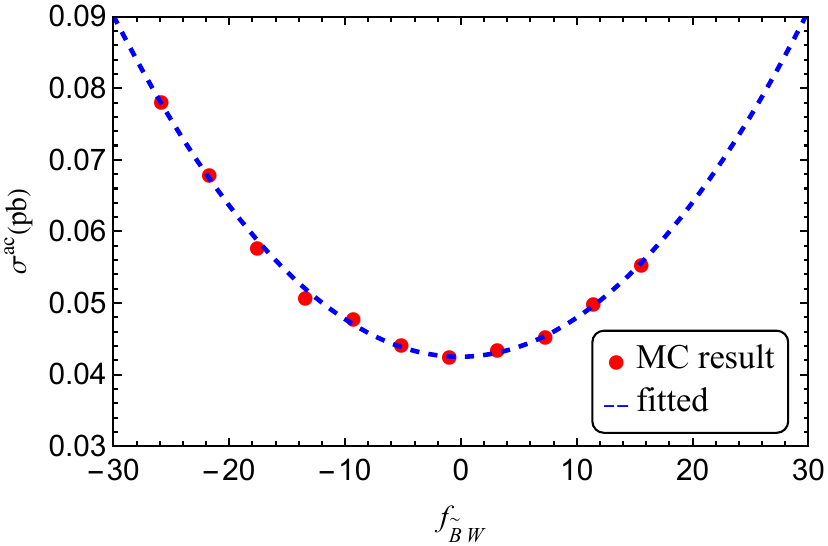}}
\subfigure[500 GeV]{\includegraphics[width=0.45\hsize]{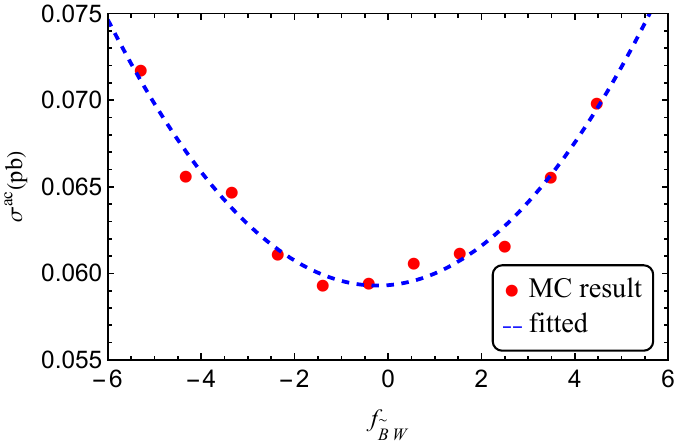}}
\subfigure[1 TeV]{\includegraphics[width=0.45\hsize]{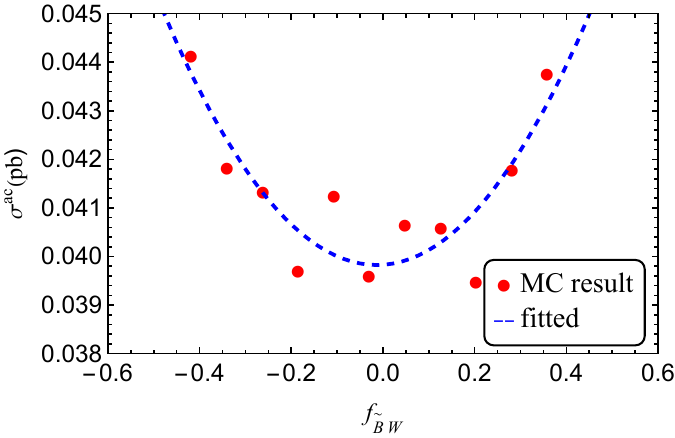}}
\subfigure[3 TeV]{\includegraphics[width=0.45\hsize]{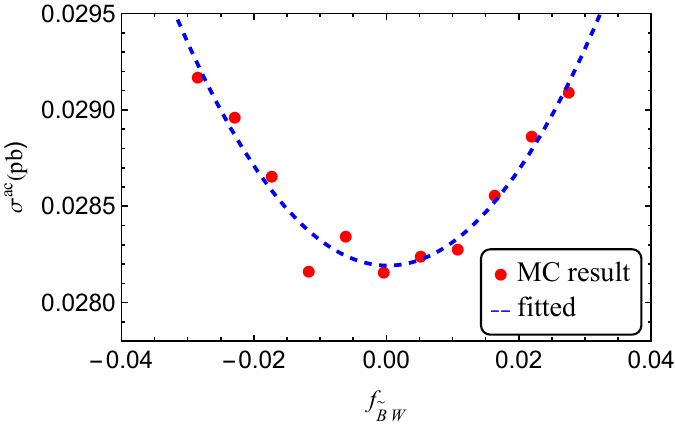}}
\caption{ Cross sections as functions of $f_{\tilde{B}W}$ ($\rm{TeV^{-4}}$) for $e^+e^-\to 4 j$.}
\label{fig:4j-csfit}
\end{figure}

The specific values of event selection strategy and the respective cut efficiencies are presented in Table~\ref{tab:4j-cutflow}.
With the cut flow applied, the total cross sections are obtained by scanning the values of coefficient with $\mathcal{O}_{\tilde{B}W}$, and the fitting functions are shown in Fig.~\ref{fig:4j-csfit}. The fitting coefficients of interference terms are given in Table~\ref{tab.4j-fit}.

\begin{table}[H]
    \centering
     \caption{\label{tab:4j-cutflow}The cross sections $\sigma_{\rm SM}$ and $\sigma_{\rm nTGC}$ of $e^+e^-\to 4 j$ (in unit of fb) and cut efficiencies after event selection strategy.}
    \footnotesize
    \scalebox{0.95}{\begin{tabular}{lllllllllll}
    \hline
    \multirow{2}{*}{~}&\multicolumn{2}{c}{$\sqrt{s}$=250 GeV}&\multicolumn{2}{c}{$\sqrt{s}$=360 GeV}&\multicolumn{2}{c}{$\sqrt{s}$=500 GeV}&\multicolumn{2}{c}{$\sqrt{s}$=1 TeV}&\multicolumn{2}{c}{$\sqrt{s}$=3 TeV}\\
    \cmidrule(r){2-3}  \cmidrule(r){4-5} \cmidrule(r){6-7} \cmidrule(r){8-9} \cmidrule(r){10-11}
    \noalign{\smallskip}
    \multirow{2}{*}{~}&SM&NP&SM&NP&SM&NP&SM&NP&SM&NP\\ \hline
    Basic Cuts &2898&2.216&251.5&23.28&2103&1.684&994.2&1.655&208.1&0.988\\
    $M_{V_1,V_2}^{\rm{rec}}\in$(60,110) GeV&103.5&1.156&14.50&14.05&78.73&1.127&42.67&0.925&52.67&0.911\\
    Efficiency $\epsilon$ &3.6\%&52\%&1.6\%&60\%&3.7\%&67\%&4.3\%&55\%&25\%&92\%\\
    \hline
    \end{tabular}}
\end{table}

\begin{table}[H]
\begin{center}
\caption{\label{tab.4j-fit}The fitted values of $\sigma^{\rm ac}_{\rm NP}$~($\rm fb\times TeV^8$) and $\hat{\sigma} _{\rm int}$~($\rm fb\times TeV^4$) for $e^+e^-\to 4 j$ at each energy point, where $\sigma^{\rm ac}_{\rm NP}$ and $\hat{\sigma} _{\rm int}$ is defined in Eq.~(\ref{eq.3.6}).}
\begin{tabular}{c|c|c|c|c|c}
\hline
$\sqrt{s}$  & 250 GeV & 360 GeV & 500 GeV & 1 TeV & 3 TeV \\
\hline
$\hat{\sigma} _{\rm int}$ & $-0.0178$ &$-0.0112$& $-0.2134$  & $-0.1332$  & $-0.5256$ \\
\hline
$\sigma^{\rm ac}_{\rm NP}$ & $0.0042$ &$0.0533$& $0.4614$  & $2.0408$  & $1268.4$ \\
\hline
\end{tabular}
\end{center}
\end{table}

The sensitivities for the $e^+e^-\to 4 j$ channel at the future $e^+e^-$ colliders and the expected constraints on the coefficients, as obtained with Eq.~\ref{eq:ss}.
The integrated luminosity $\mathcal{L}$ is based on the design values of the CEPC, ILC and CLIC at four energies.
The projected sensitivity on $f_{\tilde{B}W}$  by taking $2\sigma$, $3\sigma$ or $5\sigma$ significance levels.
The results are presented in Table~\ref{tab:constraints-4j}.
While the hadronic modes are considerably more challenging, the $4j$ signal channel at the $e^+e^-$ colliders exhibits a high degree of sensitivity to the $\mathcal{O}_{\tilde{B}W}$ operators, which are even more pronounced than the four charged leptons mode.

\begin{table}[H]
    \centering
    \caption{The excepted constraints on $f_{\tilde{B}W}$  ($\rm{TeV^{-4}}$) for $e^+e^-\to 4 j$.}
    \label{tab:constraints-4j}
    \begin{tabular}{llllll}
    \hline
    \multirow{2}{*}{$S_{stat}$}&$\sqrt{s}$ (GeV)&\\
    \cline{2-6} ~&250&360&500&1000&3000\\ \hline
     2&[$-6.4,10.7$]&[$-2.7,2.9$]&[$-1.0,1.5$]&[$-0.15,0.22$]&[$-0.011,0.011$]\\
     3&[$-8.2,12.5$]&[$-3.3,3.5$]&[$-1.3,1.8$]&[$-0.20,0.26$]&[$-0.013,0.014$]\\
     5&[$-11.2,15.4$]&[$-4.3,4.4$]&[$-1.7,2.2$]&[$-0.27,0.33$]&[$-0.017,0.017$]\\ \hline
    \end{tabular}
\end{table}

\subsection{Selection of \texorpdfstring{$e^+e^-\to 2 \ell 2 j$}{2l2j} }
\label{sec3.3}

\begin{figure}[htbp]
\centering
\includegraphics[height=0.2\hsize]{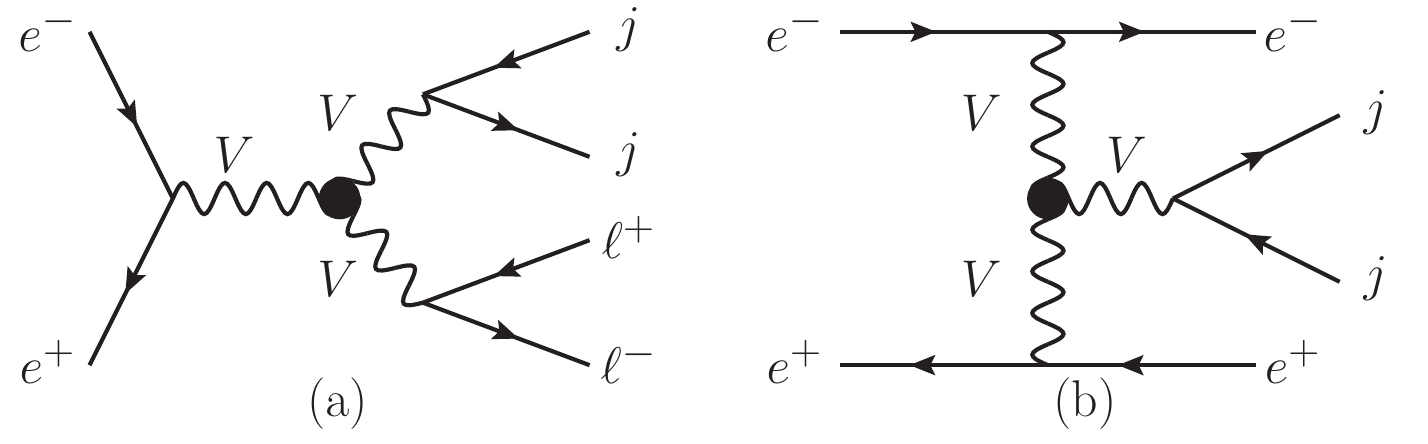}
\caption{Typical Feynman diagrams which contribute to the process $e^+e^-\to 2\ell 2j$ for signal events, where $VVV$ stands for $ZZZ$, $ZZ\gamma$ or $Z\gamma\gamma$ couplings.}
\label{fig:2l2j-sig}
\end{figure}

\begin{figure}[htbp]
\centering
\includegraphics[height=0.2\hsize]{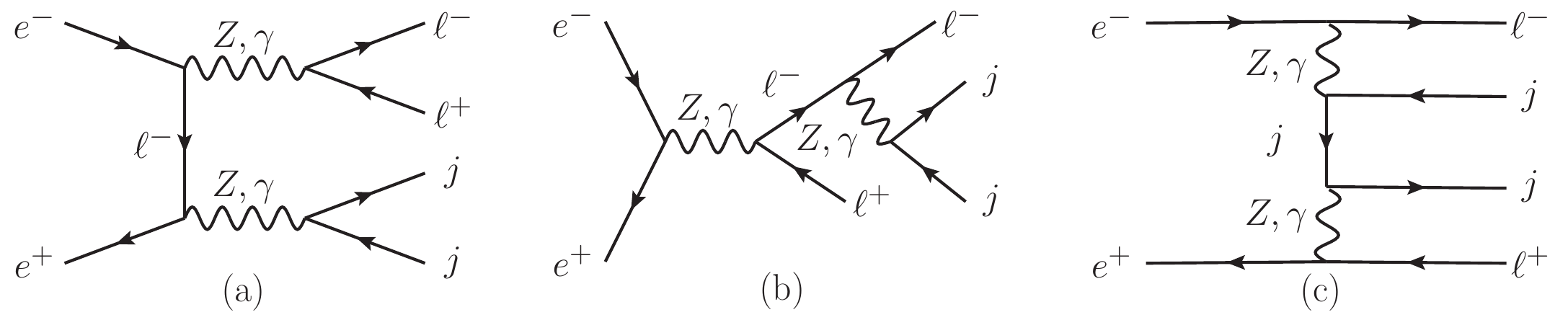}
\caption{Typical Feynman diagrams which contribute to the process $e^+e^-\to 2\ell 2j$ for background events.}
\label{fig:2l2j-bg}
\end{figure}

The signal of $2 \ell 2 j$ induced by nTGCs are composed of both VBS and di-boson production topologies with the decay of massive EW gauge bosons.
The signal from $ZZ$ production is the main component of nTGCs signal, which has a branching ratio in $ZZ$ production of $4.7\%$ per lepton flavour.
The kinematic characteristics of $ZZ$ decay products will provide the main feature of nTGC signal.
The SM background can be divided into three categories according to whether the final-state leptons and jets can reconstruct $Z$.
The typical Feynman diagrams of signal and background are shown in Fig.~\ref{fig:2l2j-sig} and Fig.~\ref{fig:2l2j-bg}.
The on-shell $ZZ$ event can be selected by applying simultaneous cuts on the masses of the $\ell^+\ell^-$ pair, on the remaining hadronic system and on their sum.

As in the previous analysis, the signal events are generated by assuming one operator at a time, using the largest coefficients from Table~\ref{Tab:region}. In the following studies, we require at least two leptons and two jets. All numerical results are presented after applying the particle number cuts: $N_{\ell^+} \geq 1$, $N_{\ell^-} \geq 1$, and $N_j \geq 2$, and the two hardest leptons be a lepton and an anti-lepton of the same flavor. This requirement is denoted as the $N_{\ell,j}$ cut.

Another important factor is $\Delta R_{\ell\ell}$ and $\Delta R_{jj}$. When the $Z$ boson is energetic, the leptons tend to be collinear, resulting in a small $\Delta R_{\ell\ell}$ and a small $\Delta R_{jj}$ for the signal events. However, since $\Delta R_{\ell\ell}$ is related to lepton isolation, we impose a basic cut of $\Delta R_{\ell\ell} > 0.2$, which can be achieved in experiments~\cite{zaexp1}.

\begin{figure}[H]
\centering
\includegraphics[width=0.45\hsize]{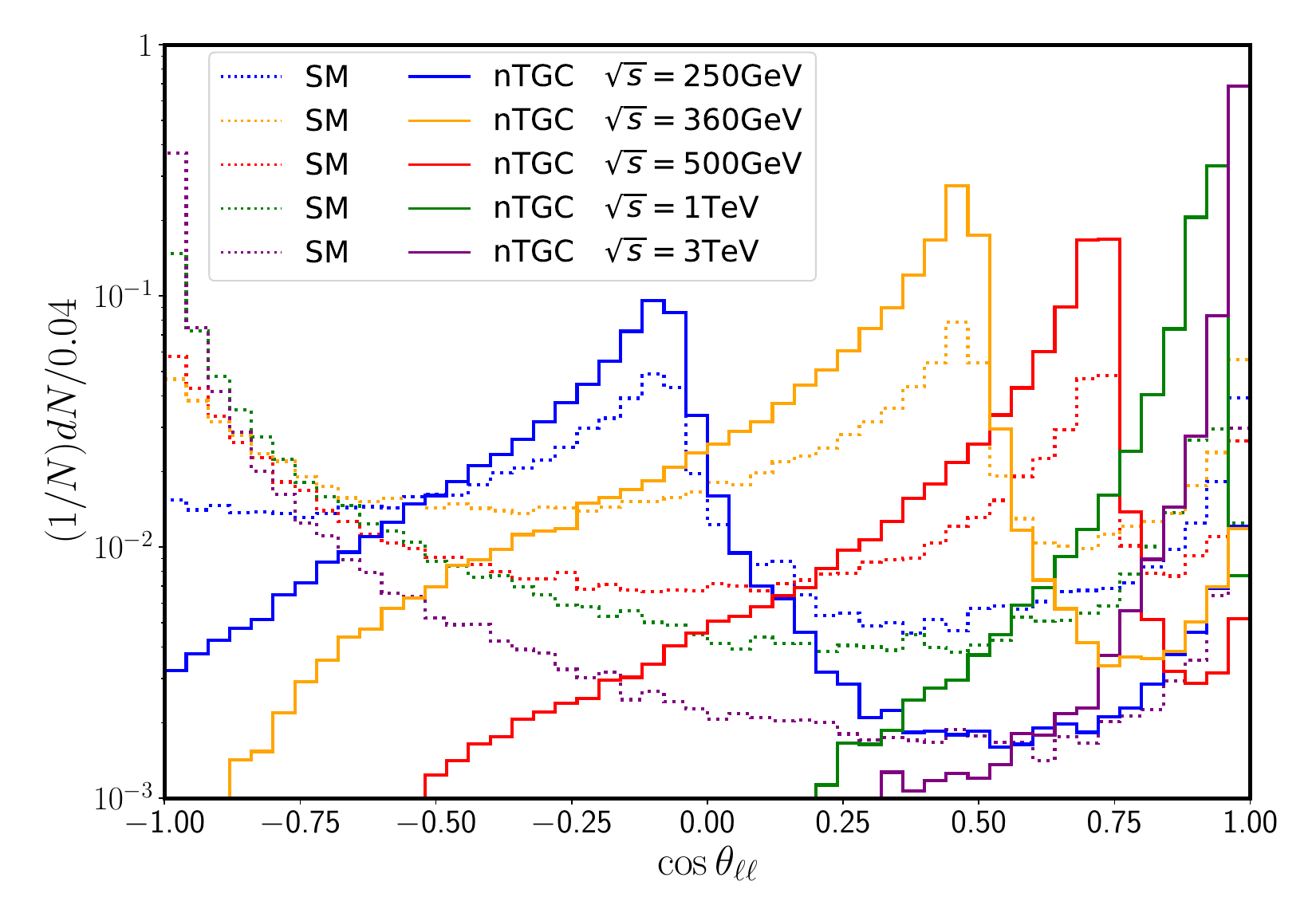}
\includegraphics[width=0.45\hsize]{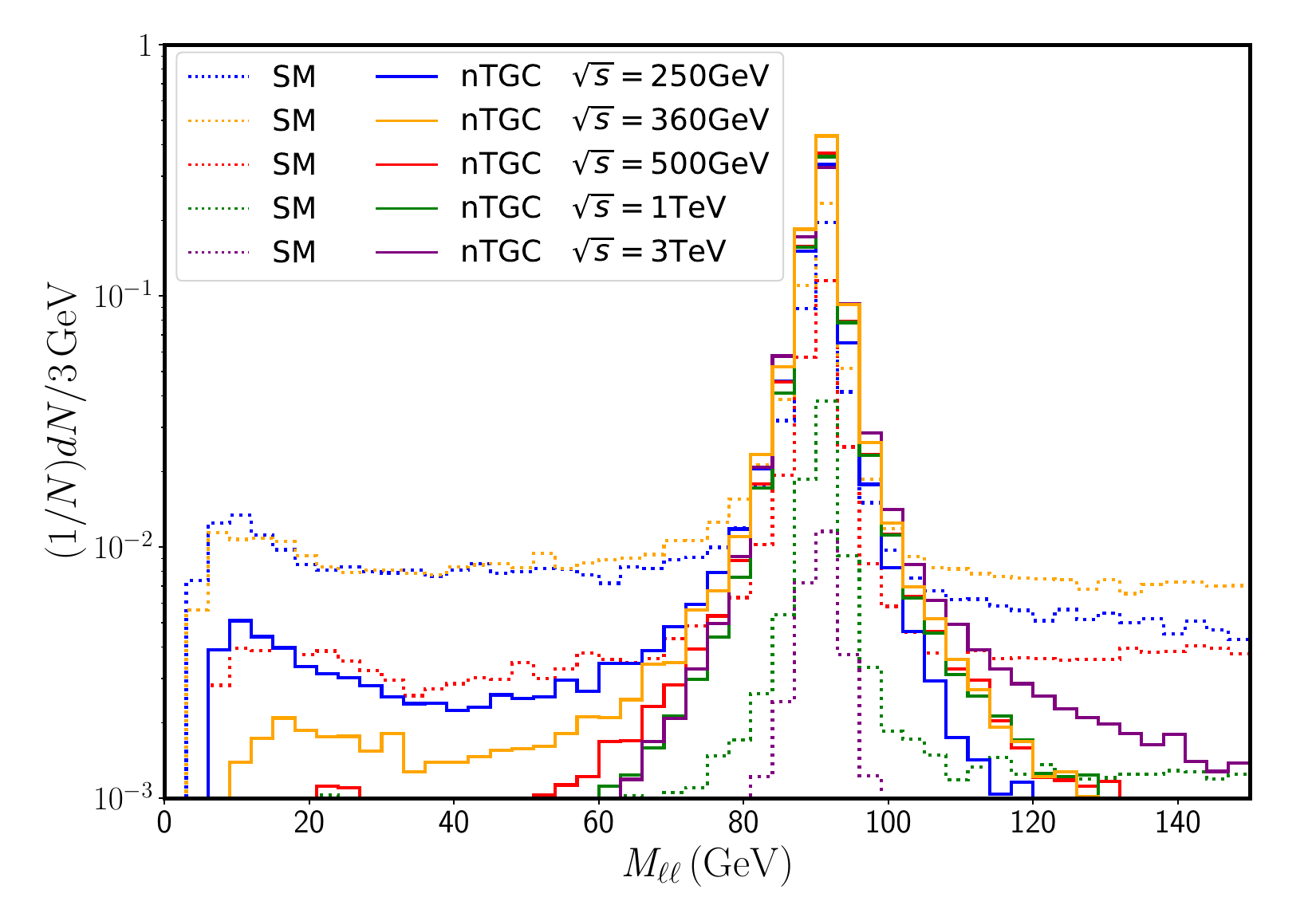}
\includegraphics[width=0.45\hsize]{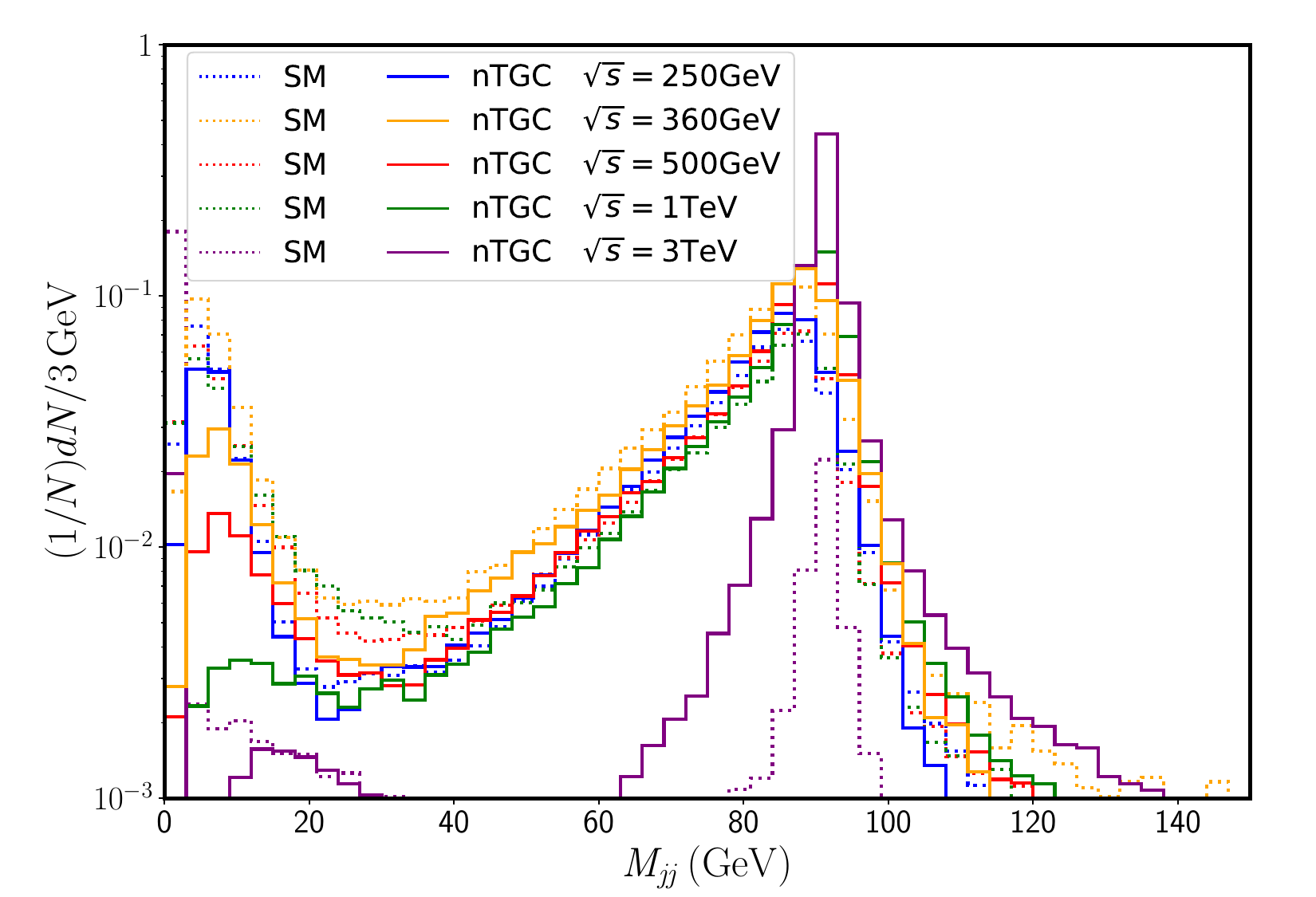}
\includegraphics[width=0.45\hsize]{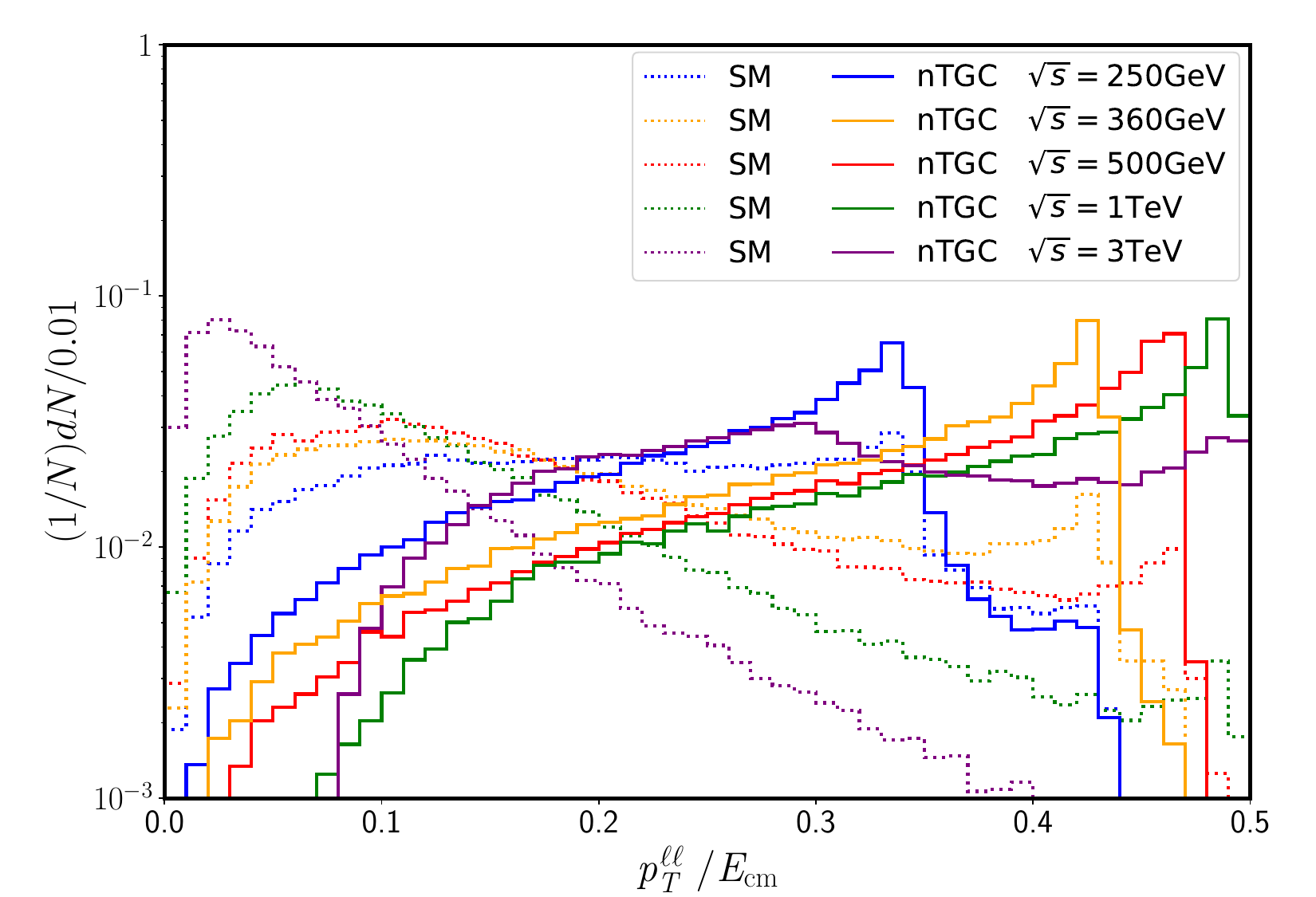}
\caption{The normalized distributions of $\cos{\theta}$,
 $M_{\ell\ell}$, $M_{jj}$, $p_T^{{\ell\ell}}/E_{\rm{cm}}$ for $e^+e^-\to 2\ell 2j$.}
\label{fig:KinematicFeature-2l2j}
\end{figure}

To remove the backgrounds without a $Z$ resonance, we require the invariant mass of the leptons and jets, denoted as $M_{\ell\ell}$ and $M_{jj}$, to be close to $M_Z$.
For events in the $\ell^+ \ell^- j j$ final states, the $Z$ direction was taken to be the direction of the reconstructed di-jet or $\ell^+ \ell^-$ pair, respectively.
The normalized distributions of $M_{\ell\ell}$ and $M_{jj}$ are shown in Fig.~\ref{fig:KinematicFeature-2l2j}.
In the distributions of the signal, the peaks at $M_Z$ are much sharper than those in the distributions of the SM backgrounds.
We cut off the events with the mass window cut $70<M_{\ell\ell}<110$ GeV.
It should be noted that in the case of $\sqrt{s}=250$ and 360 GeV, $M_{\ell\ell}$ has a resonance peak mainly from the $Z\gamma^*$ production processes at about 10 GeV.
For CEPC, the mass window cut increases the range by an additional $5<M_{\ell\ell}<20$ GeV.
The $M_{jj}$ distributions for the nTGC signal and the SM background show similarity at low energies. However, these distributions become more distinguishable at higher energies. Consequently, we included $M_{jj}$ as a selection criterion only in the 3 TeV analysis.

After applying $M_{\ell\ell,jj}$ cut, the SM backgrounds in Fig.~\ref{fig:2l2j-bg}~(b,c) can be excluded, leaving the events from Fig.~\ref{fig:2l2j-bg}~(a).
The dominate contribution of nTGCs is through $s$-channel of the di-boson diagram, Fig.~\ref{fig:2l2j-sig}~(a).
In the signal event, leptons and jets come from the decay of bosons emitted in the $s$-channel diagram.
We expect the leptons and jets in the signal to be more energetic than those in the background, and anticipate the same for their parent particle, the reconstructed $Z$ boson.
Thus $p_T^{{\ell\ell}}/{E_{\rm{cm}}}$ will work for this channel.
At high energies, the direction of the lepton or jet is approximately consistent with that of the hightly boosted $Z$.
Therefore, in the c.m. frame, the angle $\theta_{\ell\ell}$ between the two leptons in signal are different from those in background as shown in Fig.~\ref{fig:KinematicFeature-2l2j}. We cut off the events with a large $\cos{\theta_{\ell\ell}}$ at high energy scale.
This effect is not obvious on CEPC with lower c.m. energy, but the signal selection efficiency can still be slightly improved by a cut of $-0.6<\cos{\theta}<0.2$.
The specific values of event selection strategy and the respective cut efficiencies are presented in Table~\ref{tab:2l2j-cutflow}.

\begin{table}[hb]
    \centering
    \caption{The event selection strategy and cross sections of $e^+e^-\to 2\ell 2j$ (in unit of fb) after cuts. }
    \label{tab:2l2j-cutflow}
    \begin{tabular}{lllllll}
    \hline
    \hline
    \multirow{2}{*}{~}&\multicolumn{2}{c}{$\sqrt{s}$=250 GeV}&~&~&\multicolumn{2}{c}{$\sqrt{s}$=360 GeV}\\
    \cmidrule(r){2-3}   \cmidrule(r){6-7} \\
    \noalign{\smallskip}
    \multirow{2}{*}{~}&SM&NP&~&~&SM&NP\\ \hline
    Basic Cuts &86.08&0.556&~&~&63.43&7.454\\
    $-0.7<\cos{\theta_{\ell\ell}}<0.2$&53.66&0.482&~&~&~&~\\  $-0.5<\cos{\theta_{\ell\ell}}<0.7$&~&~&~&~&38.55&7.058\\
    $M_{\ell\ell}\in$\\ $(5,20)\cup(70,110)$ GeV &47.79&0.471&~&~&30.21&6.814\\
    $p_T^{{\ell}{\ell}}/E_{\rm{cm}}>0.1$&40.95&0.440&~&~&23.52&6.540\\
    Efficiency $\epsilon$ &35\%&60\%&~&~&27\%&68\%\\
    \hline
    \multirow{2}{*}{~}&\multicolumn{2}{c}{$\sqrt{s}$=500 GeV}&\multicolumn{2}{c}{$\sqrt{s}$=1 TeV}&\multicolumn{2}{c}{$\sqrt{s}$=3 TeV}\\
    \cmidrule(r){2-3}  \cmidrule(r){4-5} \cmidrule(r){6-7} \\
    \noalign{\smallskip}
    \multirow{2}{*}{~}&SM&NP&SM&NP&SM&NP\\ \hline
    Basic Cuts &52.08&0.484&25.69&0.430&11.35&0.162\\  $\cos{\theta_{\ell\ell}}>0$&25.79&0.460&6.724&0.426&~&~\\  $\cos{\theta_{\ell\ell}}>0.5$&~&~&~&~&0.898&0.157\\
    $M_{\ell\ell}\in$(70,110) GeV &17.54&0.436&3.187&0.402&~&~\\
    $M_{{\ell\ell},{jj}}\in$(70,110) GeV &~&~&~&~&0.334&0.134\\
    $p_T^{{\ell}{\ell}}/E_{\rm{cm}}>0.3$&5.305&0.312&1.084&0.305&~&~\\    $p_T^{{\ell}{\ell}}/E_{\rm{cm}}>0.1$&~&~&~&~&0.325&0.133\\
    Efficiency $\epsilon$ &7.4\%&51\%&2.9\%&54\%&2.1\%&71\%\\    \hline
    \hline
    \end{tabular}
\end{table}

\begin{figure}[H]
\centering
\subfigure[250 GeV]{\includegraphics[width=0.4\hsize]{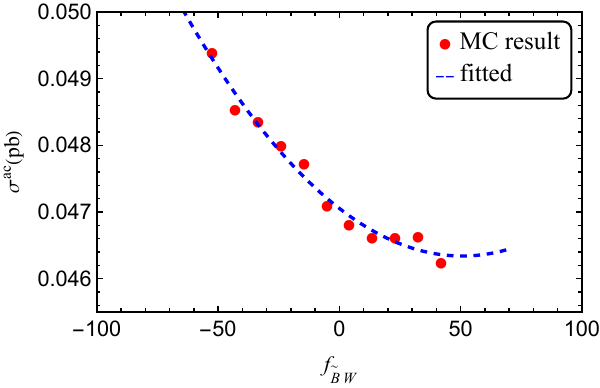}}
\subfigure[360 GeV]{\includegraphics[width=0.4\hsize]{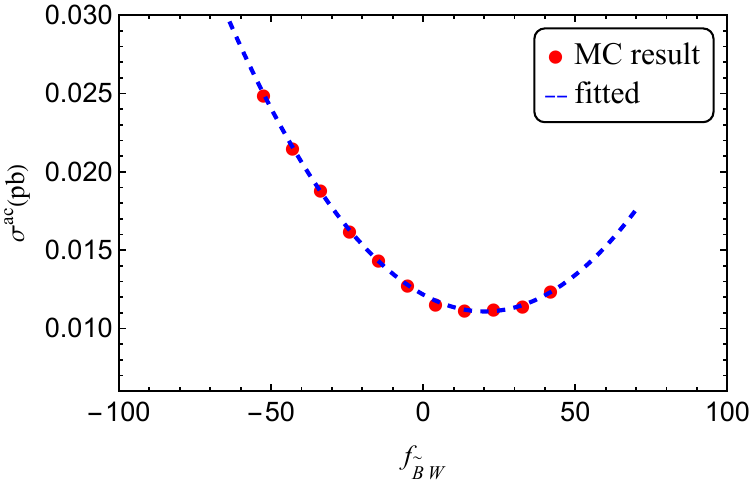}}
\subfigure[500 GeV]{\includegraphics[width=0.4\hsize]{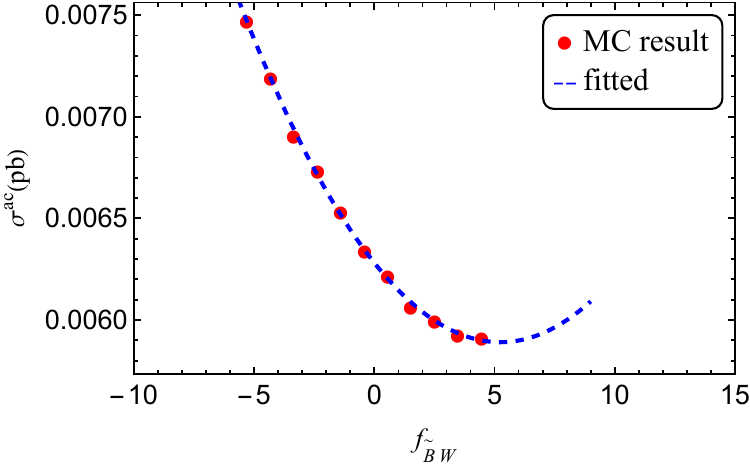}}
\subfigure[1 TeV]{\includegraphics[width=0.4\hsize]{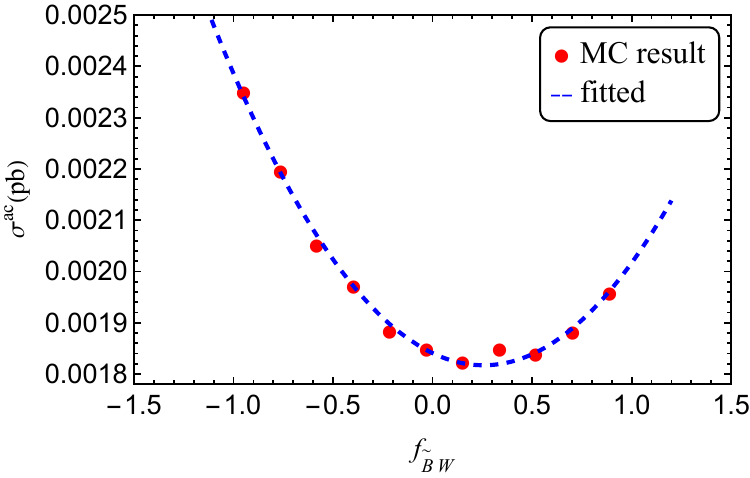}}
\subfigure[3 TeV]{\includegraphics[width=0.45\hsize]{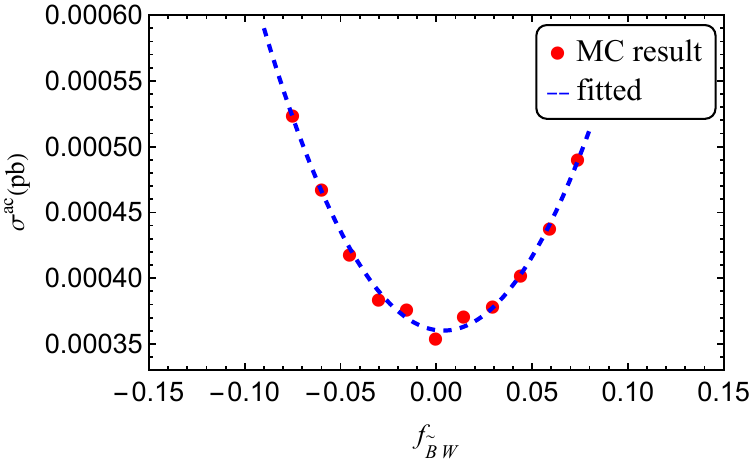}}
\caption{ Cross sections as functions of $f_{\tilde{B}W}$ ($\rm{TeV^{-4}}$) for $e^+e^-\to 2\ell2 j$.}
\label{fig:2l2j-csfit}
\end{figure}

With the event selection strategy of $e^+e^-\to 2\ell 2j$ applied, the total cross sections are obtained by scanning the values of coefficient with $\mathcal{O}_{\tilde{B}W}$, and the fitting functions are shown in Fig.~\ref{fig:2l2j-csfit}. The fitting coefficients of interference terms are given in Table~\ref{tab.2l2j-fit}.

\begin{table}[H]
\begin{center}
\caption{\label{tab.2l2j-fit}The fitted values of $\sigma^{\rm ac}_{\rm NP}$~($\rm fb\times TeV^8$) and $\hat{\sigma} _{\rm int}$~($\rm fb\times TeV^4$) for $e^+e^-\to 2\ell 2j$ at each energy point, where $\sigma^{\rm ac}_{\rm NP}$ and $\hat{\sigma} _{\rm int}$ is defined in Eq.~(\ref{eq.3.6}).}
\begin{tabular}{c|c|c|c|c|c}
\hline
$\sqrt{s}$  & 250 GeV & 360 GeV & 500 GeV & 1 TeV & 3 TeV \\
\hline
$\hat{\sigma}_{\rm int}$ & $-0.0282$ & $-0.1064$& $-0.1489$  & $-0.1848$  & $-0.1219$ \\
\hline
$\sigma^{\rm ac}_{\rm NP}$ & $0.0003$ & $0.0026$ & $0.0142$  & $0.3607$  & $24.179$ \\
\hline
\end{tabular}
\end{center}
\end{table}

The sensitivities for the $e^+e^-\to 2\ell2j$ channel and the expected constraints on the coefficients obtained with Eq.~\ref{eq:ss}.
We obtain the projected sensitivity on $f_{\tilde{B}W}$  by taking $2\sigma$, $3\sigma$ or $5\sigma$ significance.
The results are shown in Table~\ref{tab:constraints-2l2j}.

\begin{table}[H]
    \centering
    \caption{The excepted constraints on $f_{\tilde{B}W}$ ($\rm{TeV^{-4}}$) for $e^+e^-\to 2\ell 2j$.}
    \label{tab:constraints-2l2j}
    \begin{tabular}{llllll}
    \hline
    \multirow{2}{*}{$S_{stat}$}&$\sqrt{s}$ (GeV)&\\ \cline{2-6}
     ~&250&360&500&1000&3000\\ \hline
     2&[$-6.5,107.7$]&[$-2.0,42.5$]&[$-1.4,11.8$]&[$-0.30,0.81$]&[$-0.022,0.030$]\\
     3&[$-9.5,110.7$]&[$-2.9,43.4$]&[$-2.0,12.4$]&[$-0.40,0.92$]&[$-0.028,0.036$]\\
     5&[$-15.0,116.3$]&[$-4.7,45.2$]&[$-3.0,13.5$]&[$-0.58,1.09$]&[$-0.038,0.045$]\\ \hline
    \end{tabular}
\end{table}

The lepton pairs in the $2\ell 2j$ final states have a distinctive signature making possible selections with high efficiencies and low background contaminations. In comparison to previous section, the $2\ell2j$  has the most sensitive detection capability for $\mathcal{O}_{\tilde{B}W}$ operator in the high energy region.

\subsection{Selection of \texorpdfstring{$e^+e^-\to \ell^+\ell^-\nu\bar{\nu}$}{2l+missing}  }
\label{sec3.4}
\begin{figure}[H]
\centering
\includegraphics[height=0.45\hsize]{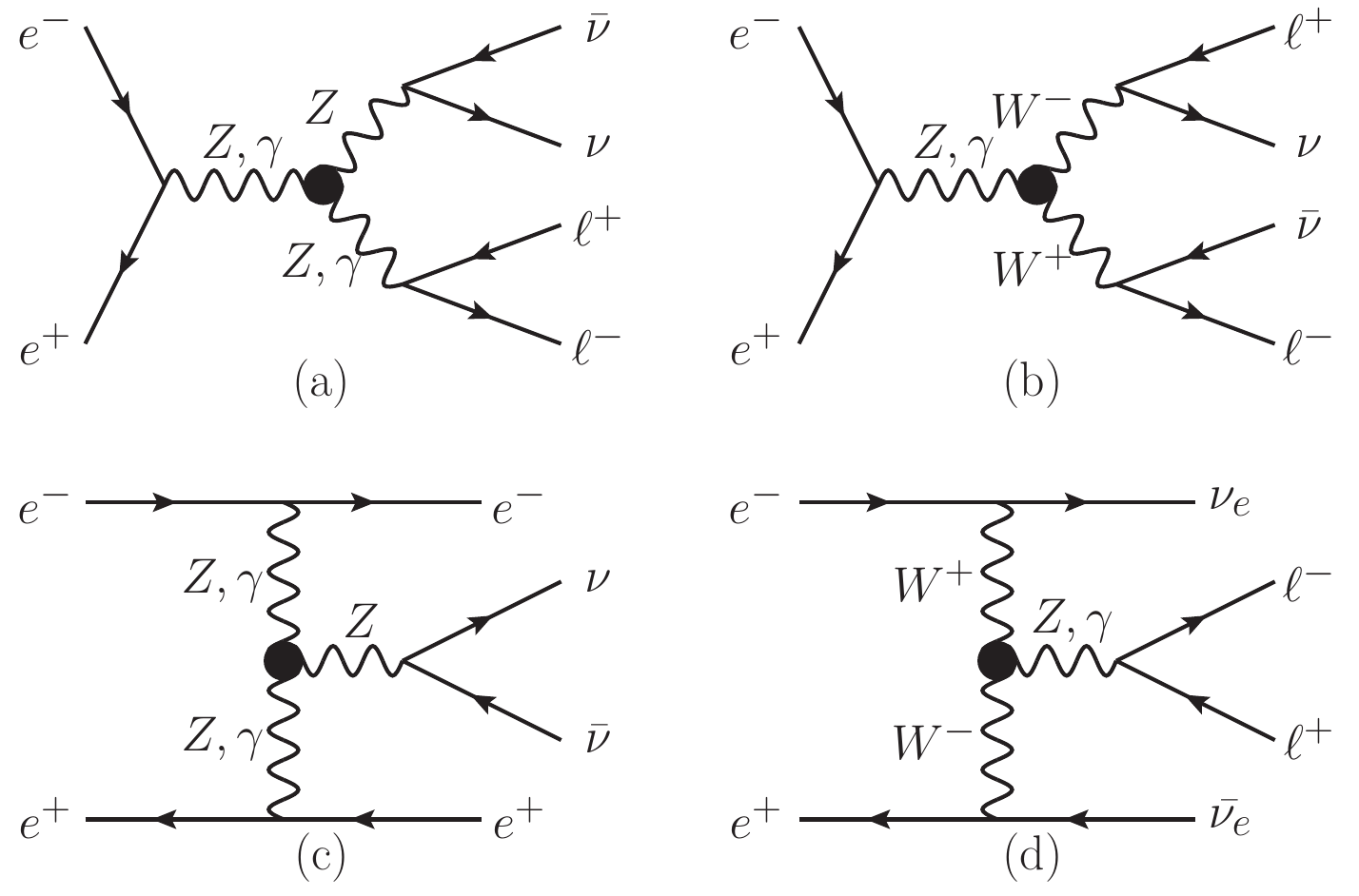}
\caption{Typical Feynman diagrams which contribute to the process $e^+e^-\to \ell^+ \ell^- \nu \bar{\nu}$ for signal events.}
\label{fig:llvv-sig}
\end{figure}

\begin{figure}[H]
\centering
\includegraphics[width=0.5\hsize]{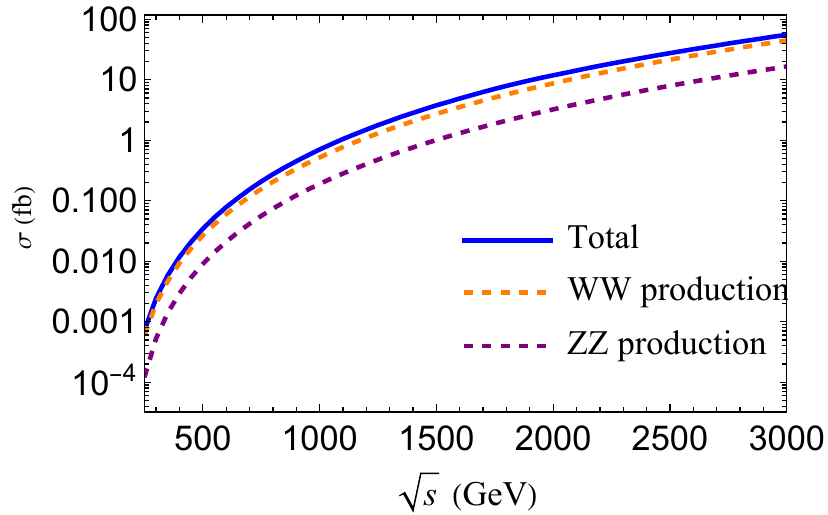}
\caption{The comparison of $\sigma_{\rm Total}$, $\sigma_{\rm WW}$ and $\sigma_{\rm ZZ}$ for $e^+e^-\to \ell\ell\nu\bar{\nu}$ as a function of c.m. energy $\sqrt{s}$ with $f_{\tilde{B}W} = 1 \ (\rm{TeV}^{-4})$.}
\label{fig:ZZ&WW}
\end{figure}

The Feynman diagrams of the contribution induced by nTGCs to the process $e^+e^-\to \ell^+\ell^-\nu\bar{\nu}$ are given in the Fig.~\ref{fig:llvv-sig}. The dominant signal is the di-boson contribution induced by nTGCs.
The contribution from the VBF process is negligible compared to the bi-bosonic process. As shown in Fig.~\ref{fig:ZZ&WW}, the contribution from $WW$ production is more significant than that from $ZZ$ production. Therefore, in the analysis of signal and background, the $\ell^+\ell^-\nu\bar{\nu}$ signal will primarily reflect the kinematic distribution characteristics of $WW$ and $ZZ$ production.

\begin{figure}[H]
\centering
\includegraphics[height=0.45\hsize]{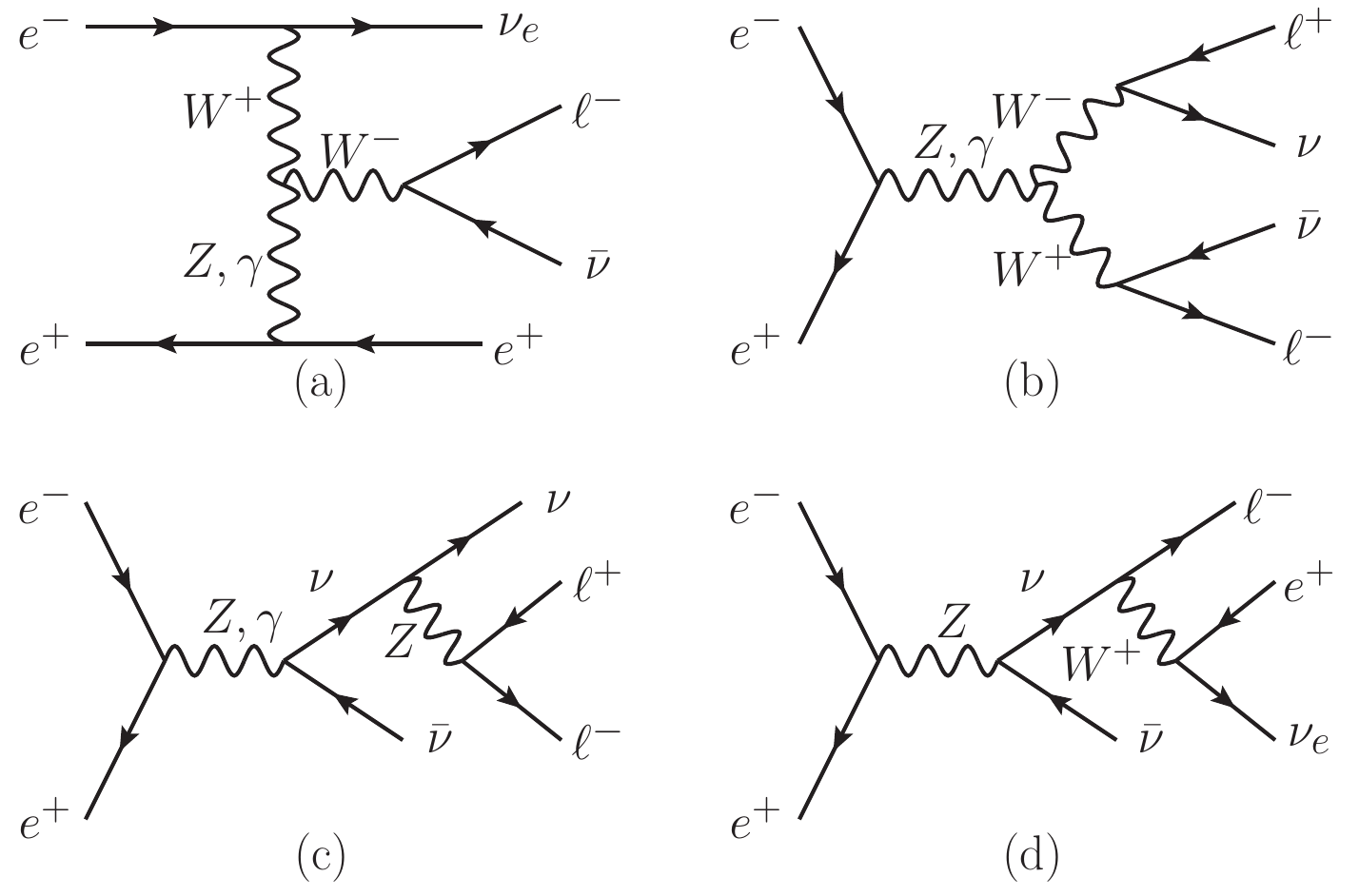}
\caption{Typical Feynman diagrams which contribute to the process $e^+e^-\to \ell^+ \ell^- \nu \bar{\nu}$ for background events.}
\label{fig:llvv-bg}
\end{figure}

The irreducible backgrounds in the $\ell^+\ell^-\nu\bar{\nu}$ channel arise from di-boson production ($WW$ and $ZZ$), electroweak VBF and neutrino production in association with a vector boson that decays leptonically.
The typical Feynman diagrams of the SM backgrounds are shown in Fig.~\ref{fig:llvv-bg}.
In order to suppress the SM backgrounds, the following basic cuts are used: $\Delta R_{\ell\ell}$ = 0.2; $p_T^\ell$ = 10 GeV; $N_{\ell} \geq 2$ and the two leptons must be a lepton and an anti-lepton of the same flavor. After these cuts, we further employ optimized kinematical cuts according to the kinematical distributions of the signal and background.

\begin{figure}[H]
\centering
\includegraphics[width=0.45\hsize]{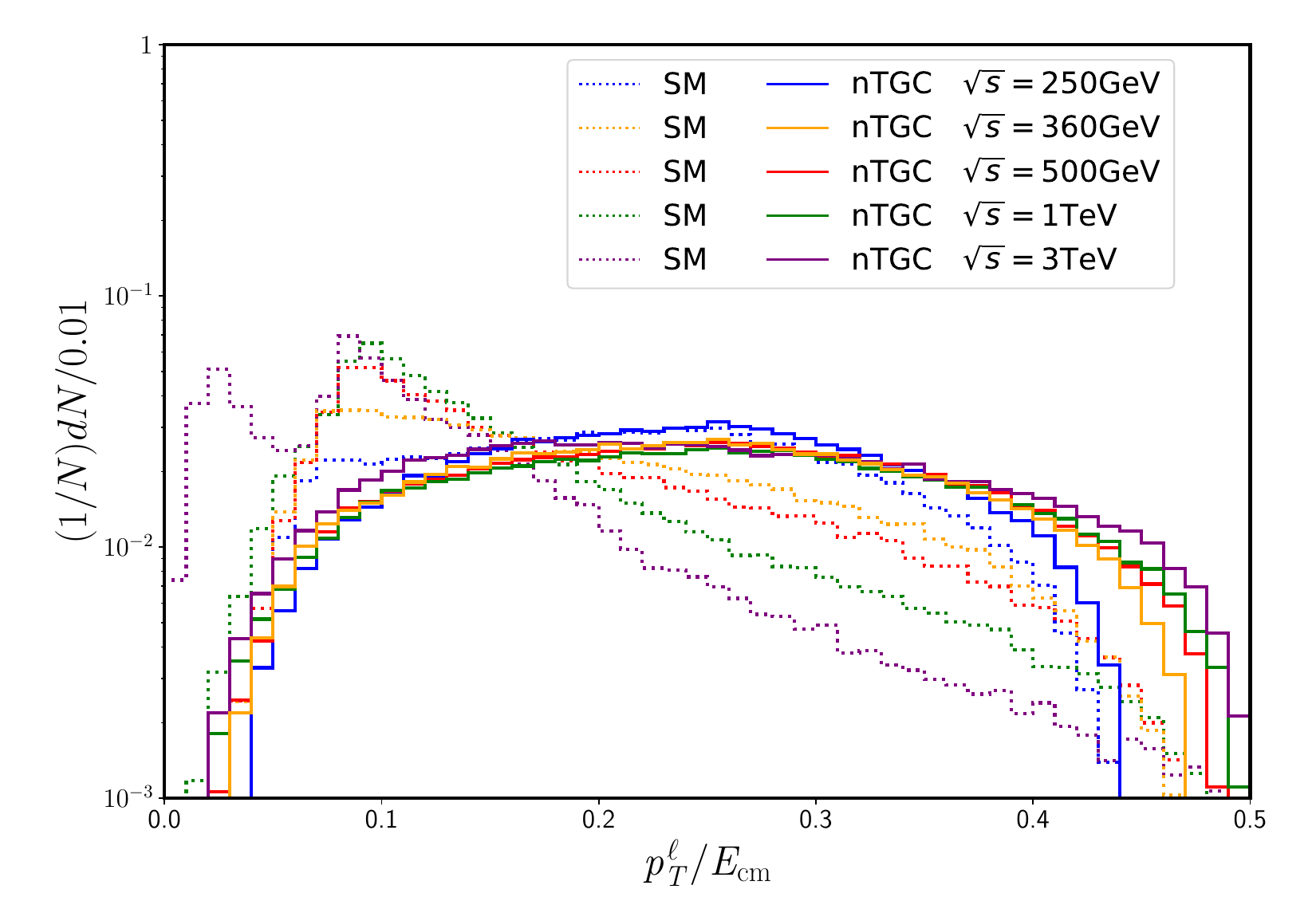}
\includegraphics[width=0.45\hsize]{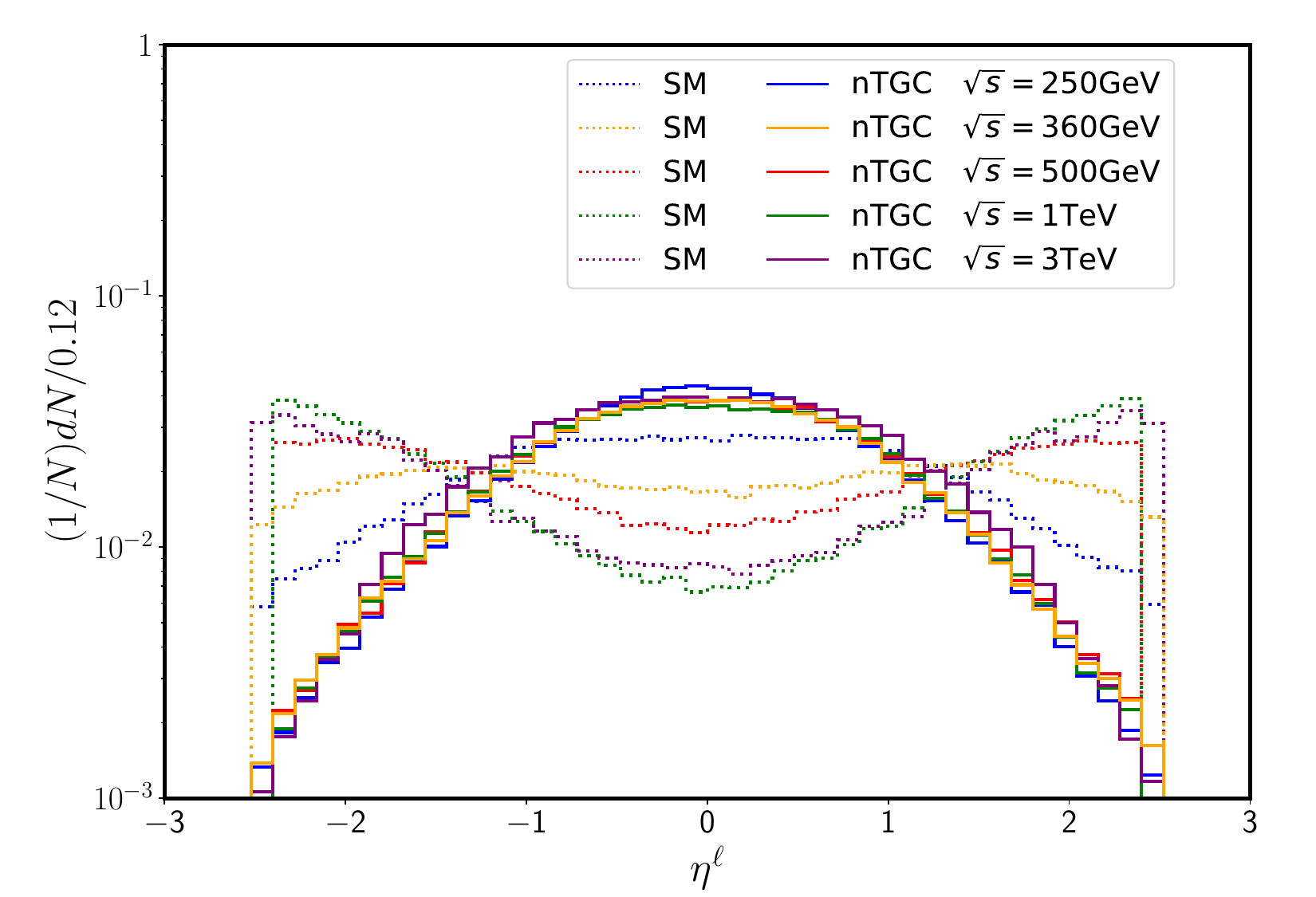}
\includegraphics[width=0.45\hsize]{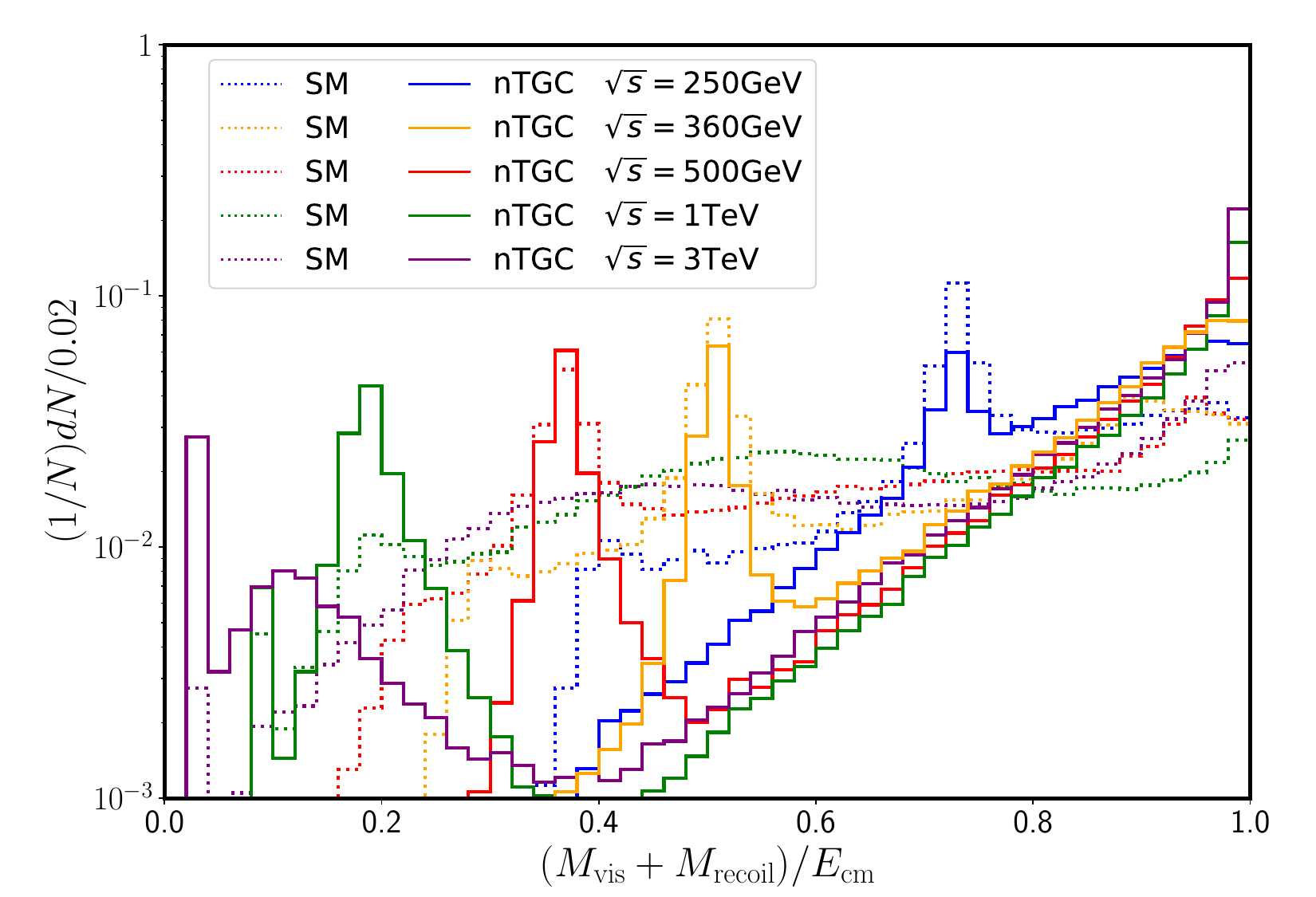}
\includegraphics[width=0.45\hsize]{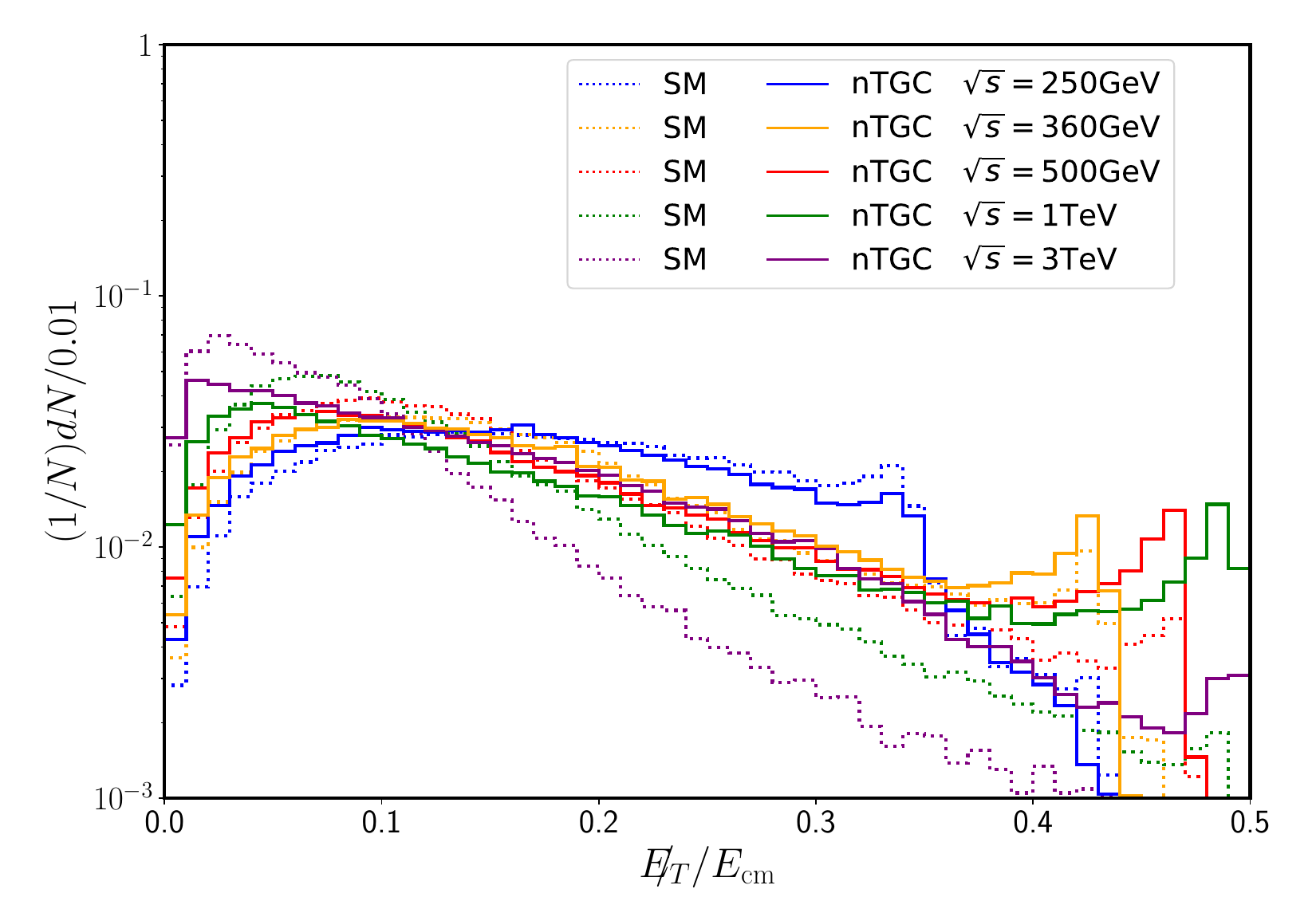}
\caption{The normalized distributions of $p_T^\ell/{E_{\rm{cm}}}$, $\eta^\ell$,  $(M_{\rm vis}+M_{\rm recoil})/\sqrt{s}$ and $\slashed{E}_T/\sqrt{s}$ for $e^+e^-\to \ell^+\ell^- \nu\bar{\nu}$.}
\label{fig:fig:KinematicFeature-2lvv}
\end{figure}

In the event of two detectable leptons and invisible particles in the final state, the invariant mass, pseudorapidity, missing energy and reconstructed invariant mass can be extracted.
We display the normalized distribution of these kinematic variables in Fig.~\ref{fig:fig:KinematicFeature-2lvv} with the largest coefficients listed in Table~\ref{Tab:region}.
As the majority of leptons in the background are produced via radiation decay, it can be observed that the transverse momentum of these leptons is comparatively smaller than that of the signal.
The $Z$ boson is typically produced in the central region in the nTGC process, and the leptons of $Z$ decay are nearly collinear with the highly boosted $Z$, resulting in a relatively small distribution of pseudorapidity for the associated lepton.
The candidate lepton is required to have $p^\ell_T > 0.2 E_{\rm cm}$, with $|\eta^\ell|<1.5$ for CEPC and $|\eta^\ell|<1$ for ILC and CLIC.

\begin{table}[htbp]
    \centering
    \caption{The event selection strategy and cross sections of $e^+e^-\to \ell^+\ell^-\nu\bar{\nu}$ (fb) after cuts. }
    \label{tab:cutflow-llvv}
    \footnotesize
    \begin{tabular}{lllllll}
    \hline
    \hline
    \multirow{2}{*}{~}&\multicolumn{2}{c}{$\sqrt{s}$=250 GeV}&~&~&\multicolumn{2}{c}{$\sqrt{s}$=360 GeV}\\
    \cmidrule(r){2-3}  \cmidrule(r){6-7}
    \noalign{\smallskip}
    \multirow{2}{*}{~}&SM&NP&~&~&SM&NP\\ \hline
    Basic Cuts &87.72&12.03&~&~&62.04&86.36\\
    $\left|{\eta^{\ell}}\right|$<1.5 &68.15&10.87&~&~&38.03&77.06\\
    $(M_{\rm vis}+M_{\rm recoil})>120 \rm{GeV}$&64.54&40.69&~&~&36.81&76.85\\
    Efficiency $\epsilon$ &59\%&72\%&~&~&46\%&70\%\\
    \hline
    \multirow{2}{*}{~}&\multicolumn{2}{c}{$\sqrt{s}$=500 GeV}&\multicolumn{2}{c}{$\sqrt{s}$=1 TeV}&\multicolumn{2}{c}{$\sqrt{s}$=3 TeV}\\
    \cmidrule(r){2-3}  \cmidrule(r){4-5} \cmidrule(r){6-7}
    \noalign{\smallskip}
    \multirow{2}{*}{~}&SM&NP&SM&NP&SM&NP\\ \hline
    Basic Cuts &49.49&5.686&32.00&3.794&16.33&2.230\\
    $p_T^\ell/{E_{\rm{cm}}}$>0.2&19.29&3.903&9.075&2.600&2.906&1.462\\
    $\left|{\eta^\ell}\right|$<1.0&12.90&3.409&4.827&2.270&1.175&1.252\\
    $(M_{\rm vis}+M_{\rm recoil})\in $\\ $(150,210){\rm GeV}\cup(0.7 E_{\rm{cm}},E_{\rm{cm}})$&9.625&3.117&2.719&2.002&0.584&1.129\\
    Efficiency $\epsilon$ &15\%&43\%&6.4\%&41\%&2.7\%&44\%\\
    \hline
    \hline
    \end{tabular}
\end{table}

\begin{figure}[H]
\centering
\subfigure[250 GeV]{\includegraphics[width=0.45\hsize]{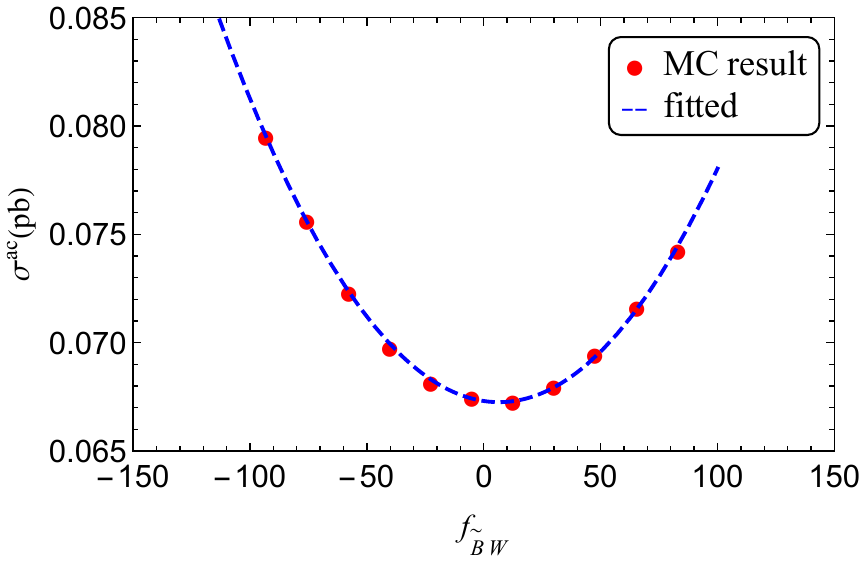}}
\subfigure[360 GeV]{\includegraphics[width=0.45\hsize]{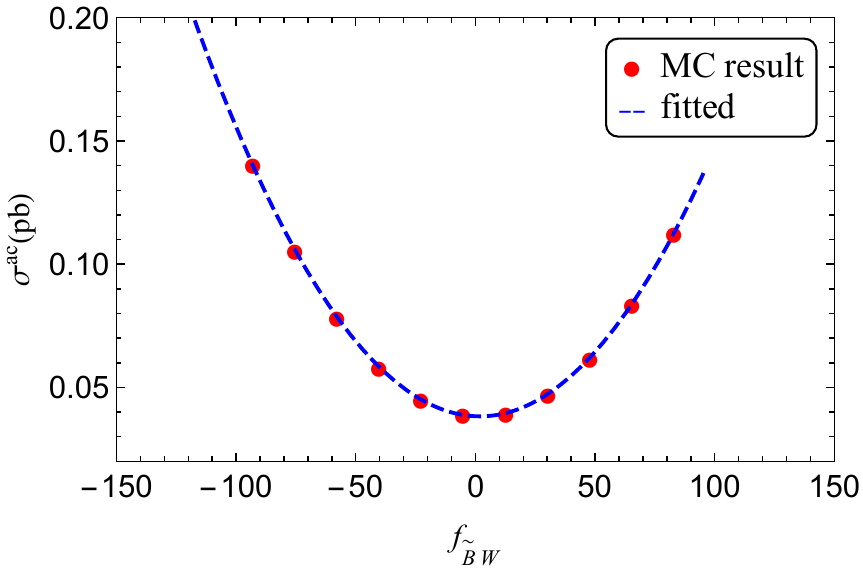}}
\subfigure[500 GeV]{\includegraphics[width=0.45\hsize]{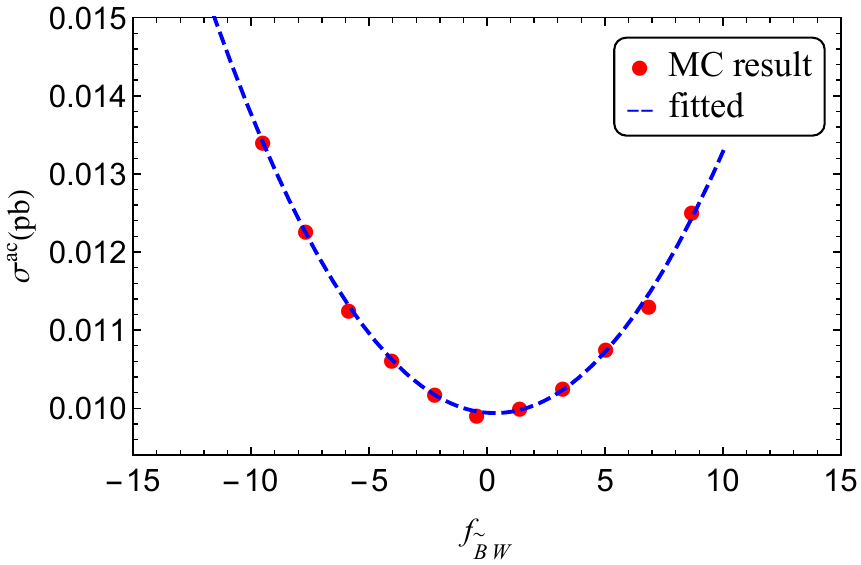}}
\subfigure[1 TeV]{\includegraphics[width=0.45\hsize]{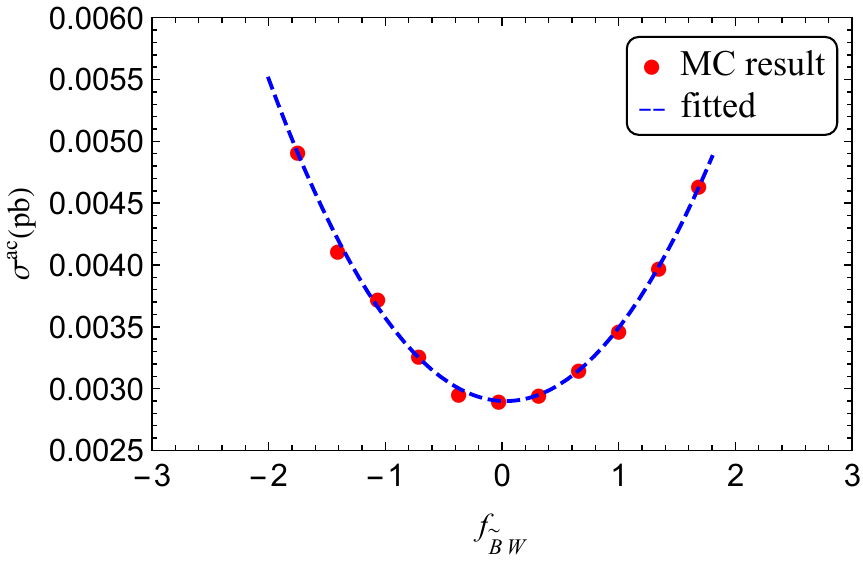}}
\subfigure[3 TeV]{\includegraphics[width=0.45\hsize]{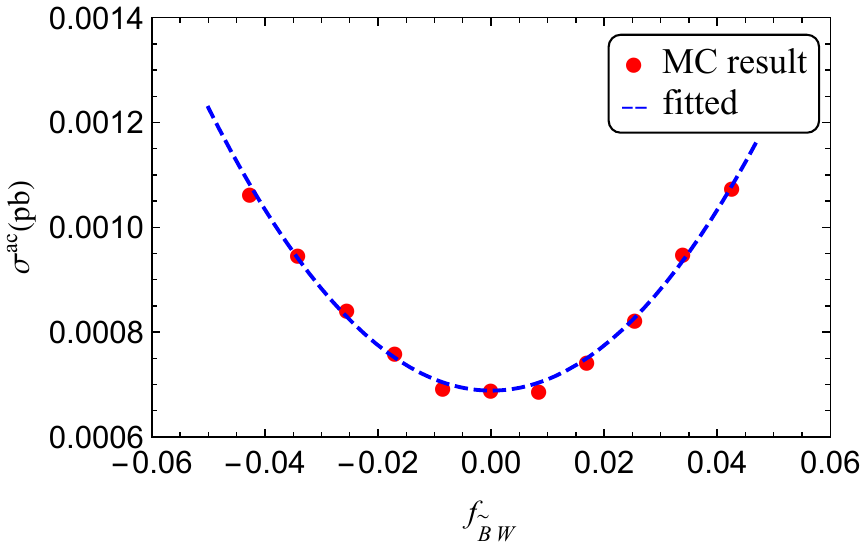}}
\caption{ Cross sections as functions of $f_{\tilde{B}W}$ ($\rm{TeV^{-4}}$) for $e^+e^-\to \ell^+\ell^-\nu\bar{\nu}$.}
\label{fig:2lvv-csfit}
\end{figure}

The most significant SM background comes from $WW$ production, where both $W$ bosons decay leptonically.
The on-shell $ZZ$ events for this final state can be selected by applying cuts on the invariant mass of the $\ell^+\ell^-$ pair and the missing transverse energy $\slashed{E}_T$.
The missing energy distributions for both the signal and background are similar.
Therefore, $\slashed{E}_T$ is not utilized in the event selection strategy.
$M_{\rm vis}$ and $M_{\rm recoil}$ are calculated from the reconstructed the $Z$ boson and invisible recoil mass obtained through energy conservation.
Then, the normalized sum of visible and recoil masses $(M_{\rm vis}+M_{\rm recoil})/\sqrt{s}$ can be used in the analysis.
Fig.~\ref{fig:fig:KinematicFeature-2lvv} shows that the distribution of $(M_{\rm vis}+M_{\rm recoil})/\sqrt{s}$ exhibits two peaks: one near
2$M_Z$, corresponding to $ZZ$ production, and another in a higher mass region, which is attributed to $WW$ production.
Given that the distribution of this observable varies with different collision energies, we propose two distinct sets of cut schemes for CEPC and ILC (CLIC) to enhance selection performance.
The specific values of event selection strategy and the respective cut efficiencies are presented in Table~\ref{tab:cutflow-llvv}.

\begin{table}[H]
\begin{center}
\caption{\label{tab.2lvv-fit}The fitted values of $\sigma^{\rm ac}_{\rm NP}$~($\rm fb\times TeV^8$) and $\hat{\sigma} _{\rm int}$~($\rm fb\times TeV^4$) for $e^+e^-\to\ell^+\ell^-\nu\bar{\nu}$ at each energy point.}
\begin{tabular}{c|c|c|c|c|c}
\hline
$\sqrt{s}$  & 250 GeV & 360 GeV & 500 GeV & 1 TeV & 3 TeV \\
\hline
$\hat{\sigma}_{\rm int}$ & $-0.0161$ & $-0.0429$ & $-0.0241$  & $-0.0403$  & $-0.0132$ \\
\hline
$\sigma^{\rm ac}_{\rm NP}$ & $0.0012$ & $0.0113$ & $0.0358$  & $0.6316$  & $215.60$ \\
\hline
\end{tabular}
\end{center}
\end{table}

With the event selection strategy applied for the final state $\ell^+\ell^-\nu\bar{\nu}$, the cross section fitting functions of $\mathcal{O}_{\tilde{B}W}$ are shown in Fig.~\ref{fig:2lvv-csfit}, and the fitting coefficients of the interference terms are provided in Table~\ref{tab.2lvv-fit}.
The sensitivity analysis and the expected constraints on the $\mathcal{O}_{\tilde{B}W}$ operator are presented in Table~\ref{tab:constraints-2l2j}.
Although the $e^+e^- \to \ell^+\ell^-\nu\bar{\nu}$ channel is not the most optimal, the expected constraints on $f_{\tilde{B}W}$ at $95\%$ confidence level (C.L.) can still reach approximately $-4 \, \text{TeV}^{-4}$ to $8 \, \text{TeV}^{-4}$ for $\mathcal{O}_{\tilde{B}W}$ at the 360 GeV CEPC.
At ILC and CLIC, with higher energy and luminosity, these constraints are further improved, achieving orders of magnitude around $10^{-1} \, \text{TeV}^{-4}$ and $10^{-2} \, \text{TeV}^{-4}$, respectively.

\begin{table}[H]
    \centering
    \caption{The excepted constraints on $f_{\tilde{B}W}$  ($\rm{TeV^{-4}}$) for $e^+e^-\to \ell^+\ell^- \nu\bar{\nu}$.}
    \label{tab:constraints-2lvv}
    \begin{tabular}{lllllll}
    \hline
    \multirow{2}{*}{$S_{stat}$}&$\sqrt{s}$(GeV)&\\ \cline{2-6}
     ~&250&360&500&1000&3000\\ \hline
     2&[$-8.7,21.7$]&[$-4.3,8.1$]&[$-2.51,3.18$]&[$-0.39,0.45$]&[$-0.010,0.011$]\\
     3&[$-11.5,24.6$]&[$-5.6,9.4$]&[$-3.15,3.83$]&[$-0.48,0.55$]&[$-0.013,0.013$]\\
     5&[$-16.2,29.2$]&[$-7.7,11.5$]&[$-4.19,4.87$]&[$-0.64,0.70$]&[$-0.017,0.017$]\\ \hline
    \end{tabular}
\end{table}

\subsection{ Selection of \texorpdfstring{$e^+e^-\to jj\nu\bar{\nu}$}{2j+missing} }
\label{sec3.5}
\begin{figure}[ht]
\centering
\includegraphics[height=0.19\hsize]{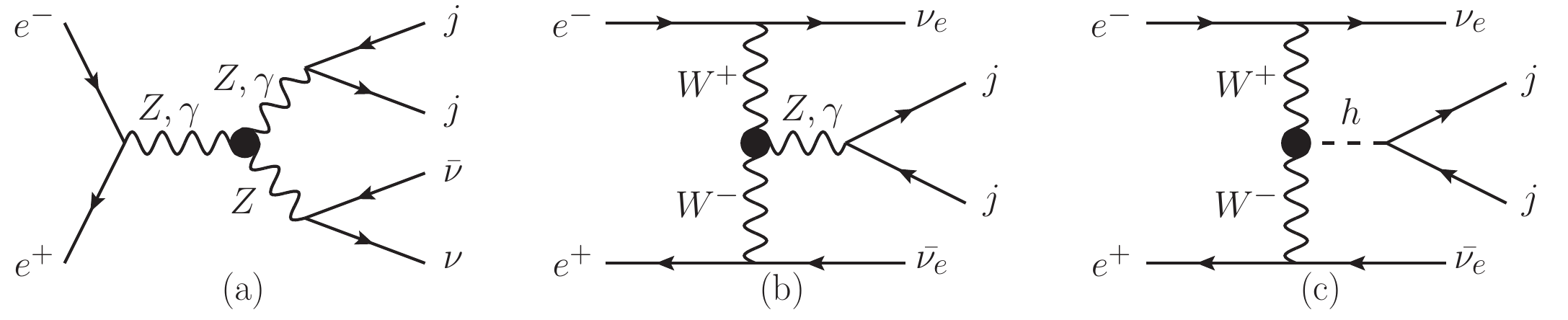}
\caption{Typical Feynman diagrams which contribute to the process $e^+e^-\to j j \nu \bar{\nu}$ for signal events.}
\label{fig:2jvv-sig}
\end{figure}

The signal $jj\nu\bar{\nu}$ from the $\mathcal{O}_{\tilde{B}W}$ operator originates from the diagrams depicted in Fig.~\ref{fig:2jvv-sig}.
In this channel, di-boson production provides the dominant contribution, with the $jj\nu\bar{\nu}$ decay mode representing approximately $28\%$ of the $ZZ$ final states.
The key signature of this decay is a large invisible energy component alongside a pair of jets with a reconstructed mass close to the $Z$ boson mass.

\begin{figure}[H]
\centering
\includegraphics[height=0.4\hsize]{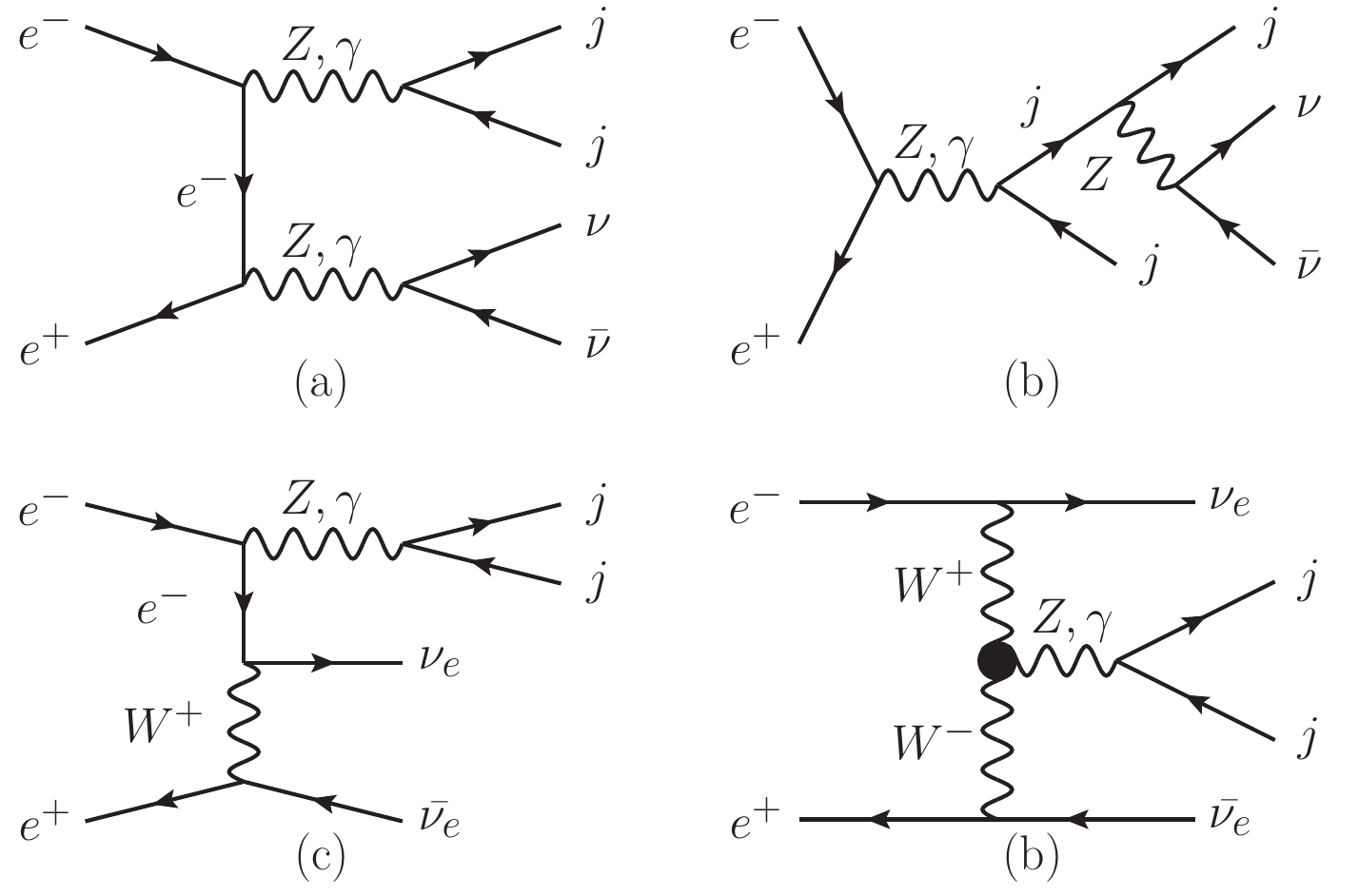}
\caption{The Feynman diagrams which contribute to the process $e^+e^-\to j j \nu \bar{\nu}$ for background events.}
\label{fig:2jvv-bg}
\end{figure}

The most challenging backgrounds arise from multiple sources: di-boson production with one boson undergoing an invisible decay, di-jet events accompanied by energetic vector boson decays, single resonant $Z\nu\bar{\nu}$ production, and the vector boson fusion (VBF) process, as illustrated in Fig.~\ref{fig:2jvv-bg}.

\begin{figure}[H]
\centering
\includegraphics[width=0.45\hsize]{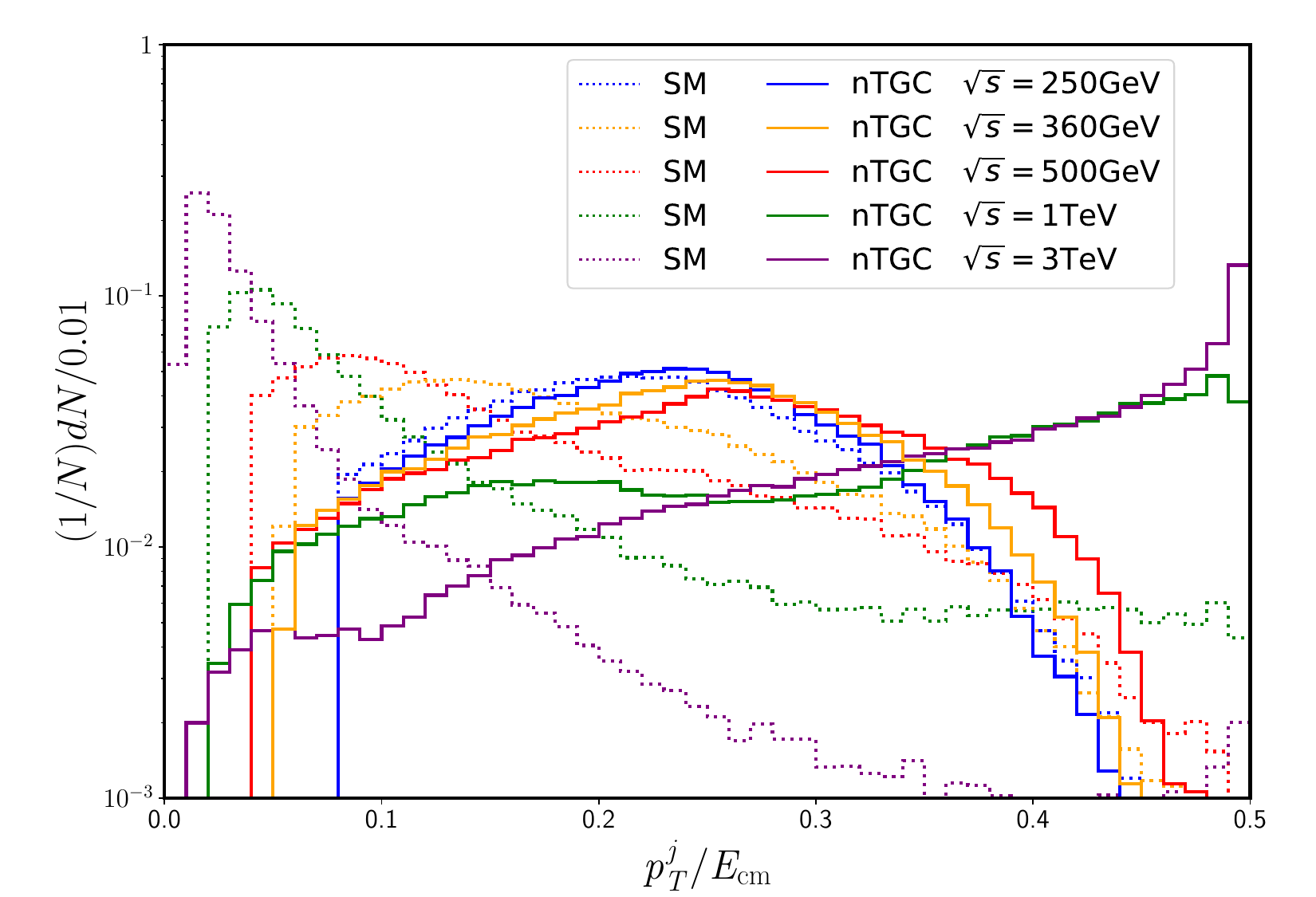}
\includegraphics[width=0.45\hsize]{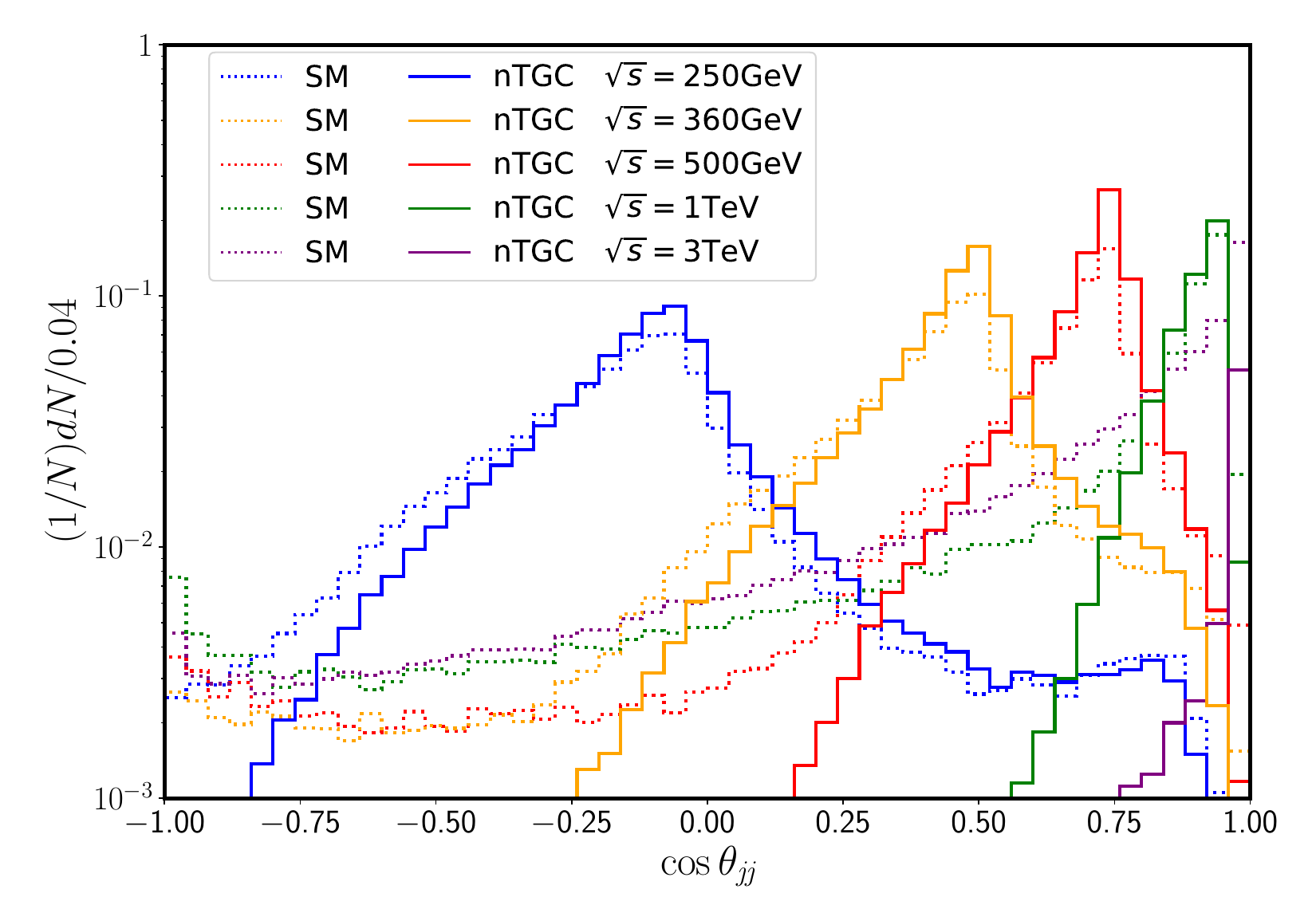}
\includegraphics[width=0.45\hsize]{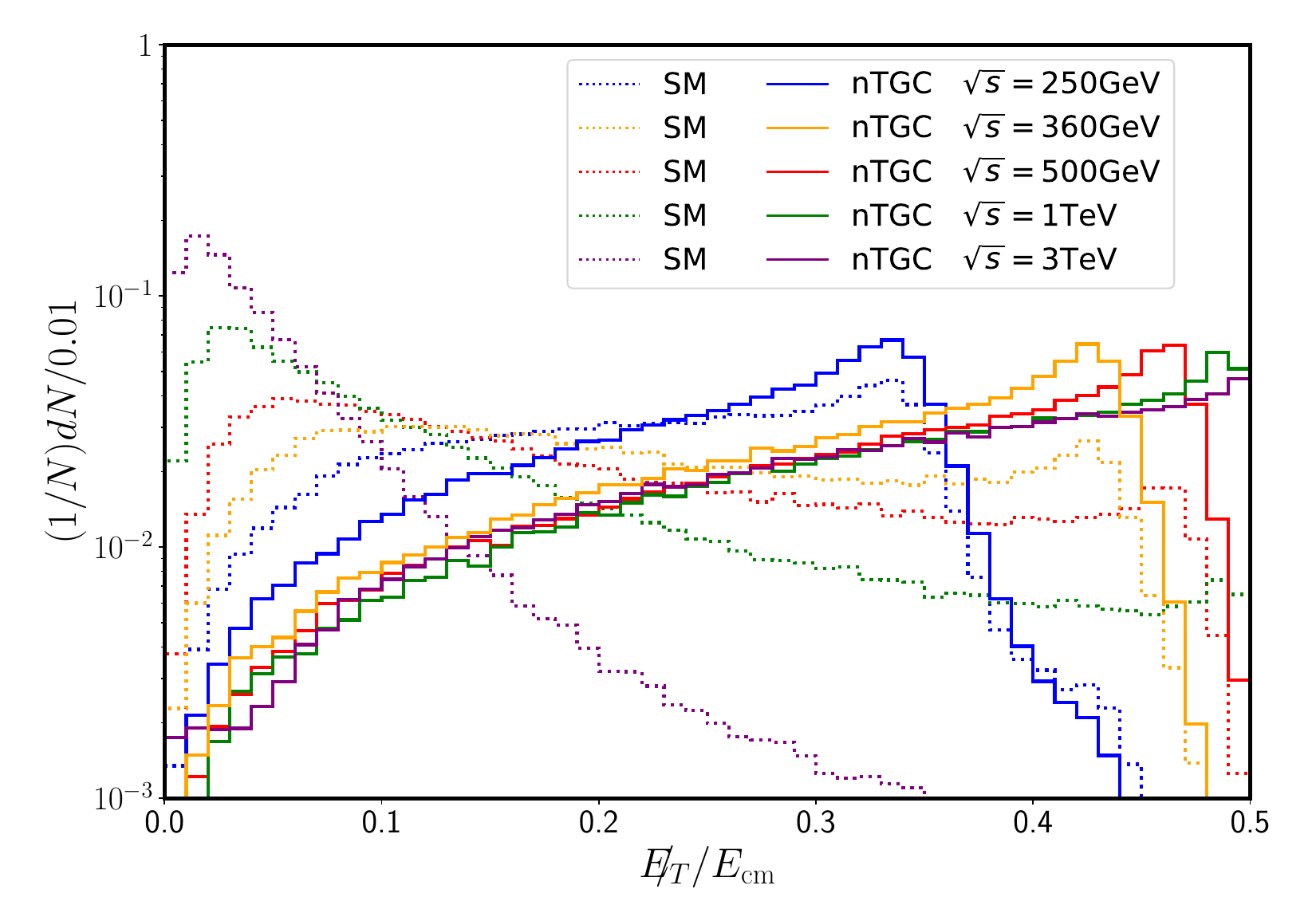}
\includegraphics[width=0.45\hsize]{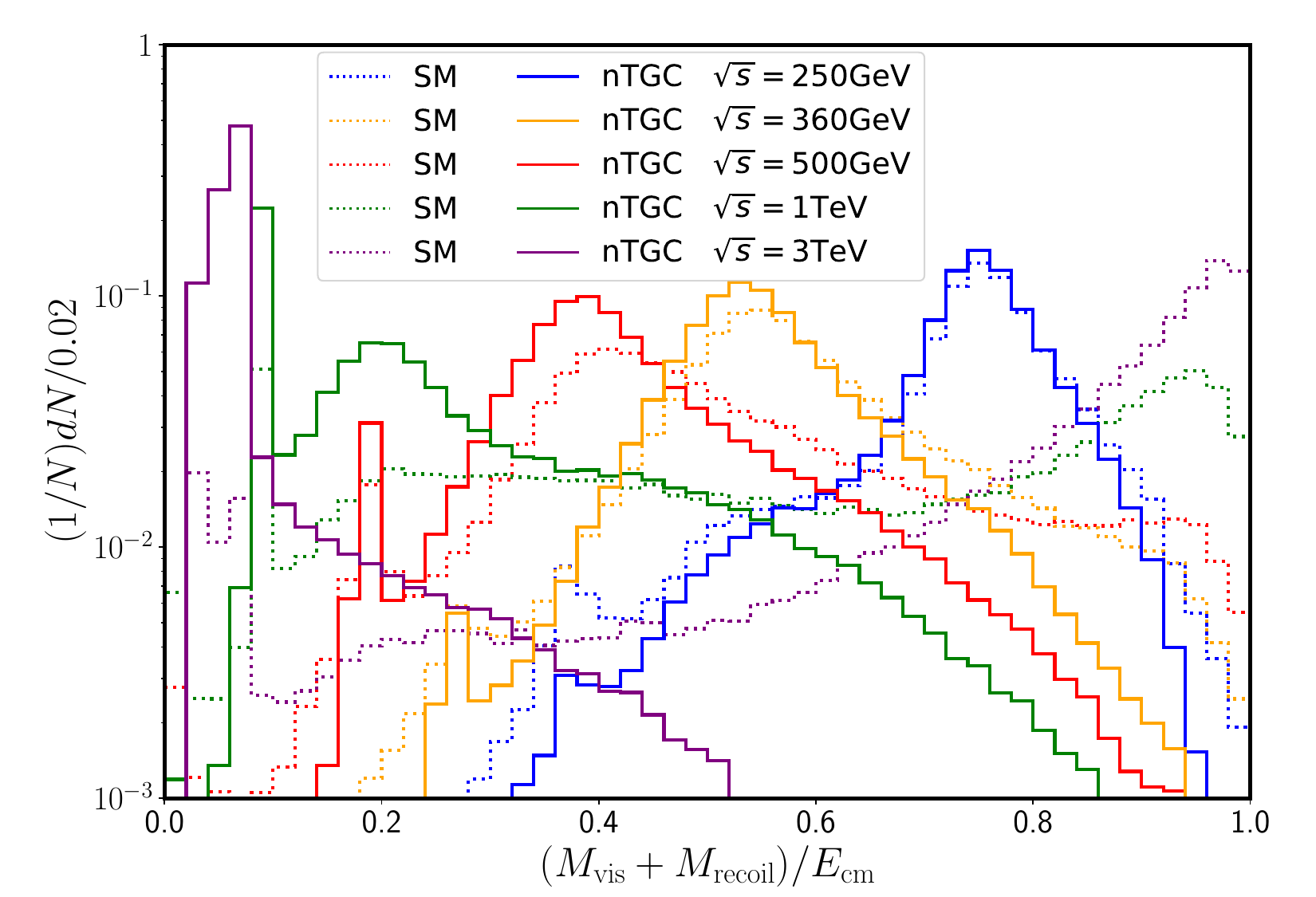}
\caption{The normalized distributions of $p_T^j/{E_{\rm{cm}}}$, $\cos{\theta_{jj}}$, $\slashed{E}_T/{E_{\rm{cm}}}$ and $(M_{\rm vis}+M_{\rm recoil})/\sqrt{s}$ for $e^+ e^- \to j j \nu\bar{\nu}$}
\label{fig:KinematicFeature-2jvv}
\end{figure}
The $\ell^+\ell^-\nu\bar{\nu}$ channel discussed in the previous section exhibit very clean signals at $e^+e^-$ colliders. However, hadronic decay mode hold certain advantages due to their larger branching ratios.
For the nTGC signal, events were generated using the largest coefficients listed in Table~\ref{Tab:region}.
The kinematic analysis initially selected events containing a hadronic jet with transverse momentum $p_T^j = 10$~GeV and pseudorapidity $|\eta^j| < 2.0$.
Following these basic cuts, we required a minimum of two jets in each event, and applied a jet separation criterion of $\Delta R_{jj} = 0.2$, as previously discussed.

The properties of hadronic and invisible decays can effectively differentiate the signal from the SM background.
Fig.~\ref{fig:KinematicFeature-2jvv} presents the normalized distributions of $p_T^j/{E_{\rm{cm}}}$, $\cos{\theta_{jj}}$, $\slashed{E}_T/{E_{\rm{cm}}}$, and $(M_{\rm vis}+M_{\rm recoil})/{E_{\rm{cm}}}$.
In signal events, the $Z$ or Higgs bosons are generally more energetic than those in background events, leading to a smaller angle between the jets and higher $p_T^j$ values in the signal, predominantly in the high-energy region.
The transverse missing energy $\slashed{E}_T$ serves to approximate the transverse energy of the $Z$ boson in its $\nu\bar{\nu}$ decay mode. As expected, $\slashed{E}_T$ in the signal is generally larger than in the background.
These distinctions become increasingly pronounced with rising collision energies, particularly at TeV-scale colliders.

\begin{table}[H]
    \centering
    \caption{The event selection strategy and cross sections of $e^+e^-\to jj\nu\bar{\nu}$ (fb) after cuts. }
    \label{tab:cutflow-2jvv}
    \begin{tabular}{lllllll}
    \hline
    \hline
    \multirow{2}{*}{~}&\multicolumn{2}{c}{$\sqrt{s}$=250 GeV}&~&~&\multicolumn{2}{c}{$\sqrt{s}$=360 GeV}\\
    \cmidrule(r){2-3}  \cmidrule(r){6-7}
    \noalign{\smallskip}
    \multirow{2}{*}{~}&SM&NP&~&~&SM&NP\\ \hline
    Basic Cuts &198.5&0.614&~&~&120.4&8.407\\
    $\slashed{E_T}/E_{\rm{cm}}>0.1$&173.1&0.574&~&~&~&~\\
    $\slashed{E_T}/E_{\rm{cm}}>0.2$&~&~&~&~&62.61&6.996\\
    $(M_{\rm vis}+M_{\rm recoil}) < 0.9E_{\rm{cm}}$ &168.6&0.567&~&~&~&~\\
    $(M_{\rm vis}+M_{\rm recoil}) < 0.8E_{\rm{cm}}$ &~&~&~&~&60.92&6.928\\     $\cos{\theta_{jj}}>-0.6$&162.5&0.55&~&~&~&~\\
    $\cos{\theta_{jj}}>-0.2$&~&~&~&~&59.93&6.894\\
    Efficiency $\epsilon$ &80.3\%&89.9\%&~&~&49\%&82\%\\
    \hline
    \multirow{2}{*}{~}&\multicolumn{2}{c}{$\sqrt{s}$=500 GeV}&\multicolumn{2}{c}{$\sqrt{s}$=1 TeV}&\multicolumn{2}{c}{$\sqrt{s}$=3 TeV}\\
    \cmidrule(r){2-3}  \cmidrule(r){4-5} \cmidrule(r){6-7}
    \noalign{\smallskip}
    \multirow{2}{*}{~}&SM&NP&SM&NP&SM&NP\\ \hline
    Basic Cuts &84.46&0.565&46.12&0.558&44.09&0.385\\    $\slashed{E_T}/E_{\rm{cm}}>0.3$&21.50&0.379&6.230&0.387&0.805&0.256\\
    $(M_{\rm vis}+M_{\rm recoil}) < 0.7E_{\rm{cm}}$ &21.20&0.377&~&~&~&~\\
    $(M_{\rm vis}+M_{\rm recoil}) < 0.6E_{\rm{cm}}$ &~&~&5.915&0.381&~&~\\
    $(M_{\rm vis}+M_{\rm recoil}) < 0.3E_{\rm{cm}}$ &~&~&~&~&0.657&0.249\\
    $\cos{\theta_{jj}}>0.2$&20.80&0.375&~&~&~&~\\
    $\cos{\theta_{jj}}>0.5$&~&~&5.787&0.379&~&~\\
    $\cos{\theta_{jj}}>0.7$&~&~&~&~&0.648&0.249\\
    Efficiency $\epsilon$ &24\%&66\%&12\%&68\%&1.4\%&65\%\\
    \hline
    \hline
    \end{tabular}
\end{table}

At several hundred GeV, where $ZZ$ production is the main contributor to both signal and background, the peak of $(M_{\rm vis}+M_{\rm recoil})$ primarily appears near $2M_Z$ on the CEPC.
As the collision energy increases, the non-$ZZ$ production contributions to the background become more significant, shifting the $(M_{\rm vis}+M_{\rm recoil})$ distribution to higher mass regions.
This shift improves the background rejection performance of $(M_{\rm vis}+M_{\rm recoil})$ at high-energy colliders.

\begin{figure}[H]
\centering
\subfigure[250 GeV]{\includegraphics[width=0.45\hsize]{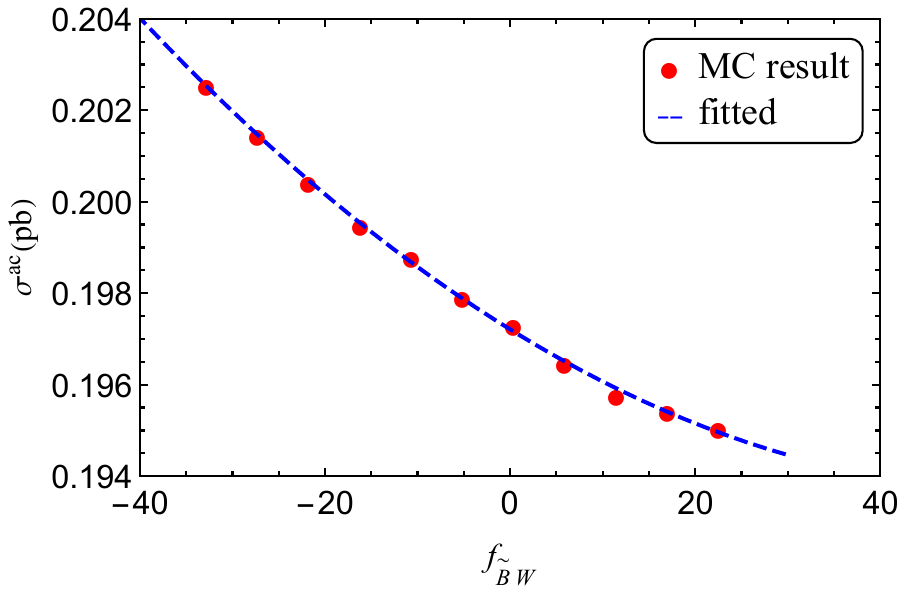}}
\subfigure[360 GeV]{\includegraphics[width=0.45\hsize]{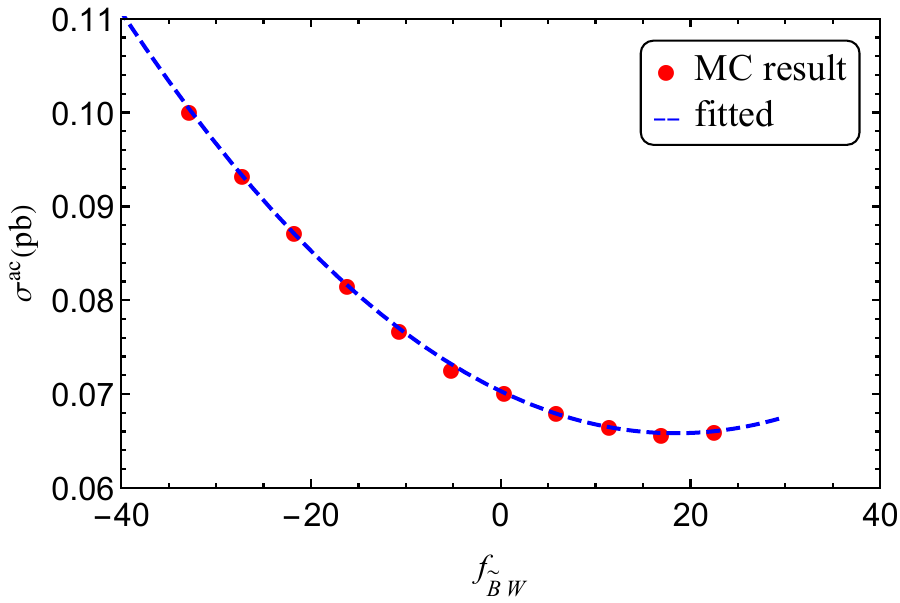}}
\subfigure[500 GeV]{\includegraphics[width=0.45\hsize]{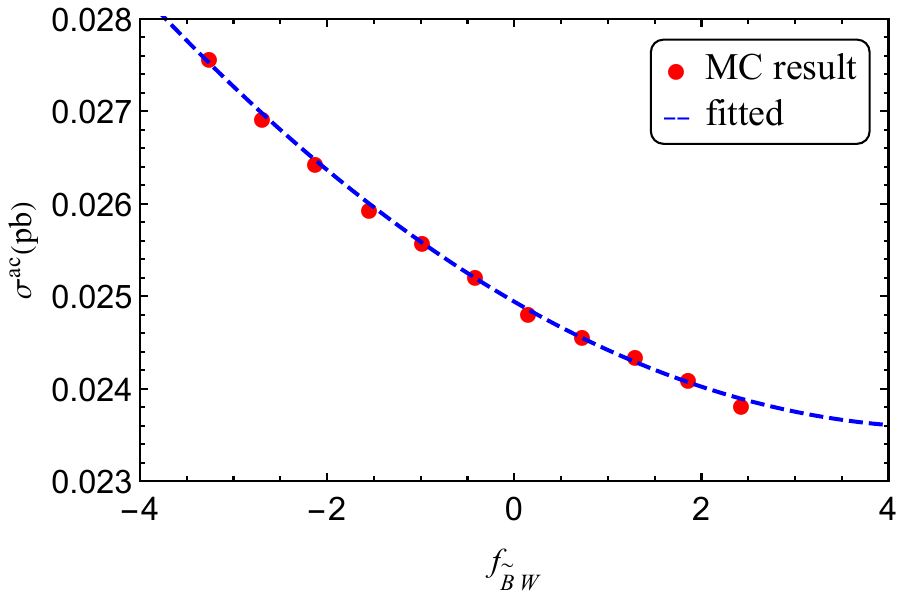}}
\subfigure[1 TeV]{\includegraphics[width=0.45\hsize]{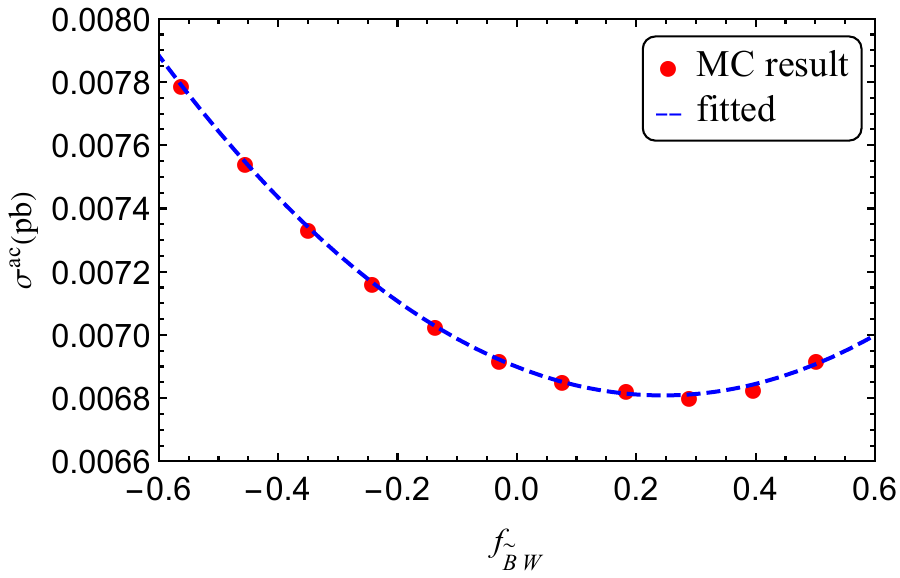}}
\subfigure[3 TeV]{\includegraphics[width=0.45\hsize]{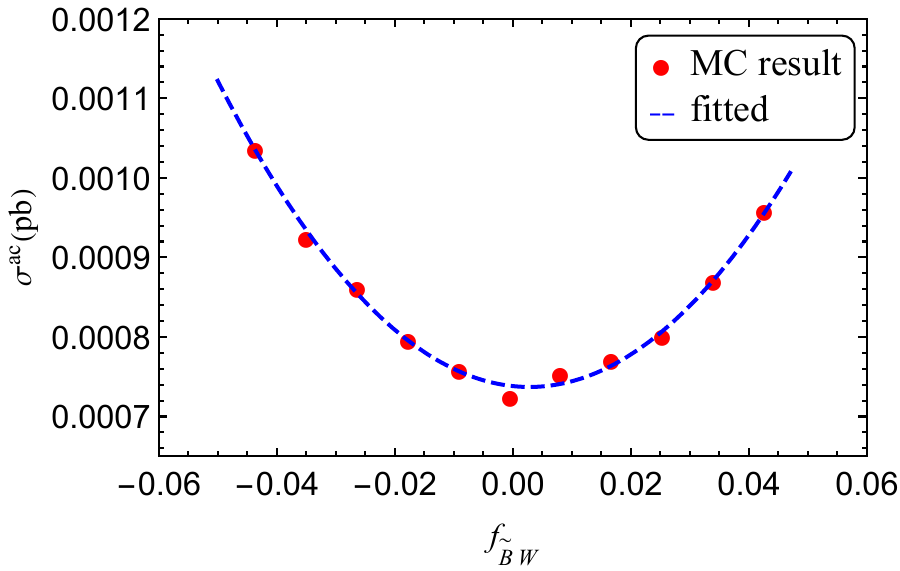}}
\caption{ Cross sections as a functions of $f_{\tilde{B}W}$ ($\text{TeV}^{-4}$) for the process $e^+e^- \to jj \nu\bar{\nu}$ at different c.m. energies.}
\label{fig:2jvv-csfit}
\end{figure}

The specific values of the event selection strategy and corresponding cut efficiencies are presented in Table~\ref{tab:cutflow-2jvv}.
For the final state $jj\nu\bar{\nu}$, the cross-section fitting functions for $\mathcal{O}_{\tilde{B}W}$ are shown in Fig.~\ref{fig:2jvv-csfit}, and the fitting coefficients for the interference terms are provided in Table~\ref{tab.2jvv-fit}.
The longitudinal polarization contribution of vector bosons dominates the SM background, and this contribution is suppressed by $\sqrt{s}$ due to an exact cancellation with the VBS process. In contrast, the cross section of the signal induced by nTGC increases with rising $\sqrt{s}$. The energy-dependent interference term depends on the combined effect of this enhancement and depression. Consequently, as shown in Fig.~\ref{fig:2jvv-csfit}, the interference between the nTGC signal and the SM background for the process $e^+e^- \to jj \nu\bar{\nu}$ decreases with increasing energy.
Table~\ref{tab:constraints-2jvv} presents the expected constraints on the nTGC operators, obtained by fitting the observed $jj \nu\bar{\nu}$ events for $2\sigma$, $3\sigma$, and $5\sigma$ significance levels at future $e^+e^-$ colliders.

\begin{table}[H]
\begin{center}
\caption{\label{tab.2jvv-fit} The fitted values of $\sigma^{\rm ac}_{\rm NP}$~($\rm fb\times TeV^8$) and $\hat{\sigma} _{\rm int}$~($\rm fb\times TeV^4$) for $e^+e^-\to jj \nu\bar{\nu}$ at each energy point.}
\begin{tabular}{c|c|c|c|c|c}
\hline
$\sqrt{s}$  & 250 GeV & 360 GeV & 500 GeV & 1 TeV & 3 TeV \\
\hline
$\hat{\sigma}_{\rm int}$ & $-0.1252$ & $-0.4851$ & $-0.6432$  & $-0.7372$  & $-0.7797$ \\
\hline
$\sigma^{\rm ac}_{\rm NP}$ & $0.0011$ & $0.0131$ & $0.0660$  & $1.5071$  & $137.88$ \\
\hline
\end{tabular}
\end{center}
\end{table}

Summarizing the expected constraints obtained for the five channels in this section, the $4j$ signal provides the strongest constraint among the five possible nTGC signals produced by the $e^+e^- \to ZZ$ process. Potential improvements in these constraints through combined analyses are discussed in section~\ref{sec3.6}.

\begin{table}[H]
    \centering
    \caption{The expected constraints on $f_{\tilde{B}W}$ ($\rm{TeV^{-4}}$) for $e^+e^- \to jj \nu\bar{\nu}$.}
    \label{tab:constraints-2jvv}
    \begin{tabular}{lllllll}
    \hline
    \multirow{2}{*}{$S_{stat}$}&$\sqrt{s}$ (GeV)&\\ \cline{2-6}
     ~&250&360&500&1000&3000\\ \hline
     2&[$-3.1,114.5$]&[$-1.1,38.0$]&[$-0.7,10.0$]&[$-0.17,0.66$]&[$-0.011,0.017$]\\
     3&[$-4.6,116.0$]&[$-1.6,38.5$]&[$-1.0,10.3$]&[$-0.23,0.72$]&[$-0.014,0.020$]\\
     5&[$-7.5,118.9$]&[$-2.6,39.5$]&[$-1.7,10.9$]&[$-0.34,0.83$]&[$-0.019,0.024$]\\ \hline
    \end{tabular}
\end{table}

\subsection{Constraints on the coefficients with combined results}
\label{sec3.6}
By combining the results obtained from the five decay modes analyzed in this section, it is possible to extract more competitive constraints beyond those provided by the each decay mode alone.
Based on the formula
\begin{equation}
\mathcal{S}_{stat}^{\rm comb} = \sqrt{\sum {S^2_{stat}(i)}},
\end{equation}
where $S_{{stat}}(i)$ is signal significance function for each decay mode of $ZZ$ production. The numerical results of expected coefficient constraints, combining data from the five decay patterns, are presented in Table~\ref{tab:Combined Result}.
These constraints provide insight into the behavior of the coefficients for the $\mathcal{O}_{\tilde{B}W}$ operator across different $\sqrt{s}$ ranging from 250 GeV to 3 TeV.

\begin{figure}[htbp]
\centering
\includegraphics[width=0.7\hsize]{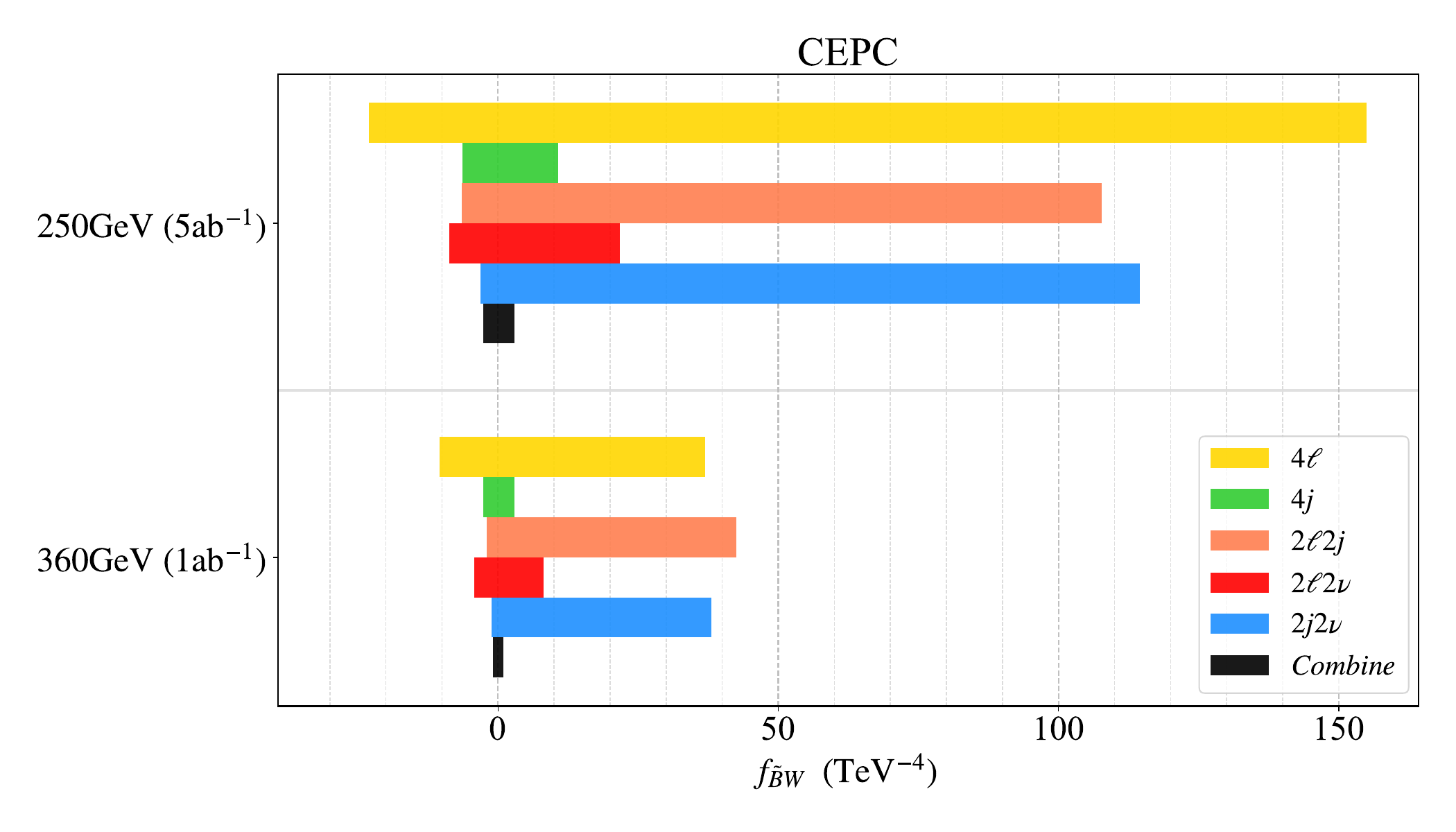}
\includegraphics[width=0.7\hsize]{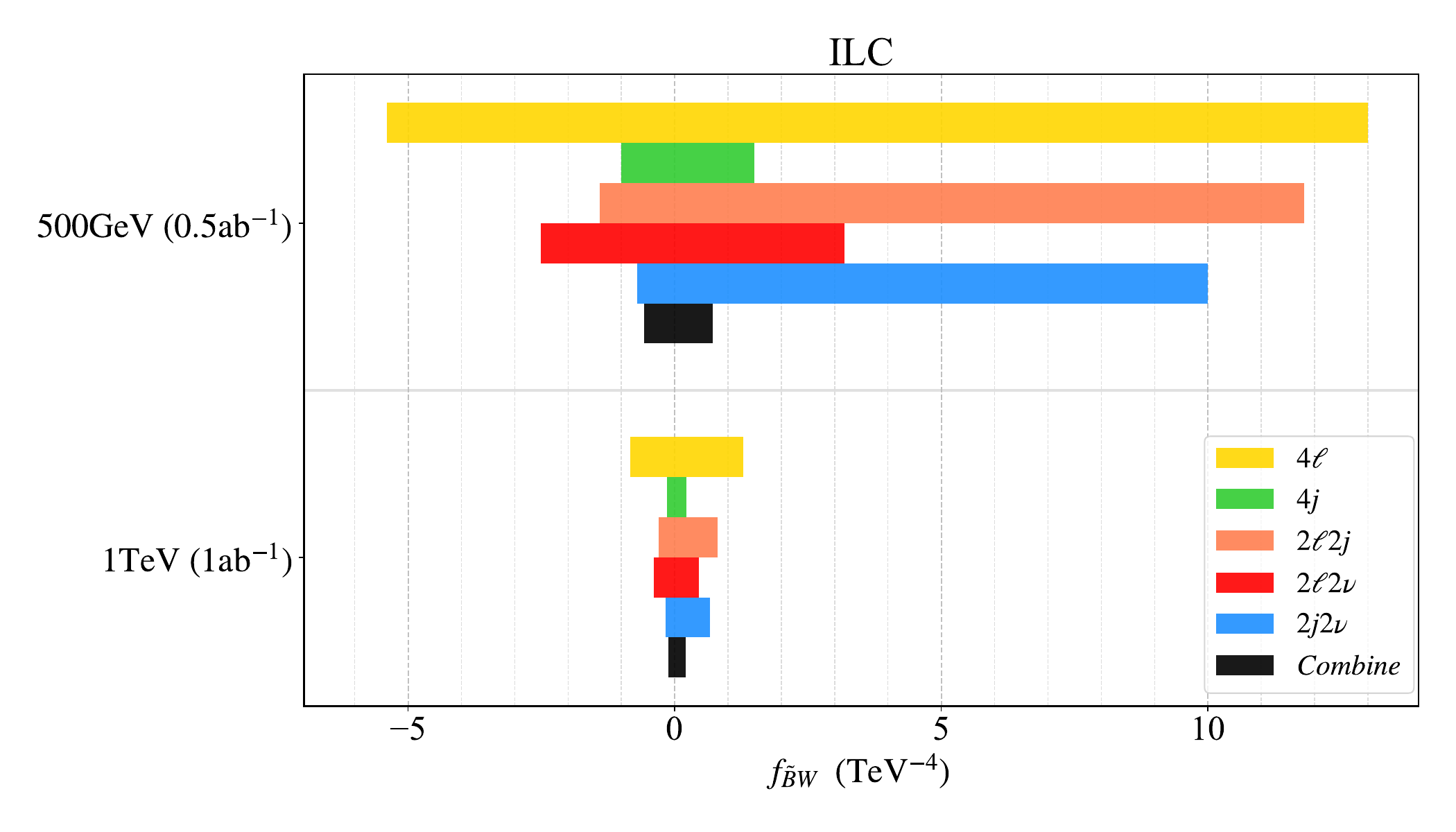}
\includegraphics[width=0.7\hsize]{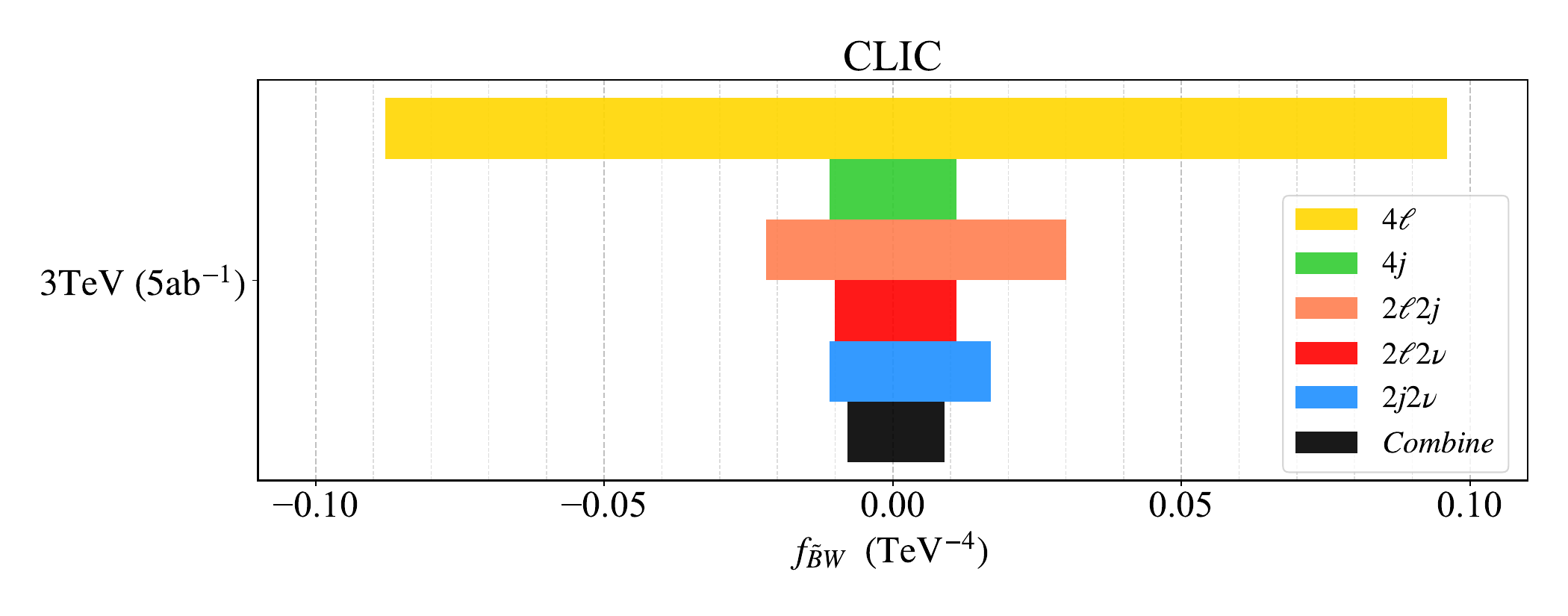}
\caption{Constraints on $f_{\tilde{B}W}$ ($\rm{TeV^{-4}}$) for different colliders with combined results at 95\% CL level.}
\label{fig:combin}
\end{figure}

We further made comparisons between the LHC constraints on $\mathcal{O}_{\tilde{B}W}$ at $\sqrt{s} = 13$ TeV~\cite{ATLAS:2018nci} and our predictions.
The Observed 95\% CL limits on the $f_{\tilde{B}W}$ is $(-1.1, 1.1)$ via measurement of the $Z\gamma\rightarrow\nu\bar{\nu}\gamma$ production.
We found that with an appropriate event selection strategy, the $e^+e^- \to ZZ$ process signal with decays at the 250 GeV CEPC is slightly less sensitive to nTGCs compared to measurements at the LHC.
However, as the CEPC collision energy increases to 360 GeV, its detection capability for nTGCs is anticipated to surpass that of the LHC.
The expected constraints on nTGCs by the ILC of $\sqrt{s} = 1$ TeV with $\mathcal{L} = 1\ {\rm ab}^{-1}$ can be improved by an order of magnitude over the LHC of 13 TeV; and that for the CLIC with $\sqrt{s} = 3$ TeV and $\mathcal{L} = 5 \ {\rm ab}^{-1}$, the expected constraints at $\mathcal{S}_{stat} = 2$ are about two orders of magnitude stronger than those at the $\sqrt{s} = 13$ TeV LHC at 95\% C.L.
We anticipate that machine learning techniques can further enhance the sensitivity of nTGC measurements.

\begin{table}[H]
    \centering
    \caption{The combined constraints on $f_{\tilde{B}W}$ ($\rm{TeV^{-4}}$) for five signal channels of $ZZ$ production.}
    \label{tab:Combined Result}
    \begin{tabular}{llllll}
    \hline
    \multirow{2}{*}{$S_{stat}$}&$\sqrt{s}$ (GeV)&\\ \cline{2-6}
     ~&250&360&500&1000&3000\\ \hline
     2&[$-2.43,2.64$]&[$-0.90,0.96$]&[$-0.52,0.61$]&[$-0.14,0.36$]&[$-0.0079,0.0093$]\\
     3&[$-3.57,4.02$]&[$-1.31,1.42$]&[$-0.75,0.92$]&[$-0.19,0.47$]&[$-0.0099,0.0112$]\\
     5&[$-5.69,6.88$]&[$-2.05,2.28$]&[$-1.14,1.49$]&[$-0.28,0.59$]&[$-0.0131,0.0145$]\\ \hline
    \end{tabular}
\end{table}

\section{\label{sec4}Conclusions}

Precision measurements of the self-couplings of electroweak gauge bosons are essential for testing the EWSB mechanism and for probing potential NP beyond the SM. nTGCs represent a possible class of anomalous multi-boson couplings within the framework of the SMEFT. Unlike the more commonly studied dimension-6 TGCs in SMEFT, nTGCs arise first from gauge-invariant dimension-8 operators.
nTGCs can contribute to the process $e^+e^- \to ZZ$ through the $ZZZ$ and $ZZ\gamma$ vertices. The potential contributions of nTGCs to this process can be investigated at future $e^+e^-$ colliders, such as CEPC, ILC and CLIC.
In this paper, we have demonstrated that these future $e^+e^-$ colliders possess the capability to measure nTGCs through $ZZ$ pair production.

We first study the influence of initial beam polarization on the measurement of $e^+e^- \to ZZ$. Specific polarization settings can enhance the cross section at an $e^+e^-$ collider. For instance, the cross section $\sigma_{+-,+-}$ of nTGCs can be twice that of the unpolarized cross section. In the case of SM contributions, the polarized cross sections $\sigma_{-+}$ and $\sigma_{+-}$ are approximately 2.5 and 1.5 times larger, respectively, than the unpolarized cross section. Consequently, we adopt an electron polarization of $+0.8$ and a positron polarization of $-0.8$ in our simulations and analyses for future $e^+e^-$ colliders.

Additionally, we present a kinematic analysis of the sensitivity of measurements of five distinct final states of $ZZ$ production induced by nTGCs at future $e^+e^-$ colliders.
To this end, we study the kinematic features of the signal and background events using Monte Carlo simulations, and propose an event selection strategy for nTGCs.
The expected constraints on the coefficients of dimension-8 operators are calculated by combining results from future $e^+e^-$ colliders with c.m. energies of 250 GeV, 360 GeV, 500 GeV, 1 TeV and 3 TeV, characteristic of CEPC, ILC, and CLIC.
As expected, higher collision energies lead to greater sensitivities.
The greatest sensitivity to $ZZ$ production is generally provided by the $ZZ \to jj \nu\bar{\nu}$ decay channel, while combining results from multiple channels further enhances the sensitivity to nTGC operators.

We also performed quantitative comparisons between the existing limits on $\mathcal{O}_{\tilde{B}W}$ from the 13 TeV LHC and our predictions.
The results indicate that the sensitivity of CEPC in detecting nTGCs is comparable to that of the LHC. Furthermore, when the collision energy increases from 250 GeV to 360 GeV, the detection capability of CEPC for nTGCs improves by approximately threefold.
On the other hand, higher-energy $e^+e^-$ colliders ($\sqrt{s} = 1 \sim 3$ TeV) are expected to achieve greater sensitivities than the LHC.
Measurements of nTGCs at the ILC and CLIC have the potential to improve current constraints on dimension-8 operators by one to two orders of magnitude. These findings further illustrate the potential for exploring and constraining dimension-8 operators within the SMEFT, complementing existing searches for aQGCs~\cite{Guo:2019agy,Guo:2020lim,Jiang:2021ytz,Yang:2021pcf,Yang:2021ukg,Yang:2022fhw,Yang:2020rjt,Zhang:2023ykh,Dong:2023nir,Zhang:2023yfg} and gQGCs~\cite{Yang:2023gos} discussed elsewhere.

\acknowledgments

This work was supported in part by the National Natural Science Foundation of China under Grants Nos. 11905093 and 12147214, the Natural Science Foundation of the Liaoning Scientific Committee No.~JYTMS20231053 and LJKZ0978.

\appendix
\section{\label{ap1}Helicity amplitudes}

The helicity amplitudes for the $e^+e^- \to ZZ$ process are given below. Here, $\theta$ and $\phi$ represent the zenith and azimuthal angles of one of the $Z$ bosons, defined in the rest frame of the $e^+e^-$ system, where the electron is moving along the $z$-axis.

\begin{equation}
\label{eq:Factor}
\mathcal{A}=\frac{1}{\sqrt{2} c_W^2 s_W^2 \left(s \left(4 M_Z^2-s\right) \cos ^2(\theta)+\left(s-2 M_Z^2\right)^2\right)},
\end{equation}
\begin{equation}
\label{eq:++00}
\mathcal{M}_{+,+,0,0}=16 \sqrt{2}\mathcal{A}\pi \alpha_{EW}  s M_Z^2 e^{i \phi} s_W^4 \sin (2 \theta),
\end{equation}
\begin{equation}
\label{eq:++0pm}
\begin{split}
&\mathcal{M}_{+,+,0,\pm}=\mp \mathcal{A}M_Z e^{i \phi} \sqrt{s} s_W^2 (\cos(\theta) \pm 1)\left(\mp s_W \left(2 M_Z^2-s\right) \left(\pm c_W^3 f_{\tilde{B}W} \left(8 M_Z^4-6 M_Z^2 s+s^2\right)\right.\right.\\\
&\quad\quad\quad\quad\left.\left.+16\pi \alpha_{EW} s_W\right)+ s_W \cos(\theta) s \left(c_W^3 f_{\tilde{B}W} \left(s-4 M_Z^2\right)^2 \cos(\theta)+16\pi \alpha_{EW} s_W\right)\right),
\end{split}
\end{equation}
\begin{equation}
\label{eq:+++0}
\begin{split}
&\mathcal{M}_{+,+,\pm,0}=\mathcal{A}M_Z e^{i \phi} \sqrt{s} s_W^2 (\cos (\theta)\mp1)\left(s_W \left(2 M_Z^2-s\right) \left(c_W^3 f_{\tilde{B}W} \left(8 M_Z^4-6 M_Z^2 s+s^2\right)\right.\right.\\
&\quad\quad\quad\quad\left.\left.+16\pi \alpha_{EW}^2 s_W\right)+ s_W \cos (\theta) s \left(c_W^3 f_{\tilde{B}W} \left(s-4 M_Z^2\right)^2 \cos (\theta)\mp16\pi \alpha_{EW}^2 s_W\right)\right),
\end{split}
\end{equation}
\begin{equation}
\label{eq:++pmpm}
\begin{split}
&\mathcal{M}_{+,+,\pm,\pm}=-8 \sqrt{2}\mathcal{A} \pi \alpha_{EW}^2 s M_Z^2 e^{i \phi} s_W^4 \sin (2 \theta),
\end{split}
\end{equation}
\begin{equation}
\label{eq:+++-}
\begin{split}
&\mathcal{M}_{+,+,+,-}=-16 \sqrt{2}\mathcal{A} \pi \alpha_{EW}^2 s e^{i \phi} s_W^4 \left(s-2 M_Z^2\right) \sin ^2\left(\frac{\theta}{2}\right) \sin (\theta),
\end{split}
\end{equation}
\begin{equation}
\label{eq:++-+}
\begin{split}
&\mathcal{M}_{+,+,-,+}=4\sqrt{2}\mathcal{A} \pi \alpha_{EW}^2 s e^{i \phi} s_W^4 \left(s-2 M_Z^2\right) \sin ^3(\theta) \csc ^2\left(\frac{\theta}{2}\right),
\end{split}
\end{equation}
\begin{equation}
\label{eq:--00}
\begin{split}
&\mathcal{M}_{-,-,0,0}=-4\sqrt{2}\mathcal{A} \left(1-2 c_W^2\right)^2 \pi \alpha_{EW}^2 s M_Z^2 e^{-i \phi} \sin (2 \theta),
\end{split}
\end{equation}
\begin{equation}
\label{eq:--0pm}
\begin{split}
&\mathcal{M}_{-,-,0,\pm}=\mathcal{A} M_Z e^{-i \phi} \sqrt{s} (\cos (\theta)\mp 1)\left(\left(2 M_Z^2-s\right)\left(-\left(1-2 c_W^2\right)^2 4\pi \alpha_{EW}^2\right.\right.\\
&\quad\quad\quad\quad\left.+c_W^3 f_{\tilde{B}W} s_W^3 \left(8 M_Z^4-6 M_Z^2 s+s^2\right)\right)+s\cos (\theta) \left(\pm\left(1-2 c_W^2\right)^2 4\pi \alpha_{EW}^2\right.\\
&\quad\quad\quad\quad\left.\left.+c_W^3 f_{\tilde{B}W} s_W^3 \left(s-4 M_Z^2\right)^2 \cos (\theta)\right)\right),
\end{split}
\end{equation}
\begin{equation}
\label{eq:--+0}
\begin{split}
&\mathcal{M}_{-,-,\pm,0}=\pm \mathcal{A} M_Z e^{-i \phi} \sqrt{s} (\cos (\theta) \pm 1)\left(\left(2 M_Z^2-s\right)\left(\pm\left(1-2 c_W^2\right)^2 4\pi \alpha_{EW}^2\right.\right.\\
&\quad\quad\quad\quad\left.\mp c_W^3 f_{\tilde{B}W} s_W^3 \left(8 M_Z^4-6 M_Z^2 s+s^2\right)\right)+s\cos (\theta) \left(\left(1-2 c_W^2\right)^2 4\pi \alpha_{EW}^2\right.\\
&\quad\quad\quad\quad\left.\left.\mp c_W^3 f_{\tilde{B}W}s_W^3 \left(s-4 M_Z^2\right)^2 \cos (\theta)\right)\right),
\end{split}
\end{equation}
\begin{equation}
\label{eq:--pmpm}
\begin{split}
&\mathcal{M}_{-,-,\pm,\pm}=4\sqrt{2}\mathcal{A} \left(1-2 c_W^2\right)^2 \pi \alpha_{EW}^2 s M_Z^2 e^{-i \phi} \sin (\theta) \cos (\theta),
\end{split}
\end{equation}
\begin{equation}
\label{eq:--+-}
\begin{split}
&\mathcal{M}_{-,-,+,-}=-2 \sqrt{2}\mathcal{A} e^{-i \phi} s \left(4\pi \alpha_{EW}-8 c_W^2 \pi \alpha_{EW}\right)^2 \left(s-2 M_Z^2\right) \sin \left(\frac{\theta}{2}\right) \cos ^3\left(\frac{\theta}{2}\right),
\end{split}
\end{equation}
\begin{equation}
\label{eq:---+}
\begin{split}
&\mathcal{M}_{-,-,-,+}=4\sqrt{2}\mathcal{A} \left(1-2 c_W^2\right)^2 \pi \alpha_{EW}^2 e^{-i \phi} s \left(s-2 M_Z^2\right) \sin ^2\left(\frac{\theta}{2}\right) \sin (\theta).\\
\end{split}
\end{equation}


\bibliography{eeZZ}
\bibliographystyle{JHEP}


\end{document}